\documentclass[
 reprint,
superscriptaddress,
nofootinbib,
 amsmath,amssymb,
 aps,
prd,
floatfix,
longbibliography
]{revtex4-2}

\usepackage{orcidlink}
\usepackage[english]{babel}
\usepackage{xcolor}
\usepackage{physics}
\usepackage{cancel}
\usepackage[normalem]{ulem}
\usepackage{graphicx}
\usepackage{dcolumn}
\usepackage{bm}
\usepackage[separate-uncertainty=true,multi-part-units=single]{siunitx}
\usepackage{hyperref}
\hypersetup{
     colorlinks=true,
     breaklinks=true,
     linkcolor=blue,
     filecolor=blue,
     citecolor=blue,      
     urlcolor=cyan,
     pdfencoding=auto
     }
\usepackage[caption=false]{subfig}

\newcommand{\Neff}{N_{\text{eff}} }

\newcommand{\dr}{ \mathrm{dr} }
\newcommand{\sync}{ {(s)} }
\newcommand{\newt}{ {(n)} }

\newcommand{\Gammadcdm}{ \Gamma_{\mathrm{dcdm}}  }
\newcommand{\fdcdm}{ f_{\mathrm{dcdm}}  }
\newcommand{\lgGamma}{ \log_{10}(\Gamma_Y/ \unit{\per\giga\year})  }
\newcommand{\Gammay}[1]{10^{#1}\unit{\per\giga\year} }
\newcommand{\MRrec}{ (\rho_m/\rho_r)_{\mathrm{rec}}  }

\DeclareSIUnit[quantity-product = {}]\parsec{\text{pc}}
\DeclareSIUnit[quantity-product = {}]\year{\text{yr}}

\graphicspath{{./}{figures/}}

\begin{document}

\title{Comprehensive Constraints on Dark Radiation Injection After BBN}

\author{Alexander C. Sobotka\,\orcidlink{0000-0002-7576-5417}}
\email{asobotka@live.unc.edu}
\affiliation{Department of Physics and Astronomy, University of North Carolina at Chapel Hill,\\
Phillips Hall CB3255, Chapel Hill, North Carolina 27599, USA }

\author{Adrienne L. Erickcek\,\orcidlink{0000-0002-0901-3591}}
\email{erickcek@physics.unc.edu}
\affiliation{Department of Physics and Astronomy, University of North Carolina at Chapel Hill,\\
Phillips Hall CB3255, Chapel Hill, North Carolina 27599, USA }

\author{Tristan L. Smith\,\orcidlink{0000-0003-2685-5405}}
\email{tsmith2@swarthmore.edu}
\affiliation{Department of Physics and Astronomy, Swarthmore College, Swarthmore,\\ Pennsylvania 19081, USA}


\begin{abstract}
We derive constraints on the injection of free-streaming dark radiation after big bang nucleosynthesis (BBN) by considering the decay of a massive hidden sector particle into dark radiation. Such a scenario has the potential to alleviate the Hubble tension by introducing a new energy component to the evolution of the early Universe. We employ observations of the cosmic microwave background (CMB) from \textit{Planck} 2018 and the South Pole Telescope (SPT-3G), measurements of the primordial deuterium abundance, Pantheon+ Type Ia supernovae data, and baryon acoustic oscillation (BAO) measurements from BOSS DR12 to constrain these decay scenarios. Pre-recombination decays are primarily restricted by observations of the CMB via their impact on the effective number of relativistic species. On the other hand, long-lived decay scenarios in which the massive particle lifetime extends past recombination tend to decrease the late-time matter density inferred from the CMB and are thus subject to constraints from Pantheon+ and BAO. We find that, when marginalizing over lifetimes of $\tau_Y = [10^{-12.08}, 10^{-1.49}]$ Gyr, the decaying particle is limited at $2\sigma$ to only contribute a maximum of 3\% of the energy density of the Universe. With limits on these decays being so stringent, neither short-lived nor long-lived scenarios are successful at substantially mitigating the Hubble tension.  
\end{abstract}
\maketitle

\section{Introduction} \label{sec:Introduction}
Measurements of cosmological parameters from observations of the cosmic microwave background (CMB) \cite{planck_collaboration_planck_2020-1}, baryon acoustic oscillations (BAO) \cite{alam_clustering_2017, eboss_collaboration_completed_2021-1}, and weak lensing \cite{heymans_kids-1000_2021,des_collaboration_dark_2022} have become significantly more precise with time. However, this increased precision has introduced apparent discrepancies between local measurements of the Hubble constant ($H_0$) \cite{yuan_consistent_2019, wong_h0licow_2020, blakeslee_hubble_2021, riess_comprehensive_2022} and matter fluctuation parameter ($S_8$) \cite{heymans_kids-1000_2021,des_collaboration_dark_2022,survey_y3_2023} and the values predicted by the standard $\Lambda$CDM model based on observations of the CMB \cite{planck_collaboration_planck_2020-1}. Many systematic uncertainties have been ruled out as the cause for these tensions \cite{efstathiou_h0_2014, addison_quantifying_2016, planck_collaboration_planck_2017,aylor_sounds_2019,shanks_gaia_2019,soltis_percent-level_2019,riess_crowded_2023}, which persist across multiple probes, potentially hinting at the need for a new cosmological model. 

Since the tension in $H_0$ can be recast as a tension in the measurement of the sound horizon at recombination \cite{aylor_sounds_2019}, it is common for extensions of $\Lambda$CDM to introduce new contributions to the energy density of the Universe prior to recombination, thereby reducing the size of the sound horizon (e.g.\ \cite{karwal_early_2016,poulin_early_2019,smith_oscillating_2020,murgia_early_2021}). One of the simplest extensions to $\Lambda$CDM that reduces the size of the sound horizon is the addition of free-streaming massless relics that alter the effective number of relativistic species, $\Neff$ \cite{bernal_trouble_2016, poulin_early_2019, di_valentino_cosmological_2020}. However, scenarios with a non-standard $\Neff$ during big bang nucleosynthesis (BBN) alter the abundance of primordial elements by changing the expansion rate during BBN and are therefore limited by measurements of primordial abundances of deuterium and helium \cite{simha_constraining_2008, hufnagel_bbn_2018-1, berlin_dark_2019, an_reconstructing_2023}. Augmenting the radiation energy density \textit{after} BBN (e.g\ \cite{aloni_dark_2023-1}) can bypass these limits, but there still exist stringent constraints on $\Neff$ from the CMB alone \cite{bashinsky_signatures_2004,hou_how_2013,follin_first_2015,planck_collaboration_planck_2020-1}. Scenarios that alter $\Neff$ after recombination by transferring energy from matter to radiation are also capable of increasing the value of $H_0$ inferred from the CMB by changing the angular diameter distance to the CMB \cite{vattis_late_2019, clark_cosmological_2021, mccarthy_converting_2023}.

In this work, we consider a massive hidden sector particle, which we call the $Y$ particle, that decays solely into dark radiation (DR) after BBN. We place bounds on the particle decay rate ($\Gamma_Y$) as well as the maximum contribution that the particle makes to the energy density of the Universe. As in \textcite{sobotka_was_2023-1}, we assume the hidden sector to be sufficiently cold such that the $Y$ particle is non-relativistic throughout BBN. The injected DR is assumed to be relativistic and free-streaming.

We only investigate decays into DR because scenarios that include decays into photons are significantly constrained \cite{sobotka_was_2023-1}. Photon injection after BBN is constrained by its influence on primordial light-element abundances, with the most stringent bounds being placed on injected photons with high enough energies to photodisintegrate primordial deuterium \cite{poulin_non-universal_2015,poulin_cosmological_2017,kawasaki_revisiting_2018,hufnagel_bbn_2018,forestell_limits_2019,kawasaki_big-bang_2020,balazs_cosmological_2022}. Even for injected photons that do not have enough energy to photodisintegrate deuterium, measurements of the abundance of deuterium still provide strict constraints via changes in the baryon-to-photon ratio during BBN \cite{sobotka_was_2023-1}. Furthermore, the injection of new photons after a redshift of $z \sim 10^6$ results in spectral distortions of the CMB energy spectrum, so bounds on spectral distortions from the COBE satellite \cite{mather_measurement_1994, fixsen_cosmic_1996} or future measurements of the CMB spectrum \cite{pixie} can be used to place limits on the time of photon injection \cite{chluba_greens_2015,bolliet_spectral_2021,chluba_new_2021}. In \textcite{sobotka_was_2023-1}, we found that COBE constraints on spectral distortions required the $Y$ particle lifetime to be less than $\sim0.032\, \unit{\year}$ if more than $\sim66\%$ of the $Y$ particle's energy was transferred to photons during its decay. Scenarios that do not inject photons are not subject to these constraints from BBN or spectral distortions and thus it is possible for the $Y$ particle to decay much later, even after recombination. 

$Y$ particle decays into DR affect the CMB anisotropies and predicted abundance of primordial elements. The primary effect that an injection of DR prior to recombination has on the CMB anisotropies is via changes in $\Neff$. Additionally, even though this decay scenario does not directly alter the baryon-to-photon ratio at BBN, CMB observations favor changes in the baryon energy density in the context of a $Y$ decay, ultimately changing the predicted abundance of primordial elements. Furthermore, the presence of the hidden sector particle slightly increases the expansion rate during BBN and thereby alters the predicted primordial abundance of elements. Finally, scenarios in which the $Y$ particle lifetime extends past the time of recombination have the potential to simultaneously increase the value of $H_0$ and decrease the cold dark matter energy density content that is preferred by CMB anisotropy data, ultimately decreasing the relative contribution of matter to the total present day energy density ($\Omega_m$) inferred by the CMB.  

The constraints derived in this work improve upon the limits derived in decaying cold dark matter (DCDM) models (e.g.\ \cite{ichiki_wmap_2004,audren_strongest_2014,poulin_fresh_2016,xiao_fractional_2020,nygaard_updated_2021-1,alvi_you_2022-1,bucko_constraining_2022}), where dark matter is composed of a stable component and a component that decays into DR. We improve upon the constraints derived in these studies with the inclusion of data from the third generation South Pole Telescope 2018 (SPT-3G) \cite{chown_maps_2018,dutcher_measurements_2021,balkenhol_measurement_2022}, Pantheon+ Type Ia supernovae measurements \cite{scolnic_pantheon_2022}, and bounds on the abundance of deuterium. More importantly, the parametrization that is commonly used in DCDM studies results in constraints that depend on the choice of prior for the particle lifetime when employing a Markov-Chain Monte Carlo (MCMC) analysis. We choose to parametrize the amount of the decaying species with the quantity $R_\Gamma$, which is a measure of the $Y$ particle energy density compared to all other species at the time when $\Gamma_Y$ equals the expansion rate. With this parametrization, derived bounds on $R_\Gamma$ are applicable to all short-lived decay scenarios and are not prior-dependent. Therefore, the $R_\Gamma$ parametrization employed in this work serves as a robust guide to how much DR can be injected prior to recombination.

Additionally, we derive adiabatic initial conditions for the DR perturbations in two common gauges: conformal Newtonian and synchronous gauge \cite{ma_cosmological_1995}. Since the DR is sourced by the decay of the $Y$ particle, the DR energy density initially scales as $a^{-1}$, where $a$ is the scale factor. Therefore, a naive application of adiabaticity would lead one to assume that $\delta_\dr = (1/4) \delta_\gamma$ in both conformal Newtonian and synchronous gauge, where $\delta_\dr$ and $\delta_\gamma$ are the fractional density perturbations in DR and photons, respectively. While this is indeed true in conformal Newtonian gauge, we find that $\delta_\dr$ is not initially equal to  $(1/4) \delta_\gamma$ in synchronous gauge; there exists an attractor solution in synchronous gauge that sets $\delta_\dr = (17/20)\delta_\gamma$. This attractor solution quickly rectifies any incorrect initial condition and so other DCDM studies that neglected to set the correct initial condition for DR in synchronous gauge are still valid. Nevertheless, we include an in-depth discussion of these DR initial conditions and provide a generalized approach that demonstrates the presence of a non-intuitive attractor solution in synchronous gauge if there is energy transfer between two species. 

This paper is organized as follows. In Sec.~\ref{sec:Decaying_Particle_Model} we describe the model and parametrization of the decaying $Y$ particle scenario, with Sec.~\ref{sec:perturbations} containing a discussion of the subtleties of the initial conditions for DR perturbations. In Sec.~\ref{sec:Effects_of_Y_Decay}, we explore the primary effects that the decay has on the CMB temperature anisotropies and primordial abundances. We motivate the choice of likelihoods and priors for our MCMC analysis in Sec.~\ref{sec:Analysis_Method}, and Sec.~\ref{sec:Results} presents the results. A summary can be found in Sec.~\ref{sec:Summary_and_Conclusions}, and we include an Appendix that contains details of our model implementation in \textsc{CLASS} \cite{blas_cosmic_2011} (Appendix \ref{sec:appendix_parametrization}), a determination of the initial condition for DR perturbations in synchronous gauge (Appendix \ref{sec:appendix_boltzmann_perturbations}), as well as a derivation of generalized adiabatic initial conditions for two cases: non-interacting fluids (Appendix \ref{sec:appendix_k2_noninteracting}) and fluids that exchange energy via a decay (Appendix \ref{sec:appendix_k2_interacting}). The Appendix also includes calculations for $\Delta \Neff$ resulting from the injection of DR (Appendix \ref{sec:appendix_Neff}). 

\section{Decaying Particle Model} \label{sec:Decaying_Particle_Model}

\subsection{Parametrization}
The equations describing the evolution of the energy density of the $Y$ particle ($\rho_Y$) and that of the injected dark radiation ($\rho_{\dr}$) are
\begin{align}
    \dot{\rho}_Y + 3H\rho_Y &= -\Gamma_Y \rho_Y, \label{eq:rho_Y} \\
    \dot{\rho}_{\dr} + 4H\rho_{\dr} &= +\Gamma_Y \rho_Y, \label{eq:rho_dr}
\end{align}
where $\Gamma_Y$ is the decay rate of the $Y$ particle, $H \equiv \dot{a}/a$ with scale factor $a$, and an over-dot denotes a proper time derivative.\footnote{Note that the right-hand sides of Eqs.~\eqref{eq:rho_Y} and \eqref{eq:rho_dr} are proportional to $m_Y n_Y$, where $m_Y$ and $n_Y$ are the mass and number density of the $Y$ particle, respectively. However, $m_Y n_Y = \rho_Y$ if the $Y$ particle equation of state is $w_Y =0$ \cite{erickcek_cannibalisms_2021}. } 

The physical quantities that we wish to constrain are the $Y$ particle's decay rate and its maximum contribution to the energy density of the Universe. The maximum contribution of $\rho_Y$ serves as a proxy for the amount of DR injected as well as a measure of the impact that the $Y$ particle has on observables while acting as extra cold dark matter. We parametrize the maximum contribution that the $Y$ particle makes to the energy density of the Universe with the quantity $R_\Gamma$ defined as
\begin{equation}
    R_\Gamma \equiv \frac{\rho_{Y,i}\left(a_i/a_\Gamma \right)^3 }{ \rho_{sr}\left(a_\Gamma \right) + \rho_{m}\left(a_\Gamma \right) + \rho_{ncdm}(a_\Gamma)}, \label{eq:R_Gamma}
\end{equation}
where $\rho_{Y,i}$ is the initial energy density of the $Y$ particle, $\rho_m$ is the combined energy density of baryons and cold dark matter, $\rho_{sr}$ is the combined energy density of photons and massless neutrinos, $\rho_{ncdm}$ is the energy density of non-cold dark matter (i.e.\ massive neutrinos), and the initial scale factor, $a_i$, is assumed to be deep in radiation domination. We assume that neutrinos are composed of two massless species and one massive species with $m_\nu = \SI{0.06}{\electronvolt}$.
%
\begin{figure}[t]
\centering
  \includegraphics[width=1\linewidth]{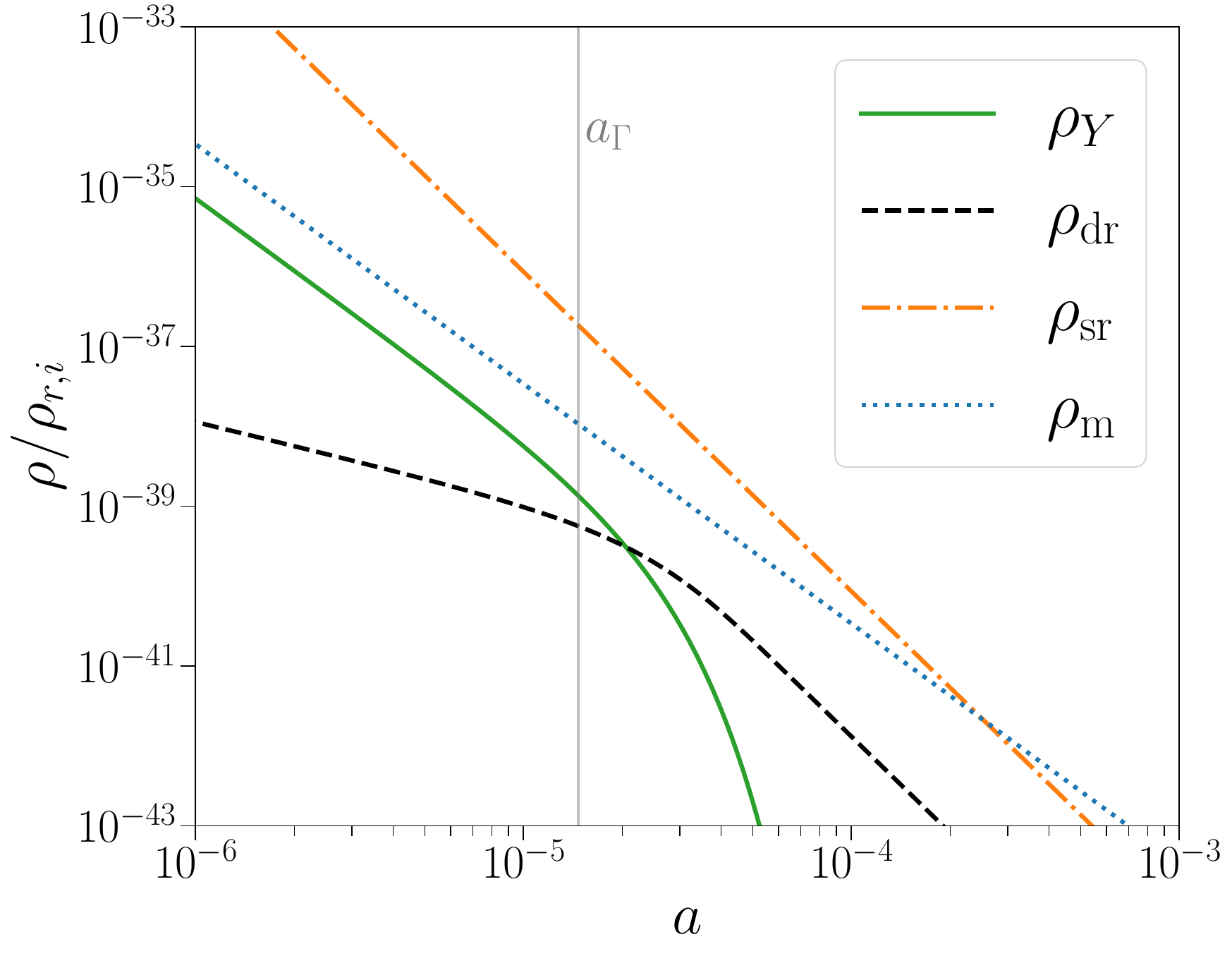}
  \caption{\footnotesize Energy densities for a decay with $\Gamma_Y = 10^{6.5}\SI{}{\per\giga\year}$ and $R_\Gamma = 0.01$. Here, $\rho_Y$ is the energy density of the $Y$ particle, $\rho_\dr$ is that of the DR, $\rho_{sr}$ is the combined energy density of photons and massless neutrinos, and $\rho_m$ is the combined energy density of baryons and cold dark matter. The $Y$ particle energy density scales as $\rho_Y \propto a^{-3}$ until around $a \approx 10^{-5}$, at which point the decay rate overcomes the expansion rate. The DR energy density initially scales as $\rho_{\dr} \propto a^{-1}$ as it is sourced by the $Y$ particle decay and then eventually transitions to scaling as $a^{-4}$. The vertical line marks $a_\Gamma$ defined by Eq.~\eqref{eq:aGamma_defintion}. }
  \label{fig:example_evolution}
\end{figure}
%
$R_\Gamma$ approximately describes the maximum ratio of $\rho_Y$ to the total energy density, $\rho_Y/\rho_{tot}$, if $a_\Gamma$ is chosen such that $\Gamma_Y \simeq H(a_\Gamma)$. Therefore, we define $a_\Gamma$ via the relation
\begin{equation}
    \Gamma_Y = H_i\sqrt{ \frac{ \rho_{Y,i}\left(\frac{a_i}{a_\Gamma} \right)^3 + \rho_{m}(a_\Gamma) + \rho_{sr}(a_\Gamma) + \rho_{ncdm}(a_\Gamma)   }{ \rho_{r,i}    }  }, \label{eq:aGamma_defintion}
\end{equation}
where $H_i \equiv H(a_i)$ and $\rho_{r,i} = \rho_{sr,i} + \rho_{ncdm,i}$ since massive neutrinos are initially relativistic. We note that Eq.~\eqref{eq:aGamma_defintion} is not an exact statement of $\Gamma_Y = H(a_\Gamma)$ because $\rho_Y$ does not evolve as $a^{-3}$ up to $a_\Gamma$ and we do not include contributions from $\rho_\dr$ in the computation of $H(a_\Gamma)$. However, for the level of $\rho_Y$ contributions that we consider in this work, Eq.~\eqref{eq:aGamma_defintion} provides a value for $a_\Gamma$ such that $\Gamma_Y \approx H(a_\Gamma)$ without numerically solving for the evolution of $\rho_Y$. 

We modify the Cosmic
Linear Anisotropy Solving System (\textsc{CLASS}) \cite{blas_cosmic_2011} public Boltzmann solver to solve Eqs.~\eqref{eq:rho_Y} and \eqref{eq:rho_dr} when given initial values for $\rho_Y$ and $\rho_{\dr}$. Determining the initial densities that generate decay scenarios with specified $R_\Gamma$ and $\Gamma_Y$ values could be accomplished with a shooting algorithm but, in the interest of simplicity, we choose to derive an analytic model that maps $R_\Gamma$ and $\Gamma_Y$ to initial conditions for $\rho_Y$ and $\rho_{\dr}$. This analytic approach is described in Appendix \ref{sec:appendix_parametrization}, and an example of the resulting evolution of $\rho_Y$ and $\rho_{\dr}$ is provided in Fig.~\ref{fig:example_evolution}. Initially, $\rho_Y$ mimics the evolution of standard non-relativistic matter ($\rho_Y \propto a^{-3}$) until $a \approx a_\Gamma$ after which $\rho_Y$ exponentially decreases. As the DR is being sourced by the $Y$ particle decay, \mbox{$\rho_\dr \propto a^{-1}$}. However, soon after $a\approx a_\Gamma$, the $Y$ particle decay no longer sources significant DR and thus $\rho_\dr$ evolves as standard radiation, $\rho_\dr \propto a^{-4}$. As we explore in the following section, the fact that $\rho_\dr$ evolves as $a^{-1}$ even though the DR equation of state parameter is 1/3 has interesting ramifications for the initial conditions of DR perturbations in synchronous gauge.  

\subsection{Perturbations} \label{sec:perturbations}

The initial conditions that are implemented in \textsc{CLASS} for baryon, cold dark matter, photon, and neutrino perturbations are described in \textcite{ma_cosmological_1995}. Here, we derive the superhorizon initial conditions during radiation domination for the scalar perturbations of the $Y$ particle and the DR that is sourced by the decay of the $Y$ particle. Since Boltzmann codes such as \textsc{CLASS} commonly evolve perturbations in synchronous gauge, we discuss the adiabatic initial conditions for the $Y$ particle and DR in both conformal Newtonian gauge and synchronous gauge. The Friedmann–Lema\^itre–Robertson–Walker (FLRW) metric in conformal Newtonian gauge is
\begin{equation}
    ds^2 = a(\tau)^2\left[-(1 + 2\Psi)d\tau^2 + \delta_{ij}(1 - 2\Phi)dx^i dx^j\right], \label{eq:Newt_metric}
\end{equation}
where $\Phi$ and $\Psi$ are scalar perturbations to the metric and $\tau$ is conformal time. The FLRW metric in synchronous gauge is
\begin{equation}
    ds^2 = a^2(\tau)\left[-d\tau^2 + (\delta_{ij} + h_{ij})dx^i dx^j   \right],
\end{equation}
where the scalar part of the perturbation $h_{ij}$ can be expressed in Fourier space as
\begin{equation}
    h_{ij}(\vec{k}, \tau) = \hat{k}_i \hat{k}_j h(\vec{k}, \tau) + \left(\hat{k}_i \hat{k}_j - \frac13 \delta_{ij}\right) 6\eta(\vec{k}, \tau),
\end{equation}
with $\vec{k}=k\hat{k}$ and $k$ being the comoving wavenumber of a perturbation mode. Throughout this discussion, a superscript $(s)$ or $(n)$ denotes synchronous or conformal Newtonian gauge, respectively.

\begin{figure}[t]
\centering
  \includegraphics[width=1\linewidth]{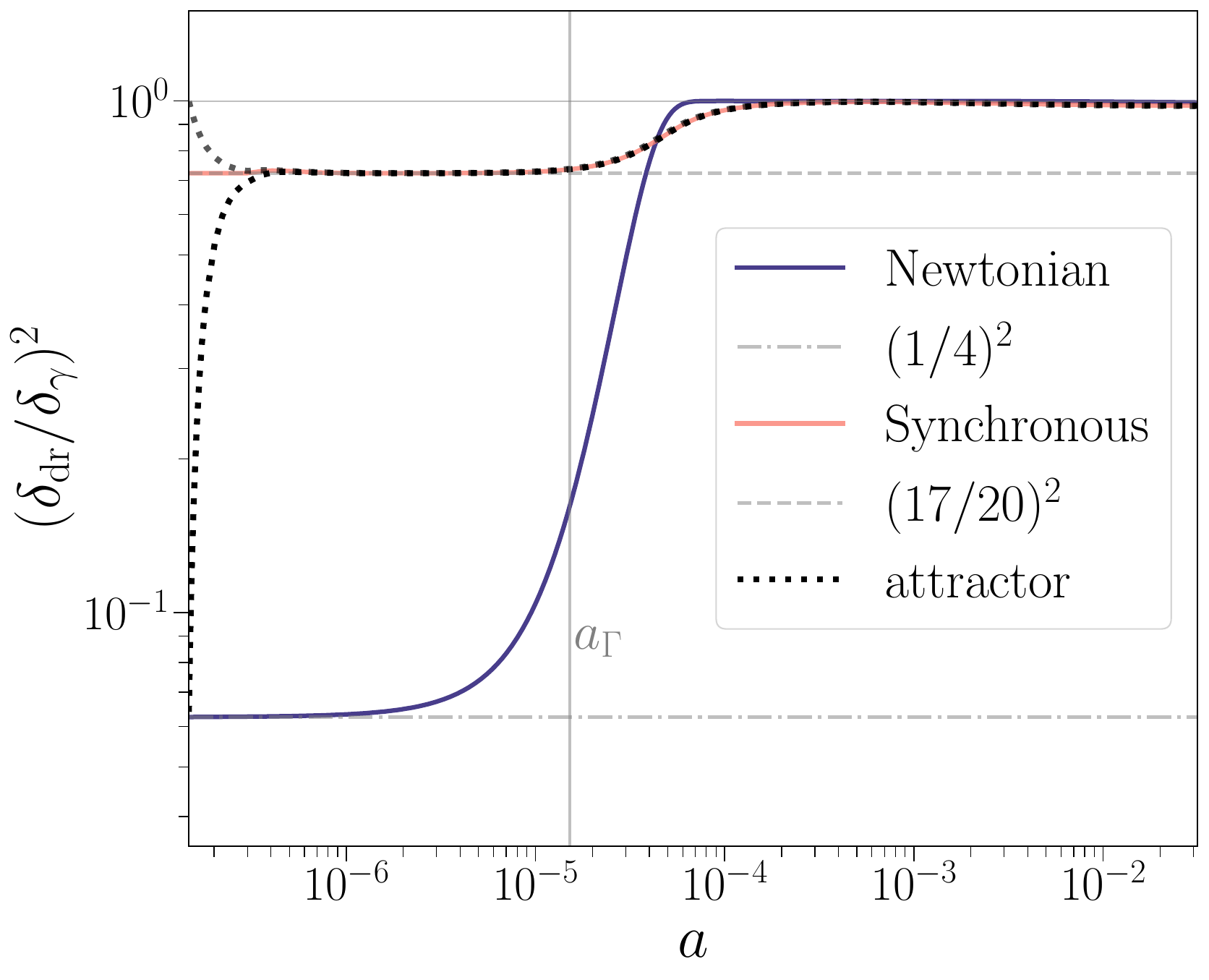}
  \caption{\footnotesize Adiabatic initial conditions for a $k=\SI{e-3}{\per\mega\parsec}$ perturbation mode and a decay with $\Gamma_Y = 10^{6.49}\SI{}{\per\giga\year}$ and $R_\Gamma = 0.1$. In Newtonian gauge, adiabaticity requires that $\delta^\newt_{\dr} = (1/4) \delta^\newt_\gamma$ initially. However, in synchronous gauge there is an attractor solution of $\delta^\sync_{\dr} = (17/20)\delta^\sync_\gamma$. This attractor quickly fixes any incorrect initial condition for $\delta^\sync_\dr$ (dotted lines). In both gauges, $\delta_{\dr} = \delta_\gamma$ once the DR is no longer being sourced and $\rho_\dr \propto a^{-4} $.}
  \label{fig:perturbations_initial_conditions}
\end{figure}

The DCDM module that is included in the public distribution of \textsc{CLASS} sets initial conditions for the fractional density perturbation of DCDM and DR in synchronous gauge as $\delta^\sync_{\mathrm{dcdm}} = (3/4)\delta^\sync_\gamma$ and $\delta^\sync_\dr = \delta^\sync_\gamma$, respectively, where $\delta^\sync_\gamma$ is the fractional density perturbation of photons. While this condition for $\delta^\sync_{\mathrm{dcdm}}$ is correct, there exists a different attractor solution for $\delta^\sync_\dr$ on superhorizon scales; the energy transfer to $\rho_\dr$ quickly pushes $\delta^\sync_\dr$ to $(17/20)\delta^\sync_\gamma$. This attractor promptly amends any incorrect initial condition for $\delta_\dr$ and therefore the incorrect default condition of $\delta^\sync_\dr = \delta^\sync_\gamma$ set by \textsc{CLASS} has no consequence (see the dotted lines in Fig.~\ref{fig:perturbations_initial_conditions}). For this work, we use $\delta_\dr^\sync = (17/20)\delta^\sync_\gamma$ as the initial condition in CLASS, and we enforce that perturbations are initialized at some initial scale factor, $a_i$, such that $\Gamma_Y/H(a_i) \leq 10^{-4}$. The factor of $\delta_\dr^\sync/\delta_\gamma^\sync = 17/20$ can be derived directly from the Boltzmann equations for DR in synchronous gauge (see Appendix \ref{sec:appendix_boltzmann_perturbations}).

In both synchronous and conformal Newtonian gauge, adiabaticity is commonly assumed to mean that the quantity $\delta \rho_i/\dot{\bar{\rho}}_i$ is the same between all fluids, where $\bar{\rho}_i$ is the background energy density of a fluid and $\delta \rho_i = \rho_i - \bar{\rho}_i$. For non-interacting fluids, this implies $\delta_i/\delta_j = (1+w_i)/(1+w_j)$ on superhorizon scales. For example, this condition leads to the familiar relation $\delta_{\mathrm{cdm}} = (3/4)\delta_\gamma$ in both conformal Newtonian and synchronous gauges for non-interacting cold dark matter and photons. 

Even for interacting fluids such as the DR sourced by the $Y$ decay, adiabatic conditions are defined by $\delta\rho_i/\dot{\bar{\rho}}_i$ being the same between all fluids in conformal Newtonian gauge \cite{weinberg_adiabatic_2003}. Since adiabatic initial conditions cannot source isocurvature initial conditions, the dynamics of these interacting fluids must preserve the equivalency of $\delta \rho_i/\dot{\bar{\rho}}_i$ between all fluids in conformal Newtonian gauge. However, because these fluids are interacting, $\dot{\bar{\rho}}_i/\bar{\rho_i} \neq -3H(1+w_i)$ and so the adiabatic condition does not result in the familiar relationship of $\delta_i/\delta_j = (1+w_i)/(1+w_j)$ on superhorizon scales. Solving Eq.~\eqref{eq:rho_dr} under the assumption of radiation domination and taking $\rho_Y \propto a^{-3}$, it follows that $\dot{\bar{\rho}}_\dr/\bar{\rho}_\dr = -1$ while the DR is being sourced, implying that $\delta^\newt_\dr = (1/4)\delta^\newt_\gamma$ initially even though DR and photons both have an equation of state parameter equal to 1/3. The solid dark line in Fig.~\ref{fig:perturbations_initial_conditions} depicts this: while $\rho_\dr \propto a^{-1}$, the correct adiabatic condition for DR is $\delta^\newt_\dr = (1/4)\delta^\newt_\gamma$. Once the $Y$ particle has sufficiently decayed away so that the DR is no longer being significantly sourced by the decay, $\rho_\dr \propto a^{-4}$ and adiabaticity leads to $\delta^\newt_\dr = \delta^\newt_\gamma$.

When transforming from conformal Newtonian to synchronous gauge, most of the familiar adiabatic conditions are preserved (e.g.\ $\delta^\sync_{Y} = \delta^\sync_{\mathrm{cdm}} = (3/4) \delta^\sync_\gamma$). However, this is not the case for DR: $\delta^\newt_\dr/\delta^\newt_\gamma \neq \delta^\sync_\dr/\delta^\sync_\gamma$. The underlying reason for this discrepancy is that the superhorizon limit means something different in synchronous gauge compared to conformal Newtonian gauge. Whereas $\delta^\newt$ approaches a constant nonzero value in the $k\tau \rightarrow 0$ limit, $\delta^\sync$ is proportional to $(k\tau)^2$ and vanishes in this limit \cite{ballesteros_dark_2010}. 

To illustrate why this difference in superhorizon limits between gauges leads to $\delta^\newt_\dr/\delta^\newt_\gamma \neq \delta^\sync_\dr/\delta^\sync_\gamma$, we perform a $k\tau$ expansion on the general gauge transformation between conformal Newtonian and synchronous gauge \cite{ma_cosmological_1995},
\begin{equation}
    \delta_i^\sync(k) = \delta_i^\newt(k) - \alpha(k)\frac{\bar{\rho}'_i}{\bar{\rho}_i}. \label{eq:gauge_transform}\\
\end{equation}
Here, $\alpha(k)\equiv (h' + 6\eta')/2k^2$ and a prime denotes differentiation with respect to conformal time. We expand $\delta_i^\newt(k)$ in $k\tau$ such that it is composed of a zeroth order piece, $\delta_i^{\newt,0}$, and a second order component, $C_i k^2\tau^2$. Similarly, $\alpha(k)$ is expanded such that \mbox{$\alpha(k) \approx \alpha^0 + C_\alpha k^2\tau^2 $}. Equation \eqref{eq:gauge_transform} then becomes
\begin{equation}
    \delta_i^\sync(k) \approx \left[\delta_i^{\newt,0} + C_i k^2\tau^2 \right] - \left[ \alpha^0 + C_\alpha k^2\tau^2  \right]\frac{\bar{\rho}'_i}{\bar{\rho}_i}. \\
\end{equation}
Since adiabatic initial conditions in synchronous gauge have leading order terms proportional to $(k\tau)^2$, \mbox{$\delta^{\newt,0}_i = \alpha^0 (\bar{\rho}'_i/\bar{\rho}_i)$}.\footnote{Note that this expression is another manifestation of adiabatic initial conditions in conformal Newtonian gauge: $\delta^{\newt,0}_i/\delta^{\newt,0}_j = (1+w_i)/(1+w_j)$. } It follows that
\begin{equation}
    \delta_i^\sync(k) = C_i k^2\tau^2 - C_\alpha k^2\tau^2  \frac{\bar{\rho}'_i}{\bar{\rho}_i} + \mathcal{O}(k^4\tau^4). \label{eq:delta_sync_approx}
\end{equation}
In other words, one cannot simply gauge transform the zeroth-order adiabatic solution in conformal Newtonian gauge in order to derive the correct adiabatic condition in synchronous gauge. The adiabatic solution in the superhorizon limit for synchronous gauge is the second order gauge transformation of the ($k\tau)^2$ terms in the adiabatic solution in conformal Newtonian gauge \cite{weinberg_adiabatic_2003} (i.e.\ the transformation of $C_i$). Taking the ratio of Eq.~\eqref{eq:delta_sync_approx} for two different species $i$ and $j$, and assuming $\delta^{\newt,0} = \alpha^0 (\bar{\rho}'/\bar{\rho})$, we have
\begin{equation}
    \frac{\delta^\sync_i}{ \delta^\sync_j } = 
     \frac{ \delta_i^{\newt,0}  \left[ \frac{C_i}{\delta_i^{\newt,0} }  - \frac{C_\alpha}{\alpha^0}  \right]  }{ \delta_j^{\newt,0}  \left[ \frac{C_j}{\delta_j^{\newt,0} } - \frac{C_\alpha}{\alpha^0}  \right]  } + \mathcal{O}(k^4\tau^4). \label{eq:sync_delta_ratio_approx}
\end{equation}
Therefore, if the coefficient of the $(k\tau)^2$ term in the adiabatic initial condition for conformal Newtonian gauge is such that $C_i/C_j \neq \delta^{\newt,0}_i/\delta^{\newt,0}_j$, then the ratio of initial conditions for $\delta$ in synchronous gauge will not be equal to the same ratio in conformal Newtonian gauge. 

In the case of non-interacting fluids, we show below in \textit{case 1} that the usual adiabatic condition of $\delta_i/\delta_j = (1+w_i)/(1+w_j)$ is preserved even to second order in $k\tau$ in conformal Newtonian gauge, which means the same condition is true in synchronous gauge: $\delta^\sync_i/\delta^\sync_j = (1+w_i)/(1+w_j) $. 

On the other hand, if there is some interaction between fluids such as energy exchange via a decay, we show in \textit{case 2} that the initial condition for $\delta^\newt$ of the species that is being sourced has a $(k\tau)^2$ term such that $C_i/C_j \neq \delta^{\newt,0}_i/\delta^{\newt,0}_j$. Thus, the initial condition for this fluid in synchronous gauge does \textit{not} result in the familiar adiabatic initial condition seen in Newtonian gauge: $\delta^\sync_i/\delta^\sync_j \neq (1+w_i)/(1+w_j)$. Similar results have been found for early dark energy models with a non-adiabatic sound speed \cite{ballesteros_dark_2010}.

\subsubsection{case 1 (no energy exchange)}
Let us consider two fluids with energy densities $\rho_1$ and $\rho_2$. Additionally, there is a third fluid, $\rho_d$, which dominates the energy density of the Universe. If these three fluids are non-interacting, the evolution of their energy densities are entirely set by their respective equation of state parameters:
\begin{align}
    &\dot{\rho}_{d} + 3H(1+w_d)\rho_d = 0, \label{eq:case1_rhod} \\
     &\dot{\rho}_{1} + 3H(1+w_1)\rho_1 = 0,  \label{eq:case1_rho1} \\
      &\dot{\rho}_{2} + 3H(1+w_2)\rho_2 = 0.  \label{eq:case1_rho2} 
\end{align}
Therefore, $H(a) \propto a^{-\frac{3}{2}(1+w_d)}$. In Appendix \ref{sec:appendix_k2_noninteracting} we solve the suite of Boltzmann equations in conformal Newtonian gauge for these three fluids using an iterative approach, and determine their respective adiabatic initial conditions to second order in $k\tau$.
The resulting density perturbation for the $j$-th fluid with equation of state parameter $w_j$ is given by
\begin{align*}
    \delta^\newt_j &= 2\left(\frac{1+w_j}{1+w_d}\right)\Phi_p \\ 
    &\hspace{0.15cm}+ \frac{2}{3}\left(\frac{1+w_j}{1+w_d}\right)\left[ \frac{7 + 39w_d + 63w_d^2 + 27w_d^3 }{28 + 36w_d}   \right] (k\tau)^2 \Phi_p,\stepcounter{equation}\tag{\theequation} \label{eq:delta_k2_non-interacting}
\end{align*}
where $\Phi_p$ is the initial value of the metric perturbation $\Phi$. Equation \eqref{eq:delta_k2_non-interacting} applies to all non-interacting fluids, including the dominant fluid. Here it can be seen that there is a zeroth-order term and a term of order $(k\tau)^2$, each having a coefficient proportional to $(1+w_j)/(1+w_d)$. The quantity in square brackets is solely dependant on $w_d$ and therefore the same for fluids $\rho_1$, $\rho_2$, and $\rho_d$. The ratio of the zeroth-order component for any two species is $\delta_i^{\newt,0}/\delta_j^{\newt,0} = (1+w_i)/(1+w_j)$. Similarly, the ratio of the second order terms for two species is $C_i/C_j = (1+w_i)/(1+w_j)$. Therefore, according to Eq.~\eqref{eq:sync_delta_ratio_approx}, since $C_i/C_j = \delta_i^{\newt,0}/\delta_j^{\newt,0}$, the ratio of initial conditions in synchronous gauge will equal that of the initial conditions in conformal Newtonian gauge. In other words, $\delta^\sync_i/\delta^\sync_j = (1+w_i)/(1+w_j) + \mathcal{O}(k^4\tau^4)$ for non-interacting fluids.

\subsubsection{case 2 (exchange via decay)}
Next we will consider a similar scenario to that of \textit{case 1}, except now we model species 1 as a massive species ($w_1 = 0$) decaying into species 2 with a decay rate $\Gamma$ such that the energy densities of each fluid are set by
\begin{align}
    &\dot{\rho}_{d} + 3H(1+w_d)\rho_d = 0, \label{eq:case2_rhod}\\
     &\dot{\rho}_{1} + 3H\rho_1 = -\Gamma \rho_1, \label{eq:case2_rho1} \\
      &\dot{\rho}_{2} + 3H(1+w_2)\rho_2 = +\Gamma \rho_1 .\label{eq:case2_rho2}
\end{align}
In a similar manner to case 1, we iteratively solve the perturbations equations for these fluids in Appendix \ref{sec:appendix_k2_interacting} to find the adiabatic initial conditions in conformal Newtonian gauge up to order $(k\tau)^2$. Since the dominant species is still non-interacting, the initial condition for $\delta^\newt_d$ is given by Eq.~\eqref{eq:delta_k2_non-interacting}. Additionally, under the assumption that  initial conditions are set sufficiently early, we take $\Gamma \rho_1 \ll H \rho_1$ initially. Under this assumption, Eq.~\eqref{eq:case2_rho1} reduces to Eq.~\eqref{eq:case1_rho1} for $w_1 = 0$ and therefore $\delta^\newt_1$ is described by Eq.~\eqref{eq:delta_k2_non-interacting} as well. Therefore, even though species 1 is decaying and exchanging energy with species 2, we find that $\delta^\newt_1/\delta^\newt_d = 1/(1+w_d)$ at second order\footnote{Note that this is equivalent to $\delta^\newt_1/\delta^\newt_d = (1+w_1)/(1+w_d)$ since $w_1 = 0$.} in $(k\tau)$.

In contrast, $\Gamma \rho_1$ is not initially negligible compared to $H \rho_2$ so the correct adiabatic initial condition for species 2 cannot be described by Eq.~\eqref{eq:delta_k2_non-interacting}. Instead, the adiabatic initial condition for species 2 is
\begin{equation}
    \delta^\newt_2 = \left(\frac{1- w_d}{1+w_d}\right)\Phi_p + \mathcal{W} (k\tau)^2 \Phi_p, \label{eq:delta2_k2_interacting}
\end{equation}
where $\mathcal{W}$ is a non-trivial constant that depends on $w_d$ and $w_2$ (see Eq.~\eqref{eq:bigW} in Appendix \ref{sec:appendix_k2_interacting}). Solving Eq.~\eqref{eq:case2_rho2} analytically shows that $\rho_2(a) \propto a^{-\frac{3}{2}(1 - w_d)}$. If we define an effective equation of state $w_{2,\mathrm{eff}} \equiv -(1/2)(w_d + 1)$ for species 2, then $\rho_2 \propto a^{-3(1+w_{2,\mathrm{eff}})}$ and it can be seen in Eq.~\eqref{eq:delta2_k2_interacting} that the zeroth-order term of $\delta^\newt_2$ is equal to $2\Phi_p(1+w_{2,\mathrm{eff}})/(1+w_d)$, as expected for adiabatic initial conditions. However, Eq.~\eqref{eq:delta2_k2_interacting} demonstrates that $\delta^\newt_2/\delta^\newt_d$ is not equal to $(1+w_{2,\mathrm{eff}})/(1+w_d)$ at second order in $k\tau$. Therefore, comparing to any non-interacting fluid $j$, we find that $\delta_2^{\newt,0}/\delta_j^{\newt,0} = (1+w_{2,\mathrm{eff}})/(1+w_j)$, but $C_2/C_j \neq \delta_2^{\newt,0}/\delta_j^{\newt,0}$. According to Eq.~\eqref{eq:sync_delta_ratio_approx}, this means that the correct adiabatic initial condition for species 2 in synchronous gauge has $\delta_2^\sync/\delta_j^\sync \neq \delta_2^\newt/\delta_j^\newt$. This discussion demonstrates why we see $\delta^\newt_Y/\delta^\newt_\gamma = \delta^\sync_Y/\delta^\sync_\gamma = 3/4$ while $\delta^\newt_\dr/\delta^\newt_\gamma =1/4$ but $\delta^\sync_\dr/\delta^\sync_\gamma = 17/20$.

\section{Effects of a \texorpdfstring{$Y$}{Y} Decay} \label{sec:Effects_of_Y_Decay}
The decaying Y particle and the injected DR influence observables such as the CMB temperature anisotropy spectrum and the abundance of primordial elements. In this section, we describe the primary effects of a $Y$ decay on such observables. Understanding these effects and how other cosmological parameters may be adjusted to mitigate them motivates our selection of observational data sets and will guide our interpretation of the MCMC results presented in  Sec.~\ref{sec:Results}.

We discuss the effects of a $Y$ decay for three different regimes: short-lived, intermediate, and long-lived. We refer to \textit{short-lived} cases as those scenarios in which the characteristic scale factor, $a_\Gamma$, is larger than the scale factor at the end of BBN but much smaller than the scale factor of matter-radiation equality. For \mbox{$R_\Gamma < 0.1$}, this corresponds to $Y$ particle lifetimes of \mbox{$10^{-12.08}\SI{}{\giga\year} \lesssim \tau_Y \lesssim 10^{-7}\SI{}{\giga\year}$}. \textit{Intermediate} cases are defined as scenarios in which $a_\Gamma$ is between the scale factor when the matter density is no longer negligible ($\rho_m/\rho_{tot} \sim 0.05$) and that of recombination. These limits correspond to \mbox{$ 10^{-7}\SI{}{\giga\year} \lesssim \tau_Y \lesssim 10^{-3.22}\SI{}{\giga\year} $}. Finally, \textit{long-lived} cases are those in which $a_\Gamma$ is greater than the scale factor at recombination, which correspond to $Y$ particle lifetimes of $\tau_Y \gtrsim 10^{-3.22}\SI{}{\giga\year}$.

\subsection{Short-lived regime}\label{sec:effects_short_lived}

When the $Y$ particles decay before recombination, the primary impact on the CMB arises from a change in the effective number of relativistic species, $\Neff$, due to the injected DR. In Appendix \ref{sec:appendix_Neff}, we determine this change to $\Neff$ as a function of both $R_\Gamma$ and $\Gamma_Y$, and find that $\Delta\Neff$ is purely a function of $R_\Gamma$ for short-lived cases ($\Gamma_Y \gtrsim 10^7 \unit{\per\giga\year}$) in which the injection of DR finishes deep in radiation domination (see Fig.~\ref{fig:Neff_appendix} in Appendix \ref{sec:appendix_Neff}). 

Therefore, bounds on $\Delta\Neff$ directly translate to limits on $R_\Gamma$ for these short-lived cases. This enables one to derive an upper bound on $R_\Gamma$ that is independent of any prior for $\Gamma_Y$ within the short-lived regime. This is in contrast to DCDM models, which are usually described by the dark matter decay rate ($\Gammadcdm$) and the fraction of the total dark matter density that is unstable ($\fdcdm$). The $\Delta\Neff$ induced by DCDM is then determined by $\Delta\Neff \propto \Gammadcdm^{-1/2}\fdcdm/(1 - \fdcdm)$ \cite{nygaard_updated_2021-1}. Therefore, constraints on $\Delta\Neff$ translate to constraints on the combination of $\Gammadcdm$ and $\fdcdm$; if $\fdcdm$ increases while $\Gammadcdm$ decreases, $\Neff$ can be kept constant. Due to this degeneracy between $\fdcdm$ and $\Gammadcdm$, derived bounds on $\fdcdm$ are entirely dependent on the prior adopted for $\Gammadcdm$. \textcite{poulin_fresh_2016} and \textcite{nygaard_updated_2021-1} only consider $10^3 \unit{\per\giga\year} < \Gammadcdm < 10^6 \unit{\per\giga\year}$ for short-lived cases, motivated by the fact that the inclusion of larger decay rates would lead to poor sampling convergence. Therefore, their reported bounds on $\fdcdm$ cannot be applied to decay rates outside of this range. Moreover, \textcite{holm_discovering_2022} employ profile likelihoods to illustrate that the degeneracy between $\fdcdm$ and $\Gammadcdm$ in DCDM models introduces volume effects when employing an MCMC analysis, resulting in the unwanted smoothing out of features in posterior distributions. Not only that, but enforcing a prior on $\Gammadcdm$ effectively places a nonphysical upper bound on $\fdcdm$, which prevents one from reporting accurate constraints on $\fdcdm$ alone. The $R_\Gamma$ parametrization used in this work bypasses these issues. 

The injection of free-streaming DR increases $\Neff$ and decreases the ratio of $\rho_m$ to $\rho_r$ at the time of recombination: $(\rho_m/\rho_r)_\mathrm{rec}$. The CMB temperature anisotropy spectrum is very sensitive to changes in $(\rho_m/\rho_r)_\mathrm{rec}$ via the early integrated Sachs-Wolfe (ISW) effect \cite{kable_deconstructing_2020,vagnozzi_consistency_2021}. Therefore, an increase in $\omega_{\mathrm{cdm}} \equiv \Omega_{\mathrm{cdm}}h^2$ is necessary to keep $(\rho_m/\rho_r)_\mathrm{rec}$ unaltered, where $\Omega_{\mathrm{cdm}} \equiv \rho_{cdm,0}/\rho_{\mathrm{crit},0}$ and $h\equiv H_0/(\SI{100}{\kilo\meter\per\second\per\mega\parsec})$. Furthermore, the addition of DR increases the pre-recombination expansion rate and thereby reduces the size of the sound horizon, $r_s$. The locations of the acoustic peaks in the CMB are set by the angular size of the sound horizon ($\theta_s$), so a decrease in the angular diameter distance ($d_A$) is required to keep $\theta_s = r_s/d_A$ fixed. A decrease in $d_A$ is partially accomplished by the enhancement in $\omega_{\mathrm{cdm}}$ that was required to fix $(\rho_m/\rho_r)_\mathrm{rec}$, while any remaining change to $d_A$ can be accomplished by increasing $H_0$. 

Once $\omega_{\mathrm{cdm}}$ and $H_0$ have been increased to keep $(\rho_m/\rho_r)_\mathrm{rec}$ and $\theta_s$ fixed, the remaining effects that a non-zero $\Delta\Neff$ has on the CMB are a change in the damping scale and a unique phase shift in the acoustic peaks \cite{bashinsky_signatures_2004, hou_how_2013, follin_first_2015, baumann_phases_2016}. Silk damping is affected because the photon diffusion length scales as $r_D \propto \sqrt{1/H}$, whereas the sound horizon scales as $r_s \propto 1/H$. Thus, enhancing $H_0$ to fix $\theta_s$ leads to an overall increase in the angular size of the diffusion length, $\theta_D$, causing extra suppression on small angular scales of the CMB \cite{hou_how_2013}. However, this suppression can be compensated for by amplifying the baryon content, $\omega_b$, or the spectral tilt, $n_s$.\footnote{ Increasing $\omega_b$ also increases the helium abundance, which would further dampen anisotropies on small scales \cite{hou_how_2013,sobotka_was_2023-1}. However, since the helium abundance is only logarithmically dependent on the baryon-to-photon ratio, increasing $\omega_b$ has the net effect of reducing small-scale damping. } While employing Planck anisotropy data will effectively enforce that $(\rho_m/\rho_r)_\mathrm{rec}$ and $\theta_s$ remain fixed, the addition of SPT-3G data further aids in constraining changes in $\omega_b$ and $n_s$ that are needed to accommodate the small-scale effects of a $\Delta\Neff$ induced by the $Y$ decay. SPT-3G data is also sensitive to the extra phase shift caused by a non-zero $\Delta\Neff$ and can help constrain this effect.

In addition to influencing the CMB, $Y$ decays in the short-lived regime have the potential to alter the abundance of primordial elements created during BBN. Even though we consider $Y$ particle lifetimes that are sufficiently long such that $\Delta\Neff = 0$ throughout BBN, the $Y$ particle can contribute a non-negligible energy density to the Universe during BBN, which increases the expansion rate during BBN and alters when reactions freeze-out in a manner that cannot be modeled with a simple $\Delta\Neff$ \cite{sobotka_was_2023-1}. The effect that the $Y$ particle has on primordial abundances will be maximized for large $R_\Gamma$ and short lifetimes. In order to ensure that $\Neff$ is unaltered throughout BBN, we choose $\Gamma_Y = 10^{12.08} \unit{\per\giga\year}$ to be the largest decay rate that we consider. Additionally, we only consider values of $R_\Gamma \leq 0.1$ since any short-lived case with $R_\Gamma > 0.1$ corresponds to $\Neff > 3.98$, which is ruled out by Planck temperature anisotropies alone. Therefore the $\rho_Y$ contribution during BBN is maximized when $R_\Gamma = 0.1$ and $\Gamma_Y = 10^{12.08}\unit{\per\giga\year}$, for which the $Y$ particle only makes up $0.9\%$ of the total energy density at a temperature of $T = \SI{0.01}{\mega\electronvolt}$. Thus, the $Y$ particle has a small influence on primordial abundances. Nevertheless, we still account for this effect by altering the BBN code \textsc{PArthENoPe-v3.0} \cite{gariazzo_parthenope_2022} to include contributions from $\rho_Y$, and we provide \textsc{CLASS} with a table that was created using this modified version of \textsc{PArthENoPe} (details can be found in \textcite{sobotka_was_2023-1}). We also include direct measurements of the deuterium abundance in our analysis. In addition to constraining possible changes due to the presence of $Y$ particles, these measurements also restrict changes in $\omega_b$ that are necessary to correct for excessive small-scale damping by the injected DR.

\subsection{Intermediate regime}\label{sec:effects_intermediate}

Changes to $\Neff$ caused by injected DR are directly proportional to $R_\Gamma$ for short-lived cases. However, once the decay rate falls below $\Gamma_Y \approx 10^7 \unit{\per\giga\year}$, the $\Delta\Neff$ that results from a $Y$ decay increases with decreasing $\Gamma_Y$ for a fixed $R_\Gamma$ (see Fig.~\ref{fig:Neff_appendix} in Appendix \ref{sec:appendix_Neff}). This dependence on $\Gamma_Y$ arises from how $R_\Gamma$ is defined; $R_\Gamma$ is a proxy for $\rho_\dr(a_\Gamma)/\rho_{tot}(a_\Gamma)$ and, as matter becomes more significant, the same DR contribution relative to the total energy density implies a greater amount of DR relative to standard radiation. Therefore, for a given value of $R_\Gamma$, scenarios in the intermediate regime of \mbox{$ 10^{3.22} \unit{\per\giga\year} \lesssim \Gamma_Y \lesssim 10^{7} \unit{\per\giga\year}$} require larger increases in $\omega_{\mathrm{cdm}}$ compared to those of short-lived cases to fix $\MRrec$. 

\begin{figure}[t]
\centering
  \includegraphics[width=\linewidth]{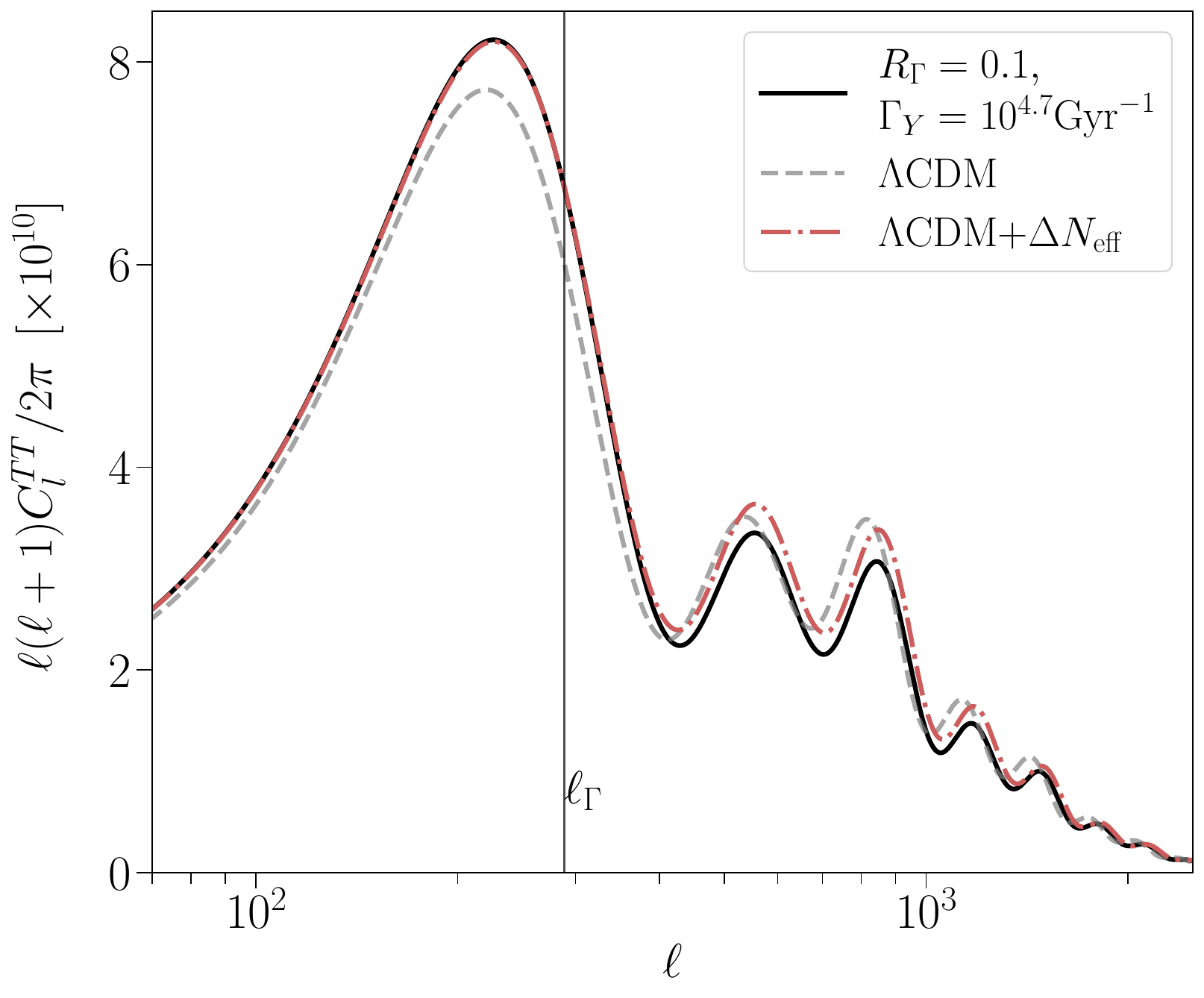}
  \caption{\footnotesize Temperature anisotropy spectrum of an example $Y$ decay in the intermediate regime (solid line). The vertical line, labeled $\ell_\Gamma$, represents a scale that is dominated by the mode entering the horizon at $a_\Gamma$ (i.e.\ $k_\Gamma = a_\Gamma H(a_\Gamma)$). The dashed curve is the spectrum of a $\Lambda$CDM model and the dot-dashed curve is the spectrum of a $\Lambda$CDM model with $\Neff$ matching the post-decay $\Neff$ calculated using Eq.~\eqref{eq:Nur_prime}. For scales with $\ell < \ell_\Gamma$, the decay spectrum is degenerate with that of $\Lambda$CDM+$\Neff$. }
  \label{fig:planck_scales}
\end{figure}

Furthermore, the impact of $Y$ decays on CMB observations in the intermediate regime cannot be mimicked by increasing $\Neff$ because perturbation modes that enter the horizon while the $Y$ particles are still present affect the CMB on scales that are observable by Planck. Let us define the mode that enters the horizon when the decay rate surpasses the Hubble rate as $k_\Gamma \equiv a_\Gamma H(a_\Gamma)$. This mode contributes the most to temperature anisotropies at an angular scale of $\ell_\Gamma \approx k_\Gamma \chi_{\mathrm{CMB}}$, where $\chi_{\mathrm{CMB}}$ is the comoving distance to the surface of last scattering. For intermediate cases that have $\Gamma_Y \lesssim 10^{6.64} \unit{\per\giga\year}$, the corresponding angular scale $\ell_\Gamma$ falls in the observable range of the Planck satellite ($\ell_\Gamma \lesssim 2500$). Figure \ref{fig:planck_scales} depicts the temperature spectrum resulting from a $Y$ decay with $\Gamma_Y = 10^{4.7} \unit{\per\giga\year}$ and $R_\Gamma = 0.1$ (solid curve). Such a scenario has a mode of wavenumber $k_\Gamma \approx 0.021 \unit{\per\mega\parsec}$ entering the horizon at $a_\Gamma$, which dominates the power at an angular scale of  $\ell_\Gamma \approx 288$ (vertical line). Large scales that enter the horizon after $a_\Gamma$ (i.e.\ $\ell < \ell_\Gamma$) are primarily influenced by the decay via $\Delta\Neff$ and so the decay spectrum is degenerate with a $\Lambda$CDM model with extra $\Neff$ (dot-dashed line) for $\ell < \ell_\Gamma$. On the other hand, scales that enter the horizon before $a_\Gamma$ are affected by the $Y$ particle contributing to the dark matter content and so the decay spectrum is suppressed compared to $\Lambda$CDM for $\ell > \ell_\Gamma$. The Wess-Zumino Dark Radiation (WZDR) model, which generates similar scale-dependent modifications to the phase and amplitude of the acoustic peaks, can reduce the Hubble tension and is slightly favored by Planck and BAO data alone \cite{aloni_step_2022-2}.

\subsection{Long-lived regime}\label{sec:effects_long_lived}

For decays in the long-lived regime, recombination occurs when no significant DR has been injected and the $Y$ particle is behaving as stable non-relativistic matter. Therefore, the $Y$ particle \textit{increases} $(\rho_m/\rho_r)_\mathrm{rec}$ and thus a reduction in $\omega_{\mathrm{cdm}}$ is needed to keep $(\rho_m/\rho_r)_\mathrm{rec}$ fixed. 

While short-lived cases significantly alter the sound horizon via increases in $\Neff$, the pre-recombination contribution of $\rho_\dr$ to the expansion is insignificant for long-lived cases. Instead, the $Y$ particle makes a non-negligible contribution to the pre-recombination expansion and thereby reduces $r_s$. However, the reduction in $\omega_{\mathrm{cdm}}$ that is required to fix $(\rho_m/\rho_r)_\mathrm{rec}$ counters this effect and keeps $r_s$ unchanged. The required reduction in $\omega_{\mathrm{cdm}}$ also amplifies the angular diameter distance to the CMB, $d_A$. Increasing $H_0$ can correct for this change to $d_A$ and maintain the observed value of $\theta_s$. The required adjustments to $H_0$ and $\omega_{\mathrm{cdm}}$ for decays in the long-lived regime both serve to decrease $\Omega_{\mathrm{cdm}} = \omega_{\mathrm{cdm}}/h^2$ and thereby decrease $\Omega_m$. Therefore, compared to $\Lambda$CDM, decays with $\Gamma_Y \lesssim 10^{3.22} \unit{\per\giga\year}$ result in a smaller value of $S_8 \equiv \sigma_8 \sqrt{\Omega_m/0.3}$, where $\sigma_8$ is the root mean square of matter fluctuations on $8 h^{-1} \SI{}{\mega\parsec}$ scales. This effect is how some studies have employed DCDM to address both the $H_0$ and $S_8$ tensions (e.g.\ \cite{pandey_alleviating_2020}). 


\section{Analysis Method} \label{sec:Analysis_Method}
We use a modified version of \textsc{CLASS-v3.2}\footnote{\url{https://github.com/alexsobotka/Class_YtoDR}} \cite{blas_cosmic_2011}, coupled with \textsc{MontePython-v3}\footnote{\url{https://github.com/brinckmann/montepython_public}} \cite{audren_conservative_2013, brinckmann_montepython_2018-1} as our MCMC engine and employ a Metropolis-Hastings algorithm. We assume a flat prior on the base six cosmological parameters \{$\omega_b$, $\omega_{\mathrm{cdm}}$, $h$, $A_s$, $n_s$, $\tau_{reio}$\} and, unless otherwise specified, we use \mbox{$R_\Gamma = [0.0, 0.1]$} and \mbox{$\lgGamma = [1.49, 12.08]$} for priors on decay parameters (see Table \ref{tab:priors}). The chosen upper limit of \mbox{$\lgGamma < 12.08$} ensures that the decay occurs after BBN has finished such that \mbox{$\Neff=3.044$} is constant throughout BBN. A lower limit of \mbox{$\lgGamma > 1.49$} ensures that $Y$ particle lifetimes that extend past the time of recombination are considered. Smaller decay rates correspond to $Y$ particle lifetimes that extend deep into matter domination such that the injected DR has no impact on observables. Such scenarios have been investigated elsewhere (e.g.\ \cite{audren_strongest_2014, poulin_fresh_2016, mccarthy_converting_2023}) and are not the primary focus of this work. 

\begin{table}[t]
\centering
    \caption{\footnotesize Priors used in MCMC analyses.  }
\begin{tabular*}{\linewidth}{l@{\extracolsep{\fill}}c}
    \hline
    \hline
    & \multicolumn{1}{l}{$ [1.49, 12.08] $ (full) }\\
    $\log_{10}(\frac{\Gamma_{Y}}{\SI{}{\giga\year}})$ 
    & \multicolumn{1}{l}{$ [3.14, 12.08] $ (pre-recombination)} \\
    &  \multicolumn{1}{l}{$ [1.49, 3.24] \,\,\,$ (post-recombination)}\\
    \hline
    $R_\Gamma$ 
    & $  [0, 0.1]    $\\
    $10^{-2}\omega_b$ 
    & $ [2.0, 2.7]  $\\
    $\omega_{\mathrm{cdm}}$ 
    & $ [0.1, 0.15]  $\\
    $h$ 
    & $ [0.5,0.85]  $\\
    $\ln 10^{10} A_s$ 
    & $ [2.9397, 3.1497] $\\
    $n_s$ 
    & $ [0.933, 0.9953]    $\\
    $\tau_{reio}$ 
    & $ [0.01, 0.1103]    $\\
    \hline
    \hline
\end{tabular*}
    \label{tab:priors}
\end{table}

To constrain the effects that a $Y$ decay has on the CMB temperature anisotropies, we employ \textit{Planck} 2018 high-$\ell$ TT, TE, EE, low-$\ell$ TT, and low-$\ell$ EE likelihood functions \cite{planck_collaboration_planck_2020-1}, as well as the the third generation South Pole Telescope 2018 (SPT-3G) TT, TE, and EE data \cite{chown_maps_2018,dutcher_measurements_2021,balkenhol_measurement_2022} adapted to the \textsc{clik} format.\footnote{\url{https://github.com/SouthPoleTelescope/spt3g_y1_dist} } To further constrain $\Omega_m$, we employ the likelihood from BOSS DR12 \cite{alam_clustering_2017} BAO analysis\footnote{While newer datasets are available, DR12 combined with CMB observations is the most effective at breaking the degeneracy between $\Omega_m$ and $r_d H_0$, where $r_d$ is the sound horizon at the end of the baryonic-drag epoch \cite{lin_early-Universe-physics_2021-1}.} as well as the Pantheon+ likelihood \cite{brout_pantheon_2022} based on uncalibrated Type Ia supernovae \cite{scolnic_pantheon_2022}. 

\begin{figure*}[t]
\centering
  \includegraphics[width=\linewidth]{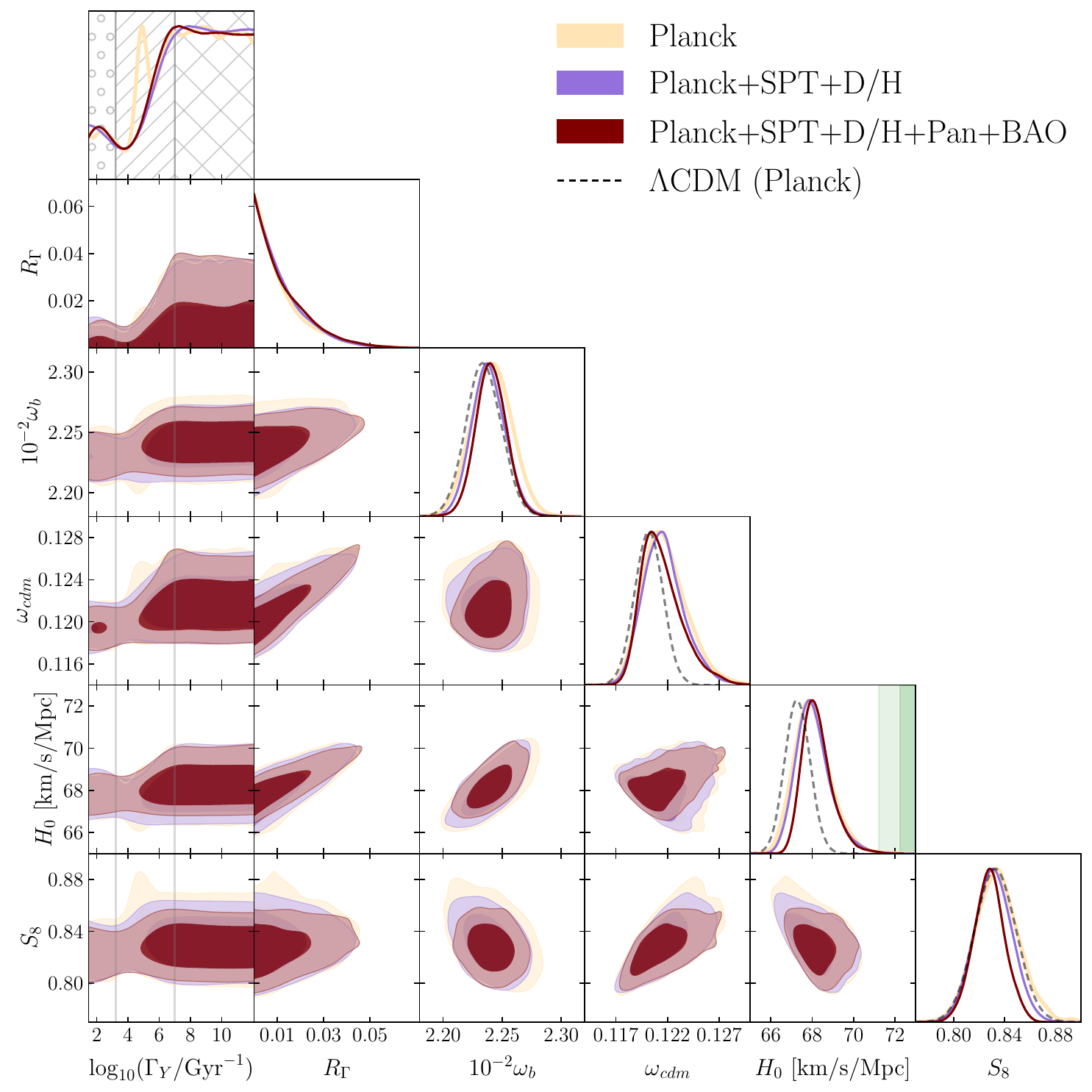}
  \caption{\footnotesize 1D and 2D posterior distributions of decay and cosmological parameters for different combinations of \textit{Planck} high-$\ell$ TT,TE,EE, low-$\ell$ TT, and low-$\ell$ EE  (\textbf{Planck}) data,  SPT-3G 2018 TT,TE, and EE data (\textbf{SPT}), bounds on the observed deuterium abundance (\textbf{D/H}), Pantheon+ data (\textbf{Pan}), and the BOSS DR12 likelihood (\textbf{BAO}). We include the 1D posteriors for $\Lambda$CDM constrained by Planck (dashed line). The vertical band shows the $1\sigma$ and $2\sigma$ bounds determined by S$H_0$ES ($H_0 = \SI{73.30 \pm 1.04}{\kilo\meter\per\second\per\mega\parsec}$). Hatches in the 1D posterior for $\lgGamma$ mark the different $Y$ particle lifetime regimes discussed in Sec.~\ref{sec:Effects_of_Y_Decay}: short-lived (cross hatch), intermediate (diagonal hatch), and long-lived (circles). }
  \label{fig:full_triangle}
\end{figure*}

As discussed in Sec.~\ref{sec:effects_short_lived}, the deuterium abundance provides additional constraints on $\rho_Y$ during BBN and also serves as an independent constraint on $\omega_b$. We include bounds on the primordial deuterium abundance from \textcite{cooke_one_2018}: $(\text{D/H}) = (2.527\pm0.030)\times 10^{-5}$. However, this bound only includes measurement uncertainty. As in \textcite{sobotka_was_2023-1}, we account for extra uncertainty in D/H from nuclear reaction rates by translating the bounds on the baryon-to-photon ratio reported by \textcite{cooke_one_2018}, $\eta = (6.119 \pm 0.100) \times 10^{-10}$, to bounds on D/H. These limits on $\eta$ include both measurement uncertainty and uncertainty associated with reaction rates. We translate these bounds to limits on the deuterium abundance by calculating D/H for a range of $5.8 \leq 10^{10}\eta \leq 6.88$ using PArthENoPe. Within this range, $\text{D/H}\propto \eta^{-1.65}$. We use this fit to determine a new fractional uncertainty for D/H ($\sigma_{\mathrm{DH}}/\text{DH} = 1.65 \times \sigma_\eta/\eta)$, resulting in $(\text{D/H}) = (2.527\pm0.068)\times 10^{-5}$. We utilize a Gaussian likelihood function for \textsc{MontePython} with a mean and standard deviation of $\mu_{\text{DH}} = 2.527\times10^{-5}$ and $\sigma_{\text{DH}} = 6.83 \times 10^{-7}$, respectively.  

We consider MCMC chains with a Gelman-Rubin \cite{gelman_inference_1992} criterion of $|R-1|<0.01$ as converged, and post-processing of all chains was done using \textsc{GetDist} \cite{lewis_getdist_2019} by removing the first 30\% of points as burn-in.

\begin{table*}[t]
\centering
    \caption{\footnotesize Results for decay and cosmological parameters from MCMC analyses with a prior of $\lgGamma = [1.49, 12.08]$, corresponding to those shown in Fig.~\ref{fig:full_triangle}. Uncertainties are reported at $68\%$ C.L. and upper limits are given at $95\%$ C.L. }
\begin{tabular*}{\linewidth}{l@{\extracolsep{\fill}}cccc}
    \hline
    \hline
     & $\Lambda$CDM (Planck) & Planck & Planck+SPT+D/H & Planck+SPT+D/H+Pan+BAO \\
    \hline
    $R_\Gamma$ 
    & ... 
    & $ < 0.0347 $
    & $ < 0.0340 $
    & $ < 0.0360     $\\
    $\log_{10}(\Gamma_{Y}/\SI{}{\per\giga\year})$ 
    & ... 
    & ...
    & ...
    & ... \\
    $H_0 \,[\SI{}{\kilo\meter\per\second\per\mega\parsec}]$ 
    & $ 67.29\pm 0.61     $ 
    & $ 68.03^{+0.73}_{-0.99}    $ 
    & $ 68.06^{+0.64}_{-0.94}     $
    & $ 68.24^{+0.52}_{-0.83}       $\\
    $S_8$ 
    & $ 0.833\pm 0.016    $ 
    & $ 0.834^{+0.016}_{-0.018}    $ 
    & $ 0.831\pm 0.015      $
    & $ 0.828\pm 0.012    $\\
    $10^{-2}\omega_b$ 
    & $ 2.234\pm 0.015    $ 
    & $ 2.241\pm 0.017      $ 
    & $ 2.239^{+0.013}_{-0.015}      $
    & $ 2.241\pm 0.013      $\\
    $\omega_{\mathrm{cdm}}$ 
    & $ 0.1202\pm 0.0014    $ 
    & $ 0.1218^{+0.0017}_{-0.0025}      $ 
    & $  0.1217^{+0.0016}_{-0.0022}     $
    & $ 0.1214^{+0.0013}_{-0.0024}    $\\
    $n_s$ 
    & $  0.9644\pm 0.0044   $ 
    & $ 0.9684^{+0.0049}_{-0.0059}    $ 
    & $ 0.9683^{+0.0045}_{-0.0058}      $
    & $ 0.9693^{+0.0044}_{-0.0052}     $\\
    \hline
    \hline
\end{tabular*}
    \label{tab:posteriors}
\end{table*}

\section{Results} \label{sec:Results}

\subsection{Full marginalized results}
Figure \ref{fig:full_triangle} shows the posterior distributions for the decay parameters $\Gamma_Y$ and $R_\Gamma$ as well as $\omega_b$, $\omega_{\mathrm{cdm}}$, $H_0$, and $S_8$ with 68\% and 95\% confidence level (C.L.) contours for various combinations of data sets. The posterior values for each parameter are reported in Table \ref{tab:posteriors}. The vertical band in the 1D posterior for $H_0$ in Fig.~\ref{fig:full_triangle} shows the $1\sigma$ and $2\sigma$ bounds reported by the S$H_0$ES collaboration ($H_0 = \SI{73.30 \pm 1.04}{\kilo\meter\per\second\per\mega\parsec}$) \cite{riess_comprehensive_2022}, and the dashed line in the 1D posteriors represents a $\Lambda$CDM model constrained with only \textit{Planck} 2018 high-$\ell$ TT, TE, EE, low-$\ell$ TT, and low-$\ell$ EE data. The 1D posterior of $\lgGamma$ also contains hatched regions marking the different $Y$ particle lifetime regimes introduced in Sec.~\ref{sec:Effects_of_Y_Decay}: short-lived (cross hatch), intermediate (diagonal hatch), and long-lived (circles).

The $\Gamma_Y$ values depicted in Fig.~\ref{fig:full_triangle} span a wide range of $Y$ particle lifetimes from right after the end of BBN to about 30 million years after recombination. For this range of decay rates, Planck anisotropy data constrains $R_\Gamma < 0.0347 $ (95\% C.L.), which translates to $\rho_Y/\rho_{tot} < 0.0204$ for the shortest lifetime we consider and \mbox{$\rho_Y/\rho_{tot} < 0.0291$} for the longest lifetime. The inclusion of SPT-3G data and bounds on the primordial deuterium abundance (Planck+SPT+D/H) tighten these bounds to $R_\Gamma < 0.0340$ (95\% C.L.), which translates to $\rho_Y/\rho_{tot} < 0.0200$ and $\rho_Y/\rho_{tot} < 0.0286$ for the shortest and longest lifetimes, respectively. Figure \ref{fig:full_triangle} shows that Planck+SPT+D/H constraints are comparable to those of Planck in the short-lived regime. However, Planck+SPT+D/H enforces the tightest constraints on $R_\Gamma$ for $\log_{10}(\Gamma_Y/ \unit{\per\giga\year}) \approx 4$. These constraints relax once $\log_{10}(\Gamma_Y/ \unit{\per\giga\year}) \lesssim 4$ because the effects of a $Y$ decay in these scenarios can be compensated for with changes in $\omega_{\mathrm{cdm}}$ and $H_0$. However, the inclusion of Pantheon+ and BAO data limits these changes in $\omega_{\mathrm{cdm}}$ and $H_0$ and thus Planck+SPT+D/H+Pan+BAO bounds on $R_\Gamma$ are stricter than those of Planck+SPT+D/H once $\log_{10}(\Gamma_Y/ \unit{\per\giga\year}) \lesssim 2.5$. As a result, a larger fraction of samples that are favored by the data fall in the short-lived regime of $\log_{10}(\Gamma_Y/ \unit{\per\giga\year}) \gtrsim 7$. Therefore, marginalizing over the entire range of $\Gamma_Y$ yields $R_\Gamma < 0.0360 $ (95\% C.L.) for Planck+SPT+D/H+Pan+BAO, which is less constrained than the corresponding bound from Planck+SPT+D/H. Being agnostic about $Y$ particle lifetimes, this bound of $R_\Gamma < 0.0360 $ (95\% C.L.) translates to $\rho_Y/\rho_{tot} < 0.0302$ over the full range of $\Gamma_Y$ considered. However, as will be discussed in subsequent sections, constraints on $R_\Gamma$ can be more stringent based on the specific regime under consideration.

As discussed in Sec.~\ref{sec:Effects_of_Y_Decay}, the addition of the $Y$ particle decreases $\theta_s$. While the manner by which the $Y$ particle decreases $\theta_s$ depends on the $Y$ particle lifetime, an increase in $H_0$ is always required to keep $\theta_s$ consistent with CMB data. For the full range of decay rates considered, $Y$ decays constrained by Planck data alone give $H_{0 } = 68.03^{+0.73}_{-0.99}\,\unit{\kilo\meter\per\second\per\mega\parsec} $ (68\% C.L.), whereas $H_0 = 67.29\pm 0.61\,\unit{\kilo\meter\per\second\per\mega\parsec}$ (68\% C.L.) for a $\Lambda$CDM model constrained by Planck. We find that the combination of Planck+SPT+D/H+Pan+BAO results in $H_0 = 68.24^{+0.52}_{-0.83} \,\unit{\kilo\meter\per\second\per\mega\parsec}$ (68\% C.L.) for $Y$ decays, only reducing the tension with S$H_0$ES from $5.2\sigma$ to $4.4\sigma$. 

Similarly, we find that $Y$ decay scenarios marginalized over the full prior for $\Gamma_Y$ does not help mitigate the $S_8$ tension. When considering the full combination of data sets (Planck+SPT+D/H+Pan+BAO), we find $S_{8 } = 0.828\pm 0.012 $ (68\% C.L.) whereas $S_{8 } = 0.825^{+0.011}_{-0.012} $ (68\% C.L.) for a $\Lambda$CDM model constrained with Planck+SPT+D/H+Pan+BAO. Therefore, these decay scenarios do not lessen the tension between the value of $S_8$ inferred from the CMB and that of local measurements. 

As discussed in Sec.~\ref{sec:Effects_of_Y_Decay}, the presence of the $Y$ particle has different effects on observables depending how the $Y$ particle lifetime compares to the times of matter-radiation equality or recombination. It is therefore more enlightening to analyze these results in the three regimes discussed in Sec.~\ref{sec:Effects_of_Y_Decay}: the short-lived ($\tau_Y \lesssim 10^{-7}\unit{\giga\year} $), intermediate ($10^{-7}\unit{\giga\year} \lesssim \tau_Y \lesssim 10^{-3.22}\unit{\giga\year}$) and long-lived ($\tau_Y \gtrsim 10^{-3.22}\unit{\giga\year}$) regimes.

\subsection{Short-lived regime} \label{sec:short-lived-results}

\begin{figure}[t]
\centering
  \includegraphics[width=\linewidth]{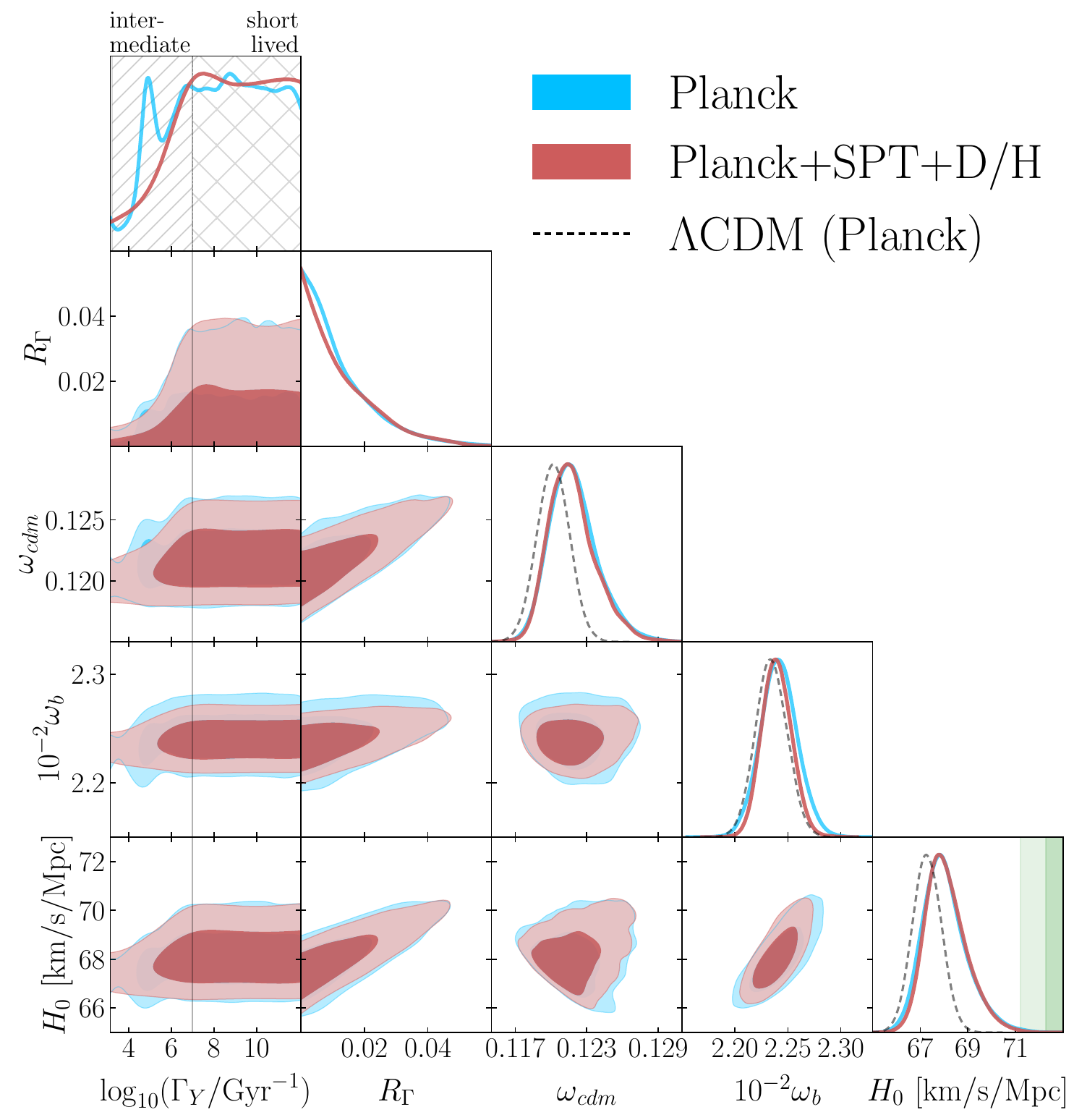}
  \caption{\footnotesize 1D and 2D posterior distributions of decays in both the short-lived and intermediate regimes i.e. $\log_{10}(\Gamma_Y/ \unit{\per\giga\year}) = [3.14, 12.08]$ . The dashed line shows the 1D posteriors for $\Lambda$CDM constrained by Planck, and the vertical band shows the $1\sigma$ and $2\sigma$ bounds determined by S$H_0$ES ($H_0 = \SI{73.30 \pm 1.04}{\kilo\meter\per\second\per\mega\parsec}$). Hatches in the 1D posterior for $\lgGamma$ mark the short-lived (cross hatch) and intermediate (diagonal hatch) regimes.}
  \label{fig:early_triangle}
\end{figure}

Figure \ref{fig:early_triangle} shows the posterior distributions for select parameters of an MCMC run with a prior of $\lgGamma = [3.14, 12.08]$, corresponding to  decay processes that finish before recombination and span the short-lived and intermediate regimes. For reference, a decay with $\Gamma_Y \approx \Gammay{3.22}$ and $R_\Gamma < 0.1$ corresponds to a scenario with $a_\Gamma \approx a_{\mathrm{rec}}$. The lower bound of $\lgGamma > 3.14$ ensures that we probe the transition from the intermediate regime to the long-lived regime. Values for select posteriors of this analysis are presented in Table \ref{tab:short-lived_posteriors}. 

Figure \ref{fig:early_triangle} confirms the expectation for decays in the short-lived regime discussed in Sec.~\ref{sec:effects_short_lived}: Planck data exhibits a preference for larger values of $H_0$ and $\omega_{\mathrm{cdm}}$ compared to those of a $\Lambda$CDM model (dashed line) to avoid altering $\theta_s$ and $\MRrec$. For the range of $\Gamma_Y$ shown in Fig.~\ref{fig:early_triangle}, corresponding to decays that finish before recombination, the combination of Planck+SPT+D/H yields $\omega_{cdm } = 0.1218^{+0.0014}_{-0.0023}$ (68\% C.L.) and  $H_{0 } = 68.10^{+0.67}_{-1.0} \,\unit{\kilo\meter\per\second\per\mega\parsec} $ (68\% C.L.) whereas a $\Lambda$CDM model constrained with Planck+SPT+D/H data lends $\omega_{cdm } = 0.1200\pm 0.0013$ (68\% C.L.) and  \mbox{$H_{0 } = 67.33\pm 0.54 \,\unit{\kilo\meter\per\second\per\mega\parsec} $} (68\% C.L.).

Increasing $\omega_{\mathrm{cdm}}$ to fix $\MRrec$ keeps the early ISW effect minimally altered. The observed Sachs-Wolfe (SW) component of the temperature spectrum is the sum of the temperature monopole, $\Theta_0$, and the gravitational perturbation, $\Psi$. As $\omega_{\mathrm{cdm}}$ increases, $\Psi$ becomes more negative, leading to a partial cancellation such that the sum of $\Theta_0 + \Psi$ is reduced. Therefore, increasing $\omega_{\mathrm{cdm}}$ causes a suppression across all peaks. However, this effect can easily be reduced by increasing the amplitude, $A_s$. Indeed, for the range of \mbox{$\lgGamma = [3.14, 12.08]$}, Planck data favors a slight increase in $A_s$: \mbox{$\ln 10^{10} A_s = 3.051\pm 0.017$} (68\% C.L.) for $Y$ decays, while \mbox{$\ln 10^{10} A_s = 3.045\pm 0.016$} (68\% C.L.) for $\Lambda$CDM. 

\begin{table}[t]
\centering
    \caption{\footnotesize Posteriors for pre-recombination decays (i.e.\ $\lgGamma = [3.14,12.08]$), corresponding to those shown in Fig.~\ref{fig:early_triangle}. Uncertainties are reported at $68\%$ C.L. and upper limits are given at $95\%$ C.L. }
\begin{tabular*}{\linewidth}{l@{\extracolsep{\fill}}cc}
    \hline
    \hline
     & Planck & Planck+SPT+D/H  \\
    \hline
    $R_\Gamma$ 
    & $   < 0.0357     $
    & $ < 0.0364      $\\
    $\log_{10}(\frac{\Gamma_{Y}}{\SI{}{\giga\year}})$ 
    & $       ...        $
    & $ ...  $ \\
    $H_0 \,[\frac{\SI{}{\kilo\meter}}{\SI{}{\second}\cdot\SI{}{\mega\parsec}}]$ 
    & $  68.05^{+0.70}_{-1.1}       $
    & $ 68.10^{+0.67}_{-1.0}      $\\
    $\omega_{\mathrm{cdm}}$ 
    & $   0.1219^{+0.0015}_{-0.0023}     $
    & $ 0.1218^{+0.0014}_{-0.0023}     $\\
    $\ln 10^{10} A_s$ 
    & $   3.051\pm 0.017             $ 
    & $ 3.046\pm 0.016   $\\
    $10^{-2}\omega_b$ 
    & $   2.242\pm 0.018             $
    & $ 2.240\pm 0.014     $\\
    $n_s$ 
    & $     0.9686\pm 0.0056           $
    & $ 0.9686^{+0.0046}_{-0.0059}      $\\
    \hline
    \hline
\end{tabular*}
    \label{tab:short-lived_posteriors}
\end{table}

Once $\MRrec$ and $\theta_s$ have been fixed and $A_s$ has been enhanced, there remains excessive Silk damping compared to the observed temperature spectrum. Therefore, as postulated in Sec.~\ref{sec:effects_short_lived}, the Planck data shows a preference for a slightly larger value of $\omega_b$ in the short-lived regime compared to $\Lambda$CDM; increasing the baryon content will decrease the free streaming length of photons and mitigate any extra small-scale damping. However, altering $\omega_b$ also affects the height ratios of odd and even peaks in the temperature anisotropy spectrum. The inclusion of SPT data aids in constraining this effect on small scales. Additionally, the increase in $\omega_b$ favored by Planck data results in a change to the predicted abundance of primordial elements and so including bounds on the deuterium abundance also limits changes in $\omega_b$. This can be seen in the $\omega_b$ vs.\ $\lgGamma$ plane of Fig.~\ref{fig:early_triangle}.

\begin{figure*}[t]
\centering
  \includegraphics[width=\linewidth]{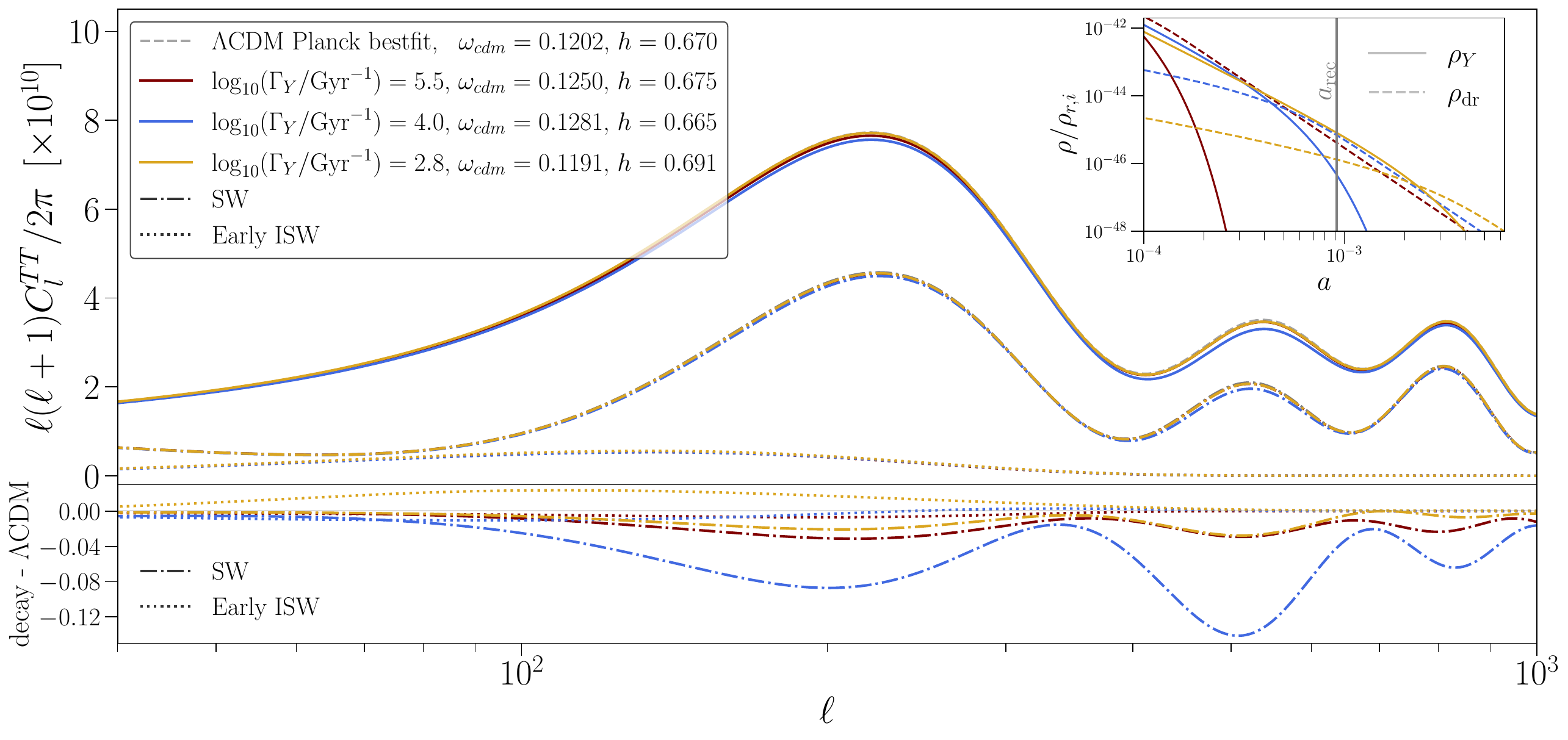}
  \caption{\footnotesize (\textit{Top panel}) Temperature anisotropy spectra of $Y$ decay scenarios with $R_\Gamma = 0.02$ and different decay rates (solid curves) compared to that of a $\Lambda$CDM model (dashed curve). Unless otherwise specified, each spectrum is created with the TT, TE, EE, low-$\ell$ EE best-fit values for the base six parameters reported by \textit{Planck} 2018 \cite{planck_collaboration_planck_2020-1}. For each decay scenario we increase $h$ so that $\theta_s$ is fixed and we adjust $\omega_{\mathrm{cdm}}$ to fix $\MRrec$. The dot-dashed and dotted curves depict the Sachs Wolfe (SW) and Early ISW components of each spectrum, respectively. (\textit{Bottom panel}) We plot the difference between the SW or early ISW component of each decay spectrum with that of the $\Lambda$CDM model. (\textit{Top right}) The evolution of energy densities $\rho_Y$ (solid) and $\rho_\dr$ (dashed) for each decay scenario, with the vertical line marking the scale factor of recombination.}
  \label{fig:SW_effects}
\end{figure*}

Figure \ref{fig:early_triangle} demonstrates the benefit of the $R_\Gamma$ parametrization used in this work compared to that commonly used in DCDM studies. As discussed in Sec.~\ref{sec:effects_short_lived}, the $\Delta\Neff$ arising from DCDM depends on the combination of $\Gammadcdm$ and $\fdcdm$. Instead, $\Delta\Neff$ is exclusively dependent on $R_\Gamma$ for short-lived decays and, as a result, there is a plateau in the $R_\Gamma$ vs.\ $\lgGamma$ plane of Fig.~\ref{fig:early_triangle}. The $2\sigma$ bound on this plateau is $R_\Gamma < 0.036$ which translates to a post-decay $\Neff \leq 3.38$. This bound is more relaxed than the reported \textit{Planck} 2018 TT,TE,EE+lowE bound of $\Neff = 2.92^{+0.36}_{-0.37}$ ($95\%$ C.L.) \cite{planck_collaboration_planck_2020-1}. However, the apparent discrepancy between these bounds stems from a difference in priors. The reported \textit{Planck} 2018 bounds result from an analysis that allows for both positive and negative $\Delta\Neff$, and Planck CMB observations exhibit a slight preference for negative $\Delta\Neff$. The injected DR considered in this work only results in a positive $\Delta\Neff$. For comparison, we perform an MCMC analysis applying Planck constraints to a $\Lambda$CDM model with a prior of $\Delta\Neff = [0, 0.5]$. In doing so, we find that $\Neff < 3.34 $ ($95\%$ C.L.). This $\Lambda$CDM+$\Delta\Neff$ model increases $\Neff$ during BBN, which results in a larger helium abundance. Simultaneous increases in both $\Neff$ and the helium abundance conspire to produce excessive damping of small scale anisotropies and are therefore tightly constrained \cite{sobotka_was_2023-1}. Therefore, the upper bound on $\Neff$ for a $\Lambda$CDM model with only positive $\Delta\Neff$ is more stringent than that of $Y$ decay scenarios in the short-lived regime.

A physically motivated extension to the $Y$ decay model would be to consider additional DR that is not sourced by the decay of a $Y$ particle. Such an ambient bath of DR would simply be modeled as a constant $\Delta\Neff$. However, this would be a trivial extension to the short-lived regime given that constraints on short-lived cases are dominated by bounds on $\Neff$; the inclusion of a preexisting bath of DR would tighten the bounds on $R_\Gamma$ that we have derived for the short-lived regime.

\subsection{Intermediate regime} \label{sec:intermediate-results}


Figure \ref{fig:early_triangle} shows constraints on $R_\Gamma$ becoming more stringent as the decay rate falls below \mbox{$\lgGamma \approx 7$}. Similar to decays in the short-lived regime, Planck data mandates an increase in both $H_0$ and $\omega_{\mathrm{cdm}}$ to effectively fix $\theta_s$ and $\MRrec$ for this intermediate regime of $\lgGamma = [3.22,7]$. However, in this intermediate regime, the $\Delta\Neff$ arising from a $Y$ decay grows with decreasing $\Gamma_Y$ for fixed $R_\Gamma$ (see Fig.~\ref{fig:Neff_appendix} in Appendix \ref{sec:appendix_Neff}). Consequently, the increase in $\omega_{\mathrm{cdm}}$ required to fix $\MRrec$ grows as $\Gamma_Y$ decreases. Therefore, for a given value of $R_\Gamma$, scenarios in the intermediate regime of $ 10^{3.22} \unit{\per\giga\year} \lesssim \Gamma_Y \lesssim 10^{7} \unit{\per\giga\year}$ require larger increases in $\omega_{\mathrm{cdm}}$ compared to those of short-lived cases to fix $\MRrec$. 

\begin{figure*}
\centering
  \includegraphics[width=\linewidth]{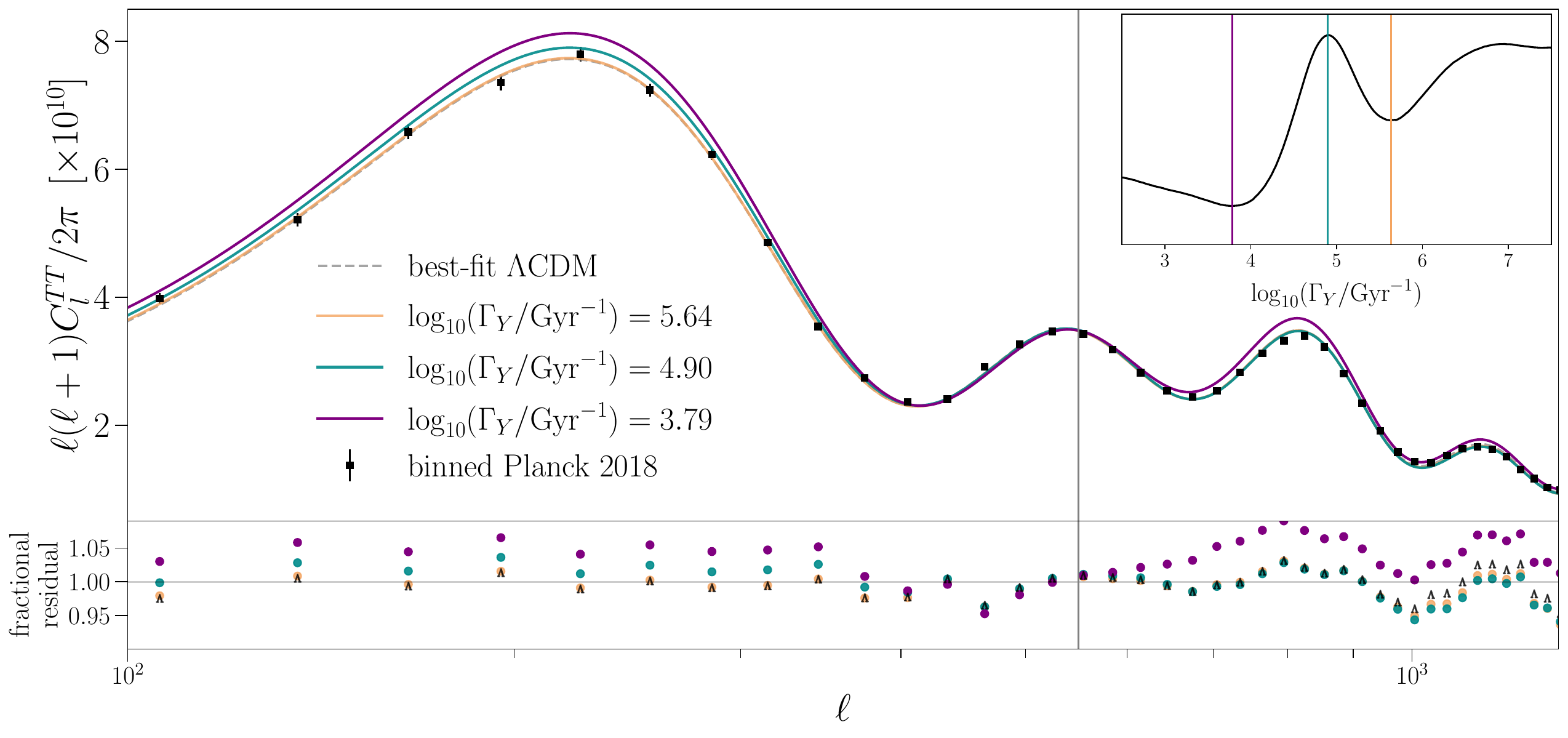}
  \caption{\footnotesize  (\textit{Top panel}) The dashed line shows the temperature anisotropy spectrum for a $\Lambda$CDM model with TT,TE,EE,low-$\ell$ EE best-fit values for the base six parameters reported by \textit{Planck} 2018 \cite{planck_collaboration_planck_2020-1}. The three solid colored curves are the resulting spectra of $Y$ decays with $R_\Gamma = 0.03$ and varying decay rates which are marked respectively as vertical lines in the top-right inset depicting the Planck 1D posterior of $\log_{10}(\Gamma_Y/ \unit{\per\giga\year})$ from Fig.~\ref{fig:full_triangle}. Each spectrum has the same \textit{Planck} 2018 best-fit values for $\omega_b$, $n_s$, and $\tau_{reio}$. For all spectra the values for $H_0$, $\omega_{\mathrm{cdm}}$, and $A_s$ have been adjusted to keep $\theta_s$, $\MRrec$, and the second peak height fixed, respectively. The \textit{Planck} 2018 binned TT data is shown by the black squares with corresponding $1\sigma$ error bars. (\textit{Bottom panel}) Fractional residuals of the best-fit $\Lambda$CDM (shown by `$\Lambda$' markers) and $Y$ decay scenario (circles) with the binned \textit{Planck} 2018 data. Of these scenarios, the decay with $\log_{10}(\Gamma_Y/ \unit{\per\giga\year}) = 4.9$ is enhanced on large scales but suppressed on small scales compared to $\Lambda$CDM and therefore it is easier for the $\log_{10}(\Gamma_Y/ \unit{\per\giga\year}) = 4.9$ decay to agree with Planck data by increasing $n_s$.}
    \label{fig:peak_spectrum}
\end{figure*}

Fixing $\MRrec$ means the early ISW effect is minimally altered but, as discussed in Sec.~\ref{sec:short-lived-results}, an increase in $\omega_{\mathrm{cdm}}$ suppresses the SW component of the temperature spectrum by enhancing gravitational potential wells. For short-lived decays, this effect can be compensated for with a increase in the amplitude, $A_s$. However, in addition to this overall suppression, intermediate cases can simultaneously affect some CMB scales with $\Delta\Neff$ while other scales are suppressed by the presence of the Y particle (see Sec.~\ref{sec:effects_intermediate}). These effects work in tandem to foster asymmetry between the peak heights of the temperature spectrum.

To demonstrate this suppression and asymmetry, Fig.~\ref{fig:SW_effects} shows the temperature spectra of three $Y$ decays with $R_\Gamma = 0.02$ and various decay rates compared the spectrum of a $\Lambda$CDM model. Each spectrum assumes the same values for $\omega_b$, $A_s$, $n_s$, and $\tau_{reio}$, and we alter $h$ and $\omega_{\mathrm{cdm}}$ such that $\theta_s$ and $(\rho_m/\rho_r)_\mathrm{rec}$ are fixed. For the decay with $\Gamma_Y = 10^{5.5} \unit{\per\giga\year}$, $\rho_Y$ is negligible by the time of recombination, and the only lasting effect is extra $\Neff$ from the injected DR. Therefore, an increase in $\omega_{\mathrm{cdm}}$ is required to fix $\MRrec$, and the early ISW component of the temperature spectrum (dotted line) is very similar to that of $\Lambda$CDM. However, increasing $\omega_{\mathrm{cdm}}$ results in deeper gravitational potential wells and so the SW component (dot-dashed line) is suppressed compared to $\Lambda$CDM. Additionally, since $\rho_Y$ is negligible by the times the modes that dominate the first and second peaks enter the horizon, there is no extra suppression from the $Y$ particle and therefore the ratio of the first and second peak heights of the SW component is minimally altered.

As $\Gamma_Y$ decreases within the intermediate regime, the required $\omega_{\mathrm{cdm}}$ to fix $\MRrec$ increases. However, once $\Gamma_Y$ falls below $10^4 \unit{\per\giga\year}$, $\rho_Y$ begins to contribute a non-negligible amount to $\MRrec$ and so less of an increase in $\omega_{\mathrm{cdm}}$ is necessary. Therefore, for a fixed $R_\Gamma$, a decay with $\Gamma_Y \approx 10^4 \unit{\per\giga\year}$ requires the maximum increase in $\omega_{\mathrm{cdm}}$ and thereby results in the maximum suppression of the SW component of the temperature spectrum compared to other values of $\Gamma_Y$. This can be seen in the bottom panel of Fig.~\ref{fig:SW_effects}: fixing $\MRrec$ for a decay with $\Gamma_Y = 10^4 \unit{\per\giga\year}$ results in a significant suppression of the SW component.

Not only is there an overall suppression in the SW component for this decay with $\Gamma_Y = 10^4 \unit{\per\giga\year}$, but also the second acoustic peak is further suppressed by the presence of the $Y$ particle; as discussed in Sec.~\ref{sec:effects_intermediate}, if the mode that dominates the second peak enters the horizon while the $Y$ particle is still present, the second acoustic peak will experience extra suppression. This ultimately results in a change in the height ratios between the first and second peak, thus making it more difficult to compensate for these changes with a simple shift in $A_s$. Instead, Fig.~\ref{fig:early_triangle} shows that Planck data exhibits a preference for $R_\Gamma \rightarrow 0$ as $\lgGamma \rightarrow 4$, ensuring both that suppression from the $Y$ particle is negligible and that a minimal $\Delta\Neff$ is produced even for small values of $\Gamma_Y$.

When only applying Planck data, Figures \ref{fig:full_triangle} and \ref{fig:early_triangle} show a peak in the posterior of $\Gamma_Y$ around \mbox{$\lgGamma \approx 4.9$}. Interestingly, a similar peak was observed in \textcite{poulin_fresh_2016}. \textcite{nygaard_updated_2021-1} observed a slight plateau rather than a peak, and attributed this to their use of a stricter Gelman-Rubin convergence criterion for MCMC sampling compared to that of \textcite{poulin_fresh_2016}. However, by employing profile likelihoods, \textcite{holm_discovering_2022} pinpoint a best-fit value of $\lgGamma = 4.763$ and explain the appearance of a plateau rather than a peak in the posteriors of \textcite{nygaard_updated_2021-1} to volume effects from Bayesian methods. We find that, with the parametrization used in this work, there is indeed a preferential peak in the $\Gamma_Y$ posterior even when enforcing the same convergence criteria as \textcite{nygaard_updated_2021-1}. However, this preference for $\lgGamma \approx 4.9$ is eliminated with the addition of SPT data and/or bounds on the primordial deuterium abundance.

To explain this preference for $\lgGamma \approx 4.9$, we consider the resulting temperature anisotropy spectra of $Y$ decay scenarios with three different decay rates. Figure \ref{fig:peak_spectrum} shows the temperature spectra for decays with $R_\Gamma = 0.03$ and $\lgGamma$ values of 5.64,  4.9, or 3.79. Included in the figure is an inset in the top right corner that depicts the Planck 1D posterior for $\lgGamma$ found in Fig.~\ref{fig:full_triangle}, where the solid vertical lines mark the decay rates of each decay scenario with corresponding colors. Figure \ref{fig:peak_spectrum} also includes the spectrum of a $\Lambda$CDM model (dashed line) with best-fit TT,TE,EE,low-$\ell$ EE values for $\omega_b$, $\omega_{\mathrm{cdm}}$, $\theta_s$, $A_s$, $n_s$, and $\tau_{reio}$ reported by \textit{Planck} 2018 \cite{planck_collaboration_planck_2020-1}. Each decay scenario has the same best-fit values for $\omega_b$, $\theta_s$, $n_s$, and $\tau_{reio}$ as this $\Lambda$CDM model. Values for $H_0$ and $\omega_{\mathrm{cdm}}$ have been adjusted for each decay scenario such that $\theta_s$ and $\MRrec$ are the same between all spectra. Additionally, the three decay spectra have been normalized to the second acoustic peak by augmenting $A_s$. Black points are the binned TT spectrum data provided by the Planck Collaboration\footnote{\url{http://pla.esac.esa.int/pla/\#cosmology}} and the lower panel of Fig.~\ref{fig:peak_spectrum} depicts the fractional residuals between the $\Lambda$CDM spectrum (shown by the `$\Lambda$' markers) or decay spectra (colored circles) and these binned data. 

Each decay scenario in Fig.~\ref{fig:peak_spectrum} injects DR before recombination and thereby requires an increase in $\omega_{\mathrm{cdm}}$ to fix $\MRrec$. Therefore, each decay spectrum in Fig.~\ref{fig:peak_spectrum} experiences suppression across all peaks. Additionally, depending on the value of $\Gamma_Y$, the presence of the $Y$ particle suppresses some scales more than others and leads to an asymmetry in peak heights between the first and second peak. By adjusting $A_s$ to normalize the second peak height between all spectra, this asymmetry effect is apparent in Fig.~\ref{fig:peak_spectrum}. The $\lgGamma = 5.64$ scenario requires the smallest increase in $\omega_{\mathrm{cdm}}$ compared to the other decay rates shown in Fig.~\ref{fig:peak_spectrum} and so the first peak is in good agreement with that of $\Lambda$CDM even after normalizing the second peak height. However, some small-scale suppression compared to $\Lambda$CDM is present for this $\lgGamma = 5.64$ scenario due to the presence of the $Y$ particle. On the other hand, the $\lgGamma = 3.79$ decay spectrum necessitates the largest increase in $\omega_{\mathrm{cdm}}$, leading to a notable suppression across all acoustic peaks. The second acoustic peak is further suppressed because the mode that dominates this peak enters the horizon while the $Y$ particle is still present. Therefore, the second acoustic peak is more suppressed compared to the first and third peaks and, by normalizing the second peak height with $A_s$, the first and third peaks of the $\lgGamma = 3.79$ spectrum are enhanced compared to $\Lambda$CDM.

\begin{figure}[t]
\centering
  \includegraphics[width=\linewidth]{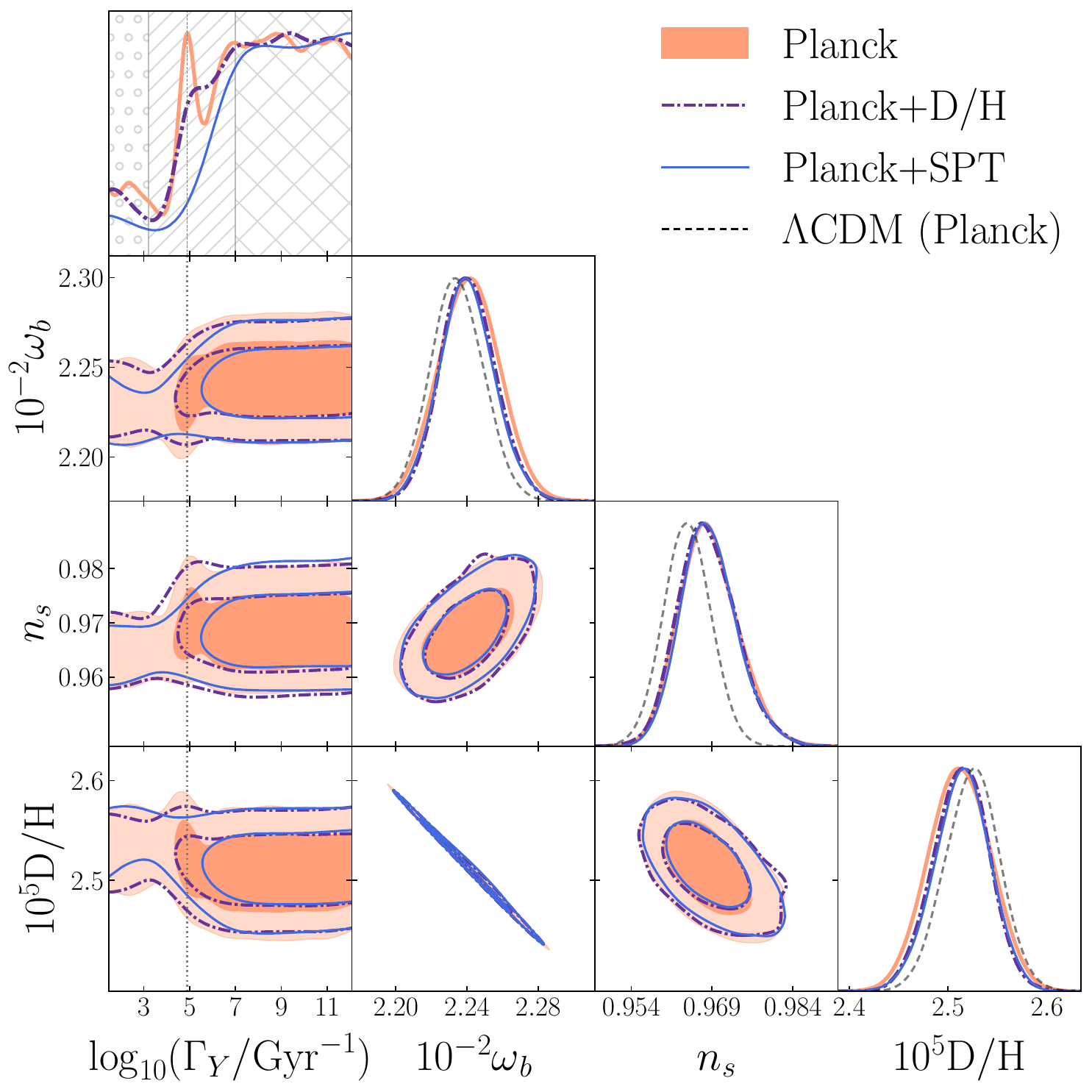}
  \caption{\footnotesize 1D and 2D posterior distributions for $Y$ decays with $\log_{10}(\Gamma_Y/ \unit{\per\giga\year}) = [1.49, 12.08]$. Hatches in the 1D posterior for $\lgGamma$ mark the different $Y$ particle lifetime regimes: short-lived (cross hatch), intermediate (diagonal hatch), and long-lived (circles). A decay rate of $\lgGamma \approx 4.9$ agrees with Planck temperature anisotropies when accompanied with changes in $\omega_b$ and $n_s$. This preference is ruled out with the addition of SPT data and/or bounds on the abundance of primordial deuterium (D/H).}
  \label{fig:peak_triangle}
\end{figure}

The decay rate of $\lgGamma = 4.9$ is more compatible with the Planck data due to the resulting spectrum after changing $H_0$ and $\omega_{\mathrm{cdm}}$ to fix $\theta_s$ and $\MRrec$, as well as adjusting $A_s$ to normalize the second peak height. Compared to $\Lambda$CDM, the resultant spectrum of this particular decay rate exhibits enhancement on scales with $\ell \lesssim 546$, but a suppression on scales with $\ell \gtrsim 546$. The scale of $\ell_p \approx 546$ (vertical line in Fig.~\ref{fig:peak_spectrum}) is the $\ell$ value at which the temperature spectrum is invariant under changes in the spectral index, $n_s$. This indicates that increasing $n_s$ is a straightforward method to align the resulting spectrum of a $\lgGamma = 4.9$ decay with Planck data. Alternatively, a decrease in $\omega_b$ could also lead to agreement with Planck data; decreasing $\omega_b$ reduces the height separation between the first and second acoustic peaks.

The preference for changes in $\omega_b$ and $n_s$ to accommodate a $\lgGamma \approx 4.9$ decay can be seen in Fig.~\ref{fig:peak_triangle}, which depicts the 1D and 2D posterior distributions for MCMC analyses with $\lgGamma = [1.49, 12.08]$ and $R_\Gamma = [0, 0.1]$. Figure \ref{fig:peak_triangle} shows that Planck data permits larger $R_\Gamma$ values for $\lgGamma \approx 4.9$ when accompanied by a decrease in $\omega_b$ and an increase in $n_s$. However, this preference for $\lgGamma \approx 4.9$ is lost with the inclusion of additional constraints from either SPT data or bounds on the abundance of deuterium. Changing $\omega_b$ alters the baryon-to-photon ratio and thereby the abundance of primordial deuterium, allowing for bounds on the deuterium abundance to constrain the changes in $\omega_b$ necessary to accommodate this special decay rate. Meanwhile, SPT data helps constrain small-scale anisotropies of the temperature spectrum and therefore limits changes to both $\omega_b$ and $n_s$. In fact, SPT data is better at ruling out the preference for $\lgGamma \approx 4.9$ since bounds on the deuterium abundance have no constraining power over $n_s$, as seen in the $n_s$ vs. $\lgGamma$ plane in Fig.~\ref{fig:peak_triangle}.

\subsection{Long-lived regime} \label{sec:long-lived-results}

\begin{figure}[t]
\centering
  \includegraphics[width=\linewidth]{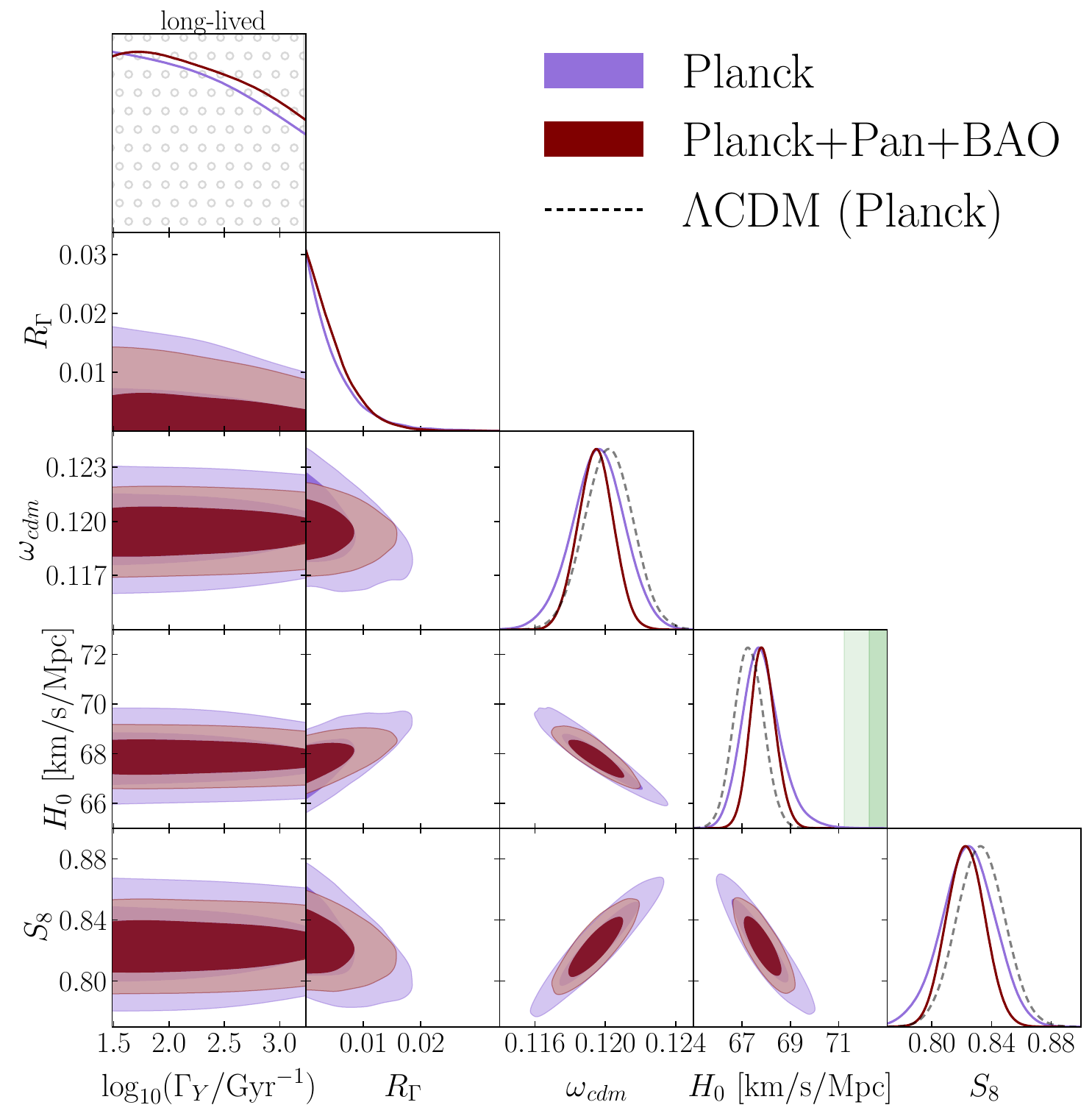}
  \caption{\footnotesize 1D and 2D posterior distributions of late decays i.e. $\log_{10}(\Gamma_Y/ \unit{\per\giga\year}) = [1.49, 3.24]$. The dashed line shows the 1D posteriors for $\Lambda$CDM constrained by Planck, and the vertical band shows the $1\sigma$ and $2\sigma$ bounds determined by S$H_0$ES ($H_0 = \SI{73.30 \pm 1.04}{\kilo\meter\per\second\per\mega\parsec}$).}
  \label{fig:late_triangle}
\end{figure}

To understand the effects of scenarios in which the $Y$ particle lifetime extends past the time of recombination, we analyze the results of an MCMC run with $\lgGamma = [1.49, 3.24]$, shown in Fig.~\ref{fig:late_triangle}. As discussed in Sec.~\ref{sec:effects_long_lived}, these long-lived scenarios require a reduction in $\omega_{\mathrm{cdm}}$ to keep $\MRrec$ fixed and also require a larger value of $H_0$ to fix $\theta_s$. Indeed, Fig.~\ref{fig:late_triangle} demonstrates the preference that Planck data has for large $H_0$ and small $\omega_{\mathrm{cdm}}$ compared to $\Lambda$CDM. 

Figure \ref{fig:late_triangle} shows $R_\Gamma$ becoming more constrained as $\lgGamma \rightarrow 3$ (shorter lifetimes), corresponding to decay scenarios in which $a_\Gamma$ approaches the scale factor of recombination. To understand why constraints tighten as $\lgGamma \rightarrow 3$, we refer back to Fig.~\ref{fig:SW_effects}. Figure \ref{fig:SW_effects} depicts a long-lived scenario with $\Gamma_Y = 10^{2.8}\unit{\per\giga\year}$ that requires a decrease in $\omega_{\mathrm{cdm}}$ to fix $\MRrec$. Here it can be seen that, even though $\MRrec$ is fixed, the early ISW effect is still somewhat enhanced compared to $\Lambda$CDM (dotted lines). This enhancement is the result of the $Y$ particle injecting DR after recombination, which results in the evolution of gravitational potentials. Additionally, the SW component of this long-lived spectrum is still suppressed compared to that of $\Lambda$CDM. For this particular decay rate of $\Gamma_Y = 10^{2.8}\unit{\per\giga\year}$, the energy density of the $Y$ particle begins to deviate from a $\rho \propto a^{-3}$ scaling before recombination. Therefore, fixing $\MRrec$ does not result in a pre-recombination expansion history equivalent to that of $\Lambda$CDM. Instead, fixing $\MRrec$ based on the value of $\rho_Y$ at recombination leads to a total pre-recombination dark matter content that is larger than that of $\Lambda$CDM. Thus, there is a suppression in the SW component of the $\Gamma_Y = 10^{2.8}\unit{\per\giga\year}$ curve in Fig.~\ref{fig:SW_effects} compared to $\Lambda$CDM. As $\Gamma_Y$ decreases, the value of $\rho_Y$ at recombination will approach $\rho_{Y,i}(a_{\mathrm{rec}}/a_i)^3$ and so, after decreasing $\omega_{\mathrm{cdm}}$ to fix $\MRrec$, the pre-recombination expansion history will mimic that of $\Lambda$CDM. Therefore, Planck data exhibits a preference for larger $R_\Gamma$ at smaller $\Gamma_Y$ (longer lifetimes), as these scenarios are easily accommodated for with a simple decrease in $\omega_{\mathrm{cdm}}$.

Figure \ref{fig:late_triangle} confirms the expected effects of decays in the long-lived regime outlined in Sec.~\ref{sec:effects_long_lived}: the increase in $H_0$ and reduction in $\omega_{\mathrm{cdm}}$ required by CMB observations for these long-lived cases result in a smaller value of $\Omega_m$, and thereby $S_8$, compared to that of $\Lambda$CDM. \textit{Planck} 2018 reported a TT,TE,EE+lowE bound of $S_8 = 0.834 \pm 0.016$ (68\% C.L.) \cite{planck_collaboration_planck_2020-1}, whereas the long-lived posterior shown in Fig.~\ref{fig:late_triangle} for Planck yields $S_8 = 0.824\pm 0.018$ (68\% C.L.). However, with the addition of data from Pantheon+ and BAO, long-lived cases are limited to $S_8 = 0.823\pm 0.013$ (68\% C.L.). Pantheon+ and BAO data constrain the reduction of $\Omega_m$ caused by decays in the long-lived regime as this reduction shifts the time of equality between matter and dark energy to earlier times. As a result, the long-lived regime is sufficiently restricted by Pantheon+ and BAO such that it does not reduce the tension between the CMB value of $S_8$ and that of local measurements. This result agrees with other studies that considered the effects of DCDM on $S_8$ (e.g.\ \cite{pandey_alleviating_2020, mccarthy_converting_2023}).

Decay cases in the long-lived regime do not significantly increase the value of $H_0$ inferred from the CMB; we find Planck+Pantheon+BAO yields $H_{0 } = 67.87^{+0.48}_{-0.54} \,\unit{\kilo\meter\per\second\per\mega\parsec} $ (68\% C.L.) for $Y$ decays in the long-lived regime. Therefore, neither the short-lived nor long-lived regimes are successful in substantially mitigating the Hubble tension. Planck+Pan+BAO posterior values for the long-lived regime are presented in Table \ref{tab:long-lived_posteriors}.

\begin{figure*}[t]
\centering
  \includegraphics[width=\linewidth]{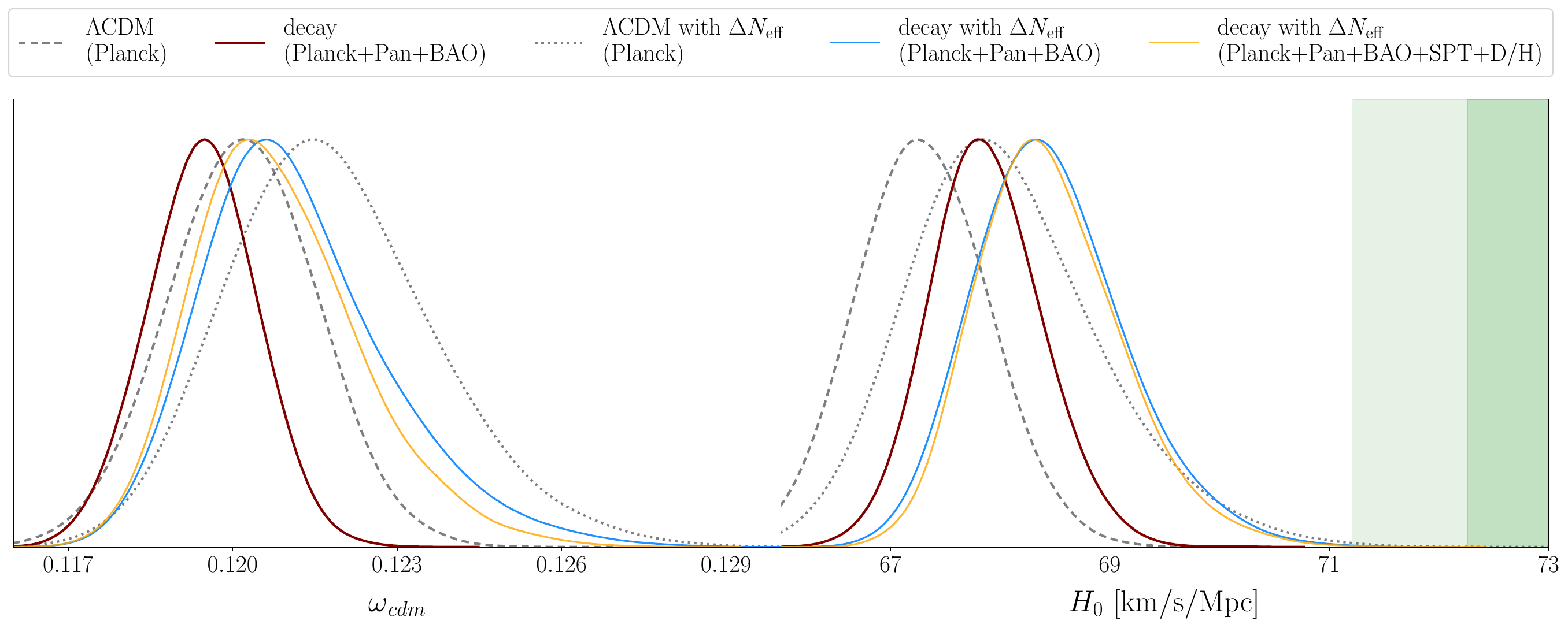}
  \caption{\footnotesize Comparison between 1D posterior distributions resulting from decay scenarios in the long-lived regime (prior of $\lgGamma = [1.49, 3.24]$). Dashed lines show resulting posteriors of a $\Lambda$CDM model with $\Delta\Neff = 0$, while the dotted lines are those of a $\Lambda$CDM model with a variable $\Delta\Neff$. The solid lines mark posteriors of either a decay model with only DR sourced by the decay (same as that shown in Fig.~\ref{fig:late_triangle}) or those of a model in which DR is sourced by the $Y$ decay as well as an ambient DR background modeled by allowing $\Delta\Neff$ to vary in the positive direction (prior of $\Delta\Neff = [0, 0.5]$). The vertical band depicts the $1\sigma$ and $2\sigma$ bounds determined by S$H_0$ES ($H_0 = \SI{73.30 \pm 1.04}{\kilo\meter\per\second\per\mega\parsec}$).}
  \label{fig:Neff_posteriors}
\end{figure*}

As mentioned in Sec.~\ref{sec:short-lived-results}, DR need not be only sourced by the decay of a $Y$ particle; it stands to reason that there could be an ambient background of DR that is not sourced by the $Y$ decay. Such a DR background serves as an interesting extension to the long-lived cases considered here. Constraints on $Y$ decays in the long-lived regime primarily derive from the necessary decrease in $\omega_{\mathrm{cdm}}$ to fix $\MRrec$, but a DR background would increase $\rho_r$ at recombination. Thus, a less significant decrease in $\omega_{\mathrm{cdm}}$ would be necessary to fix $\MRrec$ while the $Y$ particle is contributing to $\rho_m$ at recombination, potentially relaxing the constraints enforced by Pantheon+ and BAO data.

To determine the extent to which constraints on long-lived cases relax in the context of an ambient DR background, we perform an MCMC analysis comparable to that shown in Fig.~\ref{fig:late_triangle} but with the additional freedom of a variable and positive $\Delta\Neff$. We assume priors of $R_\Gamma = [0,0.1]$, $\lgGamma = [1.49, 3.24]$, and $N_{ur} = [2.0328, 2.5328]$, where $N_{ur}$ is the effective number of relativistic species excluding massive neutrinos and DR created by the decay of the $Y$ particle. The contribution that the single massive neutrino species makes to $\Neff$ is $N_{ncdm} = 1.0132$, so this prior on $N_{ur}$ corresponds to $\Delta\Neff = [0,0.5]$. This upper bound of $\Delta\Neff<0.5$ ensures that the we investigate the full extent of $\Neff$ allowed by Planck data.

\begin{table*}[t]
\centering
    \caption{\footnotesize Posteriors for post-recombination decays (i.e.\ $\lgGamma = [1.49,3.24]$), corresponding to those shown in Fig.~\ref{fig:late_triangle} and Fig.~\ref{fig:Neff_posteriors}. Uncertainties are reported at $68\%$ C.L. and upper limits are given at $95\%$ C.L. }
\begin{tabular*}{\linewidth}{l@{\extracolsep{\fill}}ccc}
    \hline
    \hline
     & Planck+Pan+BAO & Planck+Pan+BAO & Planck+Pan+BAO+SPT+D/H  \\
     & & ($\Delta\Neff$) & ($\Delta\Neff$) \\
    \hline
    $R_\Gamma$ 
    & $ < 0.0124  $
    & $< 0.0115 $
    & $ < 0.0119 $ \\
    $H_0 \,[\frac{\SI{}{\kilo\meter}}{\SI{}{\second}\cdot\SI{}{\mega\parsec}}]$ 
    & $ 67.87^{+0.48}_{-0.54}   $
    & $68.46^{+0.58}_{-0.78} $
    & $ 68.46^{+0.54}_{-0.74}   $\\
    $\omega_{\mathrm{cdm}}$ 
    & $ 0.1195\pm 0.0010    $
    & $ 0.1212^{+0.0012}_{-0.0020} $
    & $ 0.1208^{+0.0012}_{-0.0017}   $\\
    $S_8$ 
    & $ 0.823\pm 0.013   $
    & $ 0.826\pm 0.013 $
    & $ 0.825\pm 0.012    $\\
    $\Neff$ 
    & ...
    & $< 3.32 $
    & $ < 3.29  $\\
    \hline
    \hline
\end{tabular*}
    \label{tab:long-lived_posteriors}
\end{table*}

Figure \ref{fig:Neff_posteriors} shows the 1D posterior distributions for $\omega_{\mathrm{cdm}}$ and $H_0$ resulting from these analyses that include a DR background, and Table \ref{tab:long-lived_posteriors} lists posterior values for select parameters. The dashed line shows the 1D posteriors of a $\Lambda$CDM model with fixed $\Delta\Neff = 0$, while the dotted line shows those of a $\Lambda$CDM model with the prior of $\Delta\Neff = [0, 0.5]$. In the absence of a DR background, the Planck data requires a reduction in the parameter $\omega_{\mathrm{cdm}}$ for long-lived scenarios due to the presence of the $Y$ particle during recombination. With the incorporation of a DR background, Planck data favors an increase in $\omega_{\mathrm{cdm}}$ because the additional DR component significantly contributes to the radiation energy density at the time of recombination. Comparing results from the $\Lambda$CDM+$\Delta\Neff$ model to those of a $Y$ decay with variable $\Delta\Neff$, the increase in $\omega_{\mathrm{cdm}}$ required by Planck to fix $\MRrec$ is not as large for the decay as for the $\Lambda$CDM+$\Delta\Neff$ model. This is because the $Y$ particle is still contributing to the matter density at recombination for long-lived cases and thus, for the same $\Delta\Neff$, the increase in $\omega_{\mathrm{cdm}}$ required by Planck to fix $\MRrec$ is not as large compared to that needed by a $\Lambda$CDM+$\Delta\Neff$ model. This can be seen in Fig.~\ref{fig:Neff_posteriors}.

With the addition of a DR background, there is an overall reduction in the size of the sound horizon for these long-lived cases. Consequently, the Planck data favor an increase in $H_0$ to fix $\theta_s$. The decrease in $\Omega_m$ required to match the Planck data for long-lived $Y$ particles without a DR background is constrained by Pantheon+ and BAO data. However, the inclusion of an ambient DR background results in an increase in both $\omega_{\mathrm{cdm}}$ and $H_0$, so $\Omega_m$ is minimally altered and thus the constraining power of Pantheon+ and BAO data on such scenarios is diminished. Therefore, the Planck+Pan+BAO posteriors shown in Fig.~\ref{fig:Neff_posteriors} for decays with a DR background are primarily the result of Planck constraints on $\Delta\Neff$. 

Naively one would expect the bounds on $R_\Gamma$ to relax with the inclusion of a DR background since $\MRrec$ can be kept fixed even with significant contributions from $\rho_Y$ at recombination. However, we find that limits on $R_\Gamma$ become slightly more stringent with the inclusion of a DR background. For long-lived scenarios in which $\rho_Y$ becomes negligible soon after recombination (i.e.\ $\lgGamma \approx 3)$, the $Y$ decay injects new DR after recombination and enhances the early ISW effect. The addition of a DR background exacerbates these changes to the early ISW effect. Therefore, marginalizing over the full range of \mbox{$\lgGamma = [1.49, 3.24]$}, we find that $R_\Gamma < 0.115$ (95\% C.L) for long-lived cases with a DR background constrained by Planck+Pan+BAO, whereas long-lived cases without a DR background yield $R_\Gamma < 0.124$ (95\% C.L) when constrained by Planck+Pan+BAO (see Table \ref{tab:long-lived_posteriors}).

The inclusion of a DR background increases the expansion rate during BBN and thereby alters the predicted abundance of primordial elements. Therefore, including bounds on the primordial deuterium abundance should constrain this effect. However, for the level of $\Delta\Neff$ that is allowed by Planck data, the predicted abundance of deuterium does not change significantly; $\Lambda$CDM constrained by Planck yields $10^{5} \mathrm{D/H} = 2.524\pm 0.028 $ (68\% C.L.) whereas $\Lambda$CDM+$\Delta\Neff$ constrained by Planck gives $10^{5} \mathrm{D/H} = 2.545\pm 0.033 $ (68\% C.L.). Thus, the inclusion of bounds on the primordial deuterium abundance does not significantly constrain long-lived scenarios that contain a DR background. The inclusion of SPT+D/H data has a minimal impact on the constraints on $\Delta\Neff$ and thereby the value of $\omega_{\mathrm{cdm}}$ needed to fix $\MRrec$, as can be seen in Fig.~\ref{fig:Neff_posteriors} and Table \ref{tab:long-lived_posteriors}. Furthermore, the posterior for $H_0$ is not significantly altered with the addition of D/H limits. In Sec.~\ref{sec:Effects_of_Y_Decay} we note that the contribution of $\rho_Y$ to Hubble during BBN can also affect primordial element abundances. However, for the long-lived regime, any contribution from $\rho_Y$ during BBN is negligible so we neglect this effect on the deuterium abundance for long-lived scenarios that contain a DR background. 

Coupled with the presence of an ambient DR background, $Y$ particle decays in the long-lived regime only reduce the $H_0$ tension with S$H_0$ES to $4.15\sigma$; we find that Planck+SPT+D/H+Pan+BAO yields $H_0 = 68.46^{+0.54}_{-0.74}\,\unit{\kilo\meter\per\second\per\mega\parsec}$ (68\% C.L.) for decays with a DR background in the long-lived regime.

\section{Summary and Conclusions} \label{sec:Summary_and_Conclusions}
We obtain updated and comprehensive constraints on the injection of DR after the time of BBN by considering a massive hidden sector particle, called the $Y$ particle, that decays into DR. We employ a modified version of the Boltzmann solver \textsc{CLASS-v3.2} \cite{blas_cosmic_2011} to model the effects of such a $Y$ decay and, in doing so, we determine the existence of an attractor solution for DR perturbations in synchronous gauge. Under the common assumption that adiabatic initial conditions are set by equating $\delta\rho/\dot{\bar{\rho}}$ between all fluids, one would expect the fractional density perturbation of DR ($\delta_\dr$) to equal 1/4 of the fractional density perturbation of photons ($\delta_\gamma$) in both conformal Newtonian and synchronous gauge. However, in synchronous gauge, an attractor solution enforces $\delta_\dr = (17/20)\delta_\gamma$. We demonstrate that this attractor rectifies any incorrect initial condition set for $\delta_\dr$ and so all previous works remain valid (see Fig.~\ref{fig:perturbations_initial_conditions}). This attractor stems from the fact that initial conditions in conformal Newtonian gauge are dominated by zero-order terms in a $k\tau$ expansion, whereas the first non-vanishing term in the corresponding synchronous gauge initial conditions are of order $(k\tau)^2$. We derive initial conditions in conformal Newtonian gauge for species that interact via a decay and demonstrate that, for the fluid that is being sourced by the decay, the $(k\tau)^2$ term differs from that of other fluids such that the ratio of perturbations ($\delta_1/\delta_2$) is not preserved between gauges. If there is no energy exchange between fluids, then $\delta_1/\delta_2$ is preserved between gauges. 

We determine constraints on the $Y$ particle decay rate ($\Gamma_Y$) and the maximum contribution that the $Y$ particle makes to the energy density of the Universe ($R_\Gamma$) by employing observations of CMB temperature and polarization anisotropies from both \textit{Planck} 2018 \cite{planck_collaboration_planck_2020-1} and SPT-3G \cite{chown_maps_2018,dutcher_measurements_2021,balkenhol_measurement_2022}, bounds on the primordial deuterium abundance \cite{cooke_one_2018, sobotka_was_2023-1}, and data from Pantheon+ \cite{scolnic_pantheon_2022,brout_pantheon_2022} and BAO observations from BOSS DR12 \cite{alam_clustering_2017}. We consider $Y$ particle lifetimes that range from right after the end of BBN to about 30 million years after recombination ($10^{-12.08}\unit{\giga\year} < \tau_Y < 10^{-1.49}\unit{\giga\year}$), ensuring that injection of DR occurs after BBN has completed. 

In the short-lived regime of $\Gamma_Y \gtrsim \Gammay{7}$ in which the $Y$ particle decays during radiation domination, decays primarily affect the CMB via changes in $\Neff$. For these short-lived cases, the $\Delta\Neff$ that arises from injected DR is solely a function of $R_\Gamma$ (see Fig.~\ref{fig:Neff_appendix} in Appendix \ref{sec:appendix_Neff}).  In comparison, the $\Delta\Neff$ arising from DCDM models is determined by $\Delta\Neff \propto \Gammadcdm^{-1/2}\fdcdm/(1 - \fdcdm)$ \cite{nygaard_updated_2021-1}, where $\Gammadcdm$ is the dark matter decay rate and $\fdcdm$ is the fraction of total dark matter energy density that is unstable. Therefore, while $\fdcdm$ can become unbounded by CMB observations as $\Gammadcdm$ increases, the parametrization used in this work avoids these convergence issues; the same bound on $R_\Gamma$ applies to all $\Gamma_Y$ in the short-lived regime. We find a $2\sigma$ upper bound of $R_\Gamma < 0.036$ for $\Gamma_Y \gtrsim \Gammay{7}$, which translates to $\rho_Y/\rho_{tot} < 0.0211$. These bounds are similar to those derived in \textcite{sobotka_was_2023-1} for the range of $\Gammay{10.03} \lesssim \Gamma_Y \lesssim \Gammay{12.08}$, where it was found that $\rho_Y/\rho_{tot} \leq 0.0235$ (95\% C.L.) for a model in which the $Y$ particle decays into a mixture of photons and DR. 

As $\Gamma_Y$ falls below $\Gammay{7}$, the $\Delta\Neff$ that arises from a $Y$ decay increases for fixed $R_\Gamma$ and thus Planck data demands increasingly large enhancements to $\omega_{\mathrm{cdm}}$ in order to keep the early ISW effect relatively fixed. This increase in $\omega_{\mathrm{cdm}}$ results in more suppression of temperature anisotropies. Furthermore, Y decays in the intermediate regime can influence some CMB scales via $\Delta\Neff$ while simultaneously suppressing other scales with the presence of the Y particle, thereby fostering increased asymmetry between peak heights in the temperature spectrum (see Sec.~\ref{sec:effects_intermediate}). Since it is difficult to compensate for these effects with changes in $n_s$ or $\omega_b$, CMB observations yield more stringent constraints on $R_\Gamma$ in the intermediate regime of \mbox{$\Gammay{3.22} \lesssim \Gamma_Y \lesssim \Gammay{7}$}. Even so, there is a special decay rate of $\Gamma_Y \approx \Gammay{4.9}$ for which the increase in $\omega_{\mathrm{cdm}}$ required by Planck and the extra suppression from the $Y$ particle alter the ratio of peak heights in such a way that can be compensated for by increasing $n_s$ (see Sec.~\ref{sec:intermediate-results}). This same preference for $\Gamma_Y \approx \Gammay{4.9}$ was also observed in \textcite{poulin_fresh_2016} and \textcite{holm_discovering_2022}. However, the inclusion of SPT data restricts this necessary change in $n_s$ and eliminates any preference for $\Gamma_Y \approx \Gammay{4.9}$.

For cases in which the $Y$ particle decays before recombination (\mbox{$\Gammay{3.14} \lesssim \Gamma_Y \lesssim \Gammay{12.08}$}), the injection of DR decreases the size of the sound horizon and so Planck data favor an increase in $H_0$ to fix $\theta_s$. When only constraining with Planck, these pre-recombination scenarios result in $R_\Gamma < 0.0357$ (95\% C.L.). The addition of SPT data and bounds on the primordial deuterium abundance primarily limit changes in $n_s$ or $\omega_b$ needed by intermediate cases, and thus marginalizing over the full pre-recombination range of \mbox{$\Gammay{3.14} \lesssim \Gamma_Y \lesssim \Gammay{12.08}$}, we find that Planck+SPT+D/H yields $R_\Gamma < 0.0364$ (95\% C.L.). This bound on $R_\Gamma$ translates to for \mbox{$\rho_Y/\rho_{tot} < 0.0305$}. Consequently, these pre-recombination decay scenarios are not successful at mitigating the Hubble tension; Planck+SPT+D/H results in \mbox{$H_0 = 68.10^{+0.67}_{-1.0} \unit{\kilo\meter\per\second\per\mega\parsec}$}, only bringing the tension between the CMB inferred value and results from S$H_0$ES from $5.2\sigma$ \cite{planck_collaboration_planck_2020-1} to $4.4\sigma$. Even with the inclusion of SPT data and bounds on the deuterium abundance, these limits on $H_0$ are in good agreement with those derived by \textcite{nygaard_updated_2021-1} for short-lived cases.

Long-lived scenarios in which the $Y$ particle lifetime extends past the time of recombination (\mbox{$\Gamma_Y \gtrsim  \Gammay{3.24}$}) are required by Planck observations to be coupled with a decrease in $\omega_{\mathrm{cdm}}$ and an increase in $H_0$. The combination of these effects leads to an overall reduction in $\Omega_m = \omega_m/h^2$. Therefore, long-lived cases are subject to constraints from probes of $\Omega_m$ such as Pantheon+ \cite{scolnic_pantheon_2022,brout_pantheon_2022} and BAO data from BOSS DR12 \cite{alam_clustering_2017}. When applying Planck, Pantheon+, and BAO data, these long-lived scenarios are constrained such that \mbox{$R_\Gamma < 0.0124$} (95\% C.L.), corresponding to \mbox{$\rho_Y/\rho_{tot} < 0.011$} for \mbox{$\Gammay{1.49} \lesssim \Gamma_Y \lesssim \Gammay{3.24}$}. In comparison, \textcite{mccarthy_converting_2023} derived bounds on generic models that convert some form of dark matter to DR after recombination by employing likelihoods for Planck, CMB lensing \cite{planck_collaboration_lensing}, BAO \cite{beutler_6df_2011,ross_clustering_2015,alam_clustering_2017}, Pantheon Type Ia supernovae \cite{scolnic_complete_2018}, and matter clustering from DES-Y3 \cite{des_collaboration_dark_2022}. \textcite{mccarthy_converting_2023} parametrize the amount of dark matter that converts into DR with the parameter $\zeta$ and report a $2\sigma$ upper bound of $\zeta < 0.0748$. We translate
\footnote{We analytically relate $R_\Gamma$ and $\Gamma_Y$ to $\zeta$ by employing Eq.~\eqref{eq:rho_Yi} and taking $a_i = 10^{-14}$ and $\rho_Y(a_\Gamma) \approx \rho_{Y,i}(a_i/a_\Gamma)^3$. We evaluate $\rho_{r,i}$ assuming a CMB temperature of \SI{2.7255}{\kelvin} and that 3 neutrino species are initially relativistic. $\rho_{m,i}$ is determined using our posterior distribution for $\omega_m$. It follows that $\zeta$ is determined by Eq.~(9) of \textcite{mccarthy_converting_2023} by setting their model parameters of $\kappa$ and $a_t$ equal to $2$ and $a_\Gamma$, respectively, and determining $\rho_{cdm,0}$ from our posterior distribution for $\omega_{\mathrm{cdm}}$. This transformation is applied to our MCMC chains to obtain a marginalized distribution for $\zeta$.}
our Planck+Pan+BAO posteriors for $Y$ decays in the long-lived regime to this parametrization and find a more stringent bound of $\zeta < 0.0321$ (95\% C.L.). This discrepancy can be attributed to a difference in priors for the particle lifetime. Whereas the post-recombination prior of \mbox{$\Gammay{1.49} \lesssim \Gamma_Y \lesssim \Gammay{3.24}$} used here corresponds to $ 10^{-3.05} \lesssim a_\Gamma \lesssim 10^{-1.92}$, \textcite{mccarthy_converting_2023} consider $ 10^{-4} < a_t < 10^{4}$ where $a_t \approx a_\Gamma$. \textcite{mccarthy_converting_2023} see a peak in the posterior of $a_t$ at $a_t \approx 1$ that broadens the marginalized posterior for $\zeta$, leading to a bound on $\zeta$ that is much less stringent than that derived here. 

We find that decays in the long-lived regime result in $S_8 = 0.823 \pm 0.013$ (68\% C.L.) for Planck+Pan+BAO, only reducing the tension between DES-Y3 reported bounds and those inferred from the CMB from $2.5\sigma$ \cite{planck_collaboration_planck_2020-1} to $2.3\sigma$. Furthermore, compared to pre-recombination decay cases, these long-lived scenarios are not as effective at mitigating the Hubble tension; Planck+Pan+BAO results in $H_0 = 67.87^{+0.48}_{-0.54}\,\unit{\kilo\meter\per\second\per\mega\parsec}$ (68\% C.L.) for the long-lived regime, which agrees with results of comparable long-lived DCDM studies (\cite{nygaard_updated_2021-1, mccarthy_converting_2023}).

Since Pantheon+ and BAO constraints on long-lived scenarios are driven by the need to decrease $\omega_{\mathrm{cdm}}$ such that the early ISW effect is relatively fixed, introducing more radiation at recombination can relax these bounds. Including an ambient background of DR in addition to the DR sourced by the $Y$ decay can enhance the resulting posterior for $H_0$ compared to long-lived cases that lack a DR background. We model a DR background by allowing a constant positive $\Delta\Neff$ and, with a prior of $\Delta\Neff = [0,0.5]$, we find that long-lived cases constrained by Planck+SPT+D/H+Pan+BAO yield \mbox{$H_0 = 68.46^{+0.54}_{-0.74} \,\unit{\kilo\meter\per\second\per\mega\parsec}$} (68\% C.L.). These long-lived scenarios with a DR background are more successful at mitigating the Hubble tension than pre-recombination decay scenarios without a DR background. 

Overall, Planck+SPT+D/H+Pan+BAO demonstrates less constraining power for $Y$ decays in the short-lived regime, reflecting the fact that scenarios in which the $Y$ particle lifetime is comparable to or greater than the time of recombination are strongly disfavored by CMB data and probes of $\Omega_m$. Marginalizing over the full range of $Y$ particle lifetimes considered in this work, Planck+SPT+D/H+Pan+BAO data yields $R_\Gamma < 0.0360$ (95\% C.L.) which implies $\rho_Y/\rho_{tot} < 0.0302$. Therefore, the $Y$ particle is restricted to only contribute a maximum of about $3\%$ of the energy density of the Universe. As a result, $Y$ decay scenarios are sufficiently constrained such that they are not effective at mitigating the $H_0$ or $S_8$ tensions. Considering a DR background in addition to the DR sourced by $Y$ decays in the long-lived regime can aid in bolstering $H_0$ further, but still fall far short at resolving the tension with measurements from S$H_0$ES \cite{riess_comprehensive_2022}.

This work affirms the notion that the production of relativistic particles of any kind after BBN is strongly constrained. The injection of photons and/or relativistic electrons has been thoroughly investigated (\cite{poulin_non-universal_2015,poulin_cosmological_2017,kawasaki_revisiting_2018,hufnagel_bbn_2018,forestell_limits_2019,kawasaki_big-bang_2020,balazs_cosmological_2022}) and these situations are subject to stringent constraints from measurements of primordial element abundances given their ability to photodisintegrate deuterium. Furthermore, the production of photons that do not photodisintegrate deuterium can introduce spectral distortions in the CMB and therefore the injection of new photons is restricted to be within the first $\sim10^{-11}\unit{\giga\year}$ after the big bang \cite{sobotka_was_2023-1}. Such photon injection reduces the $\Delta\Neff$ associated with DR injection and relaxes CMB constraints on the amount of DR present during recombination. However, to maintain consistency with the precisely measured present-day CMB temperature, such scenarios must alter the baryon-to-photon ratio during BBN and are therefore strongly constrained by measurements of the primordial deuterium abundance \cite{sobotka_was_2023-1}. While the injection of DR alone is not significantly constrained by BBN, we find that the bounds placed on such scenarios by CMB anisotropy observations and probes of $\Omega_m$ are similar to the stringent bounds placed on scenarios that inject photons. In light of these results, it is evident that the production of any new free-streaming relativistic particles in the early Universe is highly constrained by cosmological observations. As a result, such extensions to $\Lambda$CDM cosmology are not capable of resolving the Hubble and $S_8$ tensions. This exemplifies the challenges faced when trying to resolve these tensions with early Universe modifications alone \cite{clark_h_0_2021-1, jedamzik_why_2021, vagnozzi_seven_2023}. Even so, examples of models that succeed in addressing the $H_0$ and $S_8$ tensions with only pre-recombination modifications to cosmic evolution do exist \cite{schoneberg_h_0_2022, poulin_ups_and_downs, aloni_step_2022-2, joseph_step_2023, Allali:2021azp, cruz_cold_2023-1}.

\section{Acknowledgements} \label{sec:Acknowledgements}
This work utilized the Longleaf Computing Cluster owned by the University of North Carolina at Chapel Hill. ACS and ALE received support from NSF Grants PHY-1752752 and PHY-2310719 during this investigation. TLS is supported by NSF Grants No.~2009377 and No.~2308173 and thanks the Center for Cosmology and Particle Physics (NYU) where part of this work was completed.

\clearpage
\appendix
\vspace{1cm}
\section{Model Implementation} \label{sec:appendix_parametrization}
We modify \textsc{CLASS} to solve Eqs.~\eqref{eq:rho_Y} and \eqref{eq:rho_dr} when given initial values for $\rho_Y$ and $\rho_{\dr}$. To determine these values, we derive a mapping from $R_\Gamma$ and $\Gamma_Y$ to $\rho_{Y,i}$ and $\rho_{\dr,i}$. Combining Eqs.~\eqref{eq:R_Gamma} and \eqref{eq:aGamma_defintion} determines $a_\Gamma/a_i$ as a function of $\Gamma_Y$ and $R_\Gamma$, which can then be inserted back into Eq.~\eqref{eq:R_Gamma} to solve for $\rho_{Y,i}$. However, the function for $a_\Gamma/a_i$ must be determined separately for the regimes in which the massive neutrinos are relativistic or non-relativistic at $a_\Gamma$. The evolution of $\rho_{ncdm}$ can be approximated by a broken power law that pivots from scaling as $a^{-4}$ to $a^{-3}$ at some pivot scale factor $a_p$:
\begin{equation}
\rho_{ncdm}(a)=
    \begin{cases}
        \rho_{ncdm,i}\left( \frac{a_i}{a} \right)^4 & \text{if } a < a_p\\
        \rho_{ncdm,i}\left( \frac{a_i}{a_p} \right)^4\left( \frac{a_p}{a} \right)^3 & \text{if } a > a_p \label{eq:rho_ncdm} \\
    \end{cases}.
\end{equation}
Here, $a_p = a_0(T_0/T_p)(4/11)^{1/3}$ and $T_p = m_\nu/3.15$ \cite{ganjoo_effects_2023}, with $m_\nu$ being the mass of the massive neutrino species. We take the minimal assumption that neutrinos are composed of two massless species and one massive species with $m_\nu = \SI{0.06}{\electronvolt}$. Combining Eq.~\eqref{eq:rho_ncdm} and Eq.~\eqref{eq:R_Gamma}, we find that
\begin{equation}
\frac{\rho_{Y,i}}{\rho_{r,i}}= R_\Gamma \times
    \begin{cases}
        \left(\frac{a_i}{a_\Gamma} + \frac{\rho_{m,i}}{\rho_{r,i}} \right), & a_\Gamma < a_p\\
        \left[\epsilon\left(\frac{a_i}{a_\Gamma}\right) + \frac{\rho_{m,i}}{\rho_{r,i}} + (1-\epsilon)\left(\frac{a_i}{a_p}\right)\right], & a_\Gamma > a_p\\ 
    \end{cases} \label{eq:rho_Yi}
\end{equation}
where $\epsilon \equiv \rho_{sr,i}/\rho_{r,i}$ and $\rho_{r,i} = \rho_{sr,i} + \rho_{ncdm,i}$. Inserting this initial condition back into Eq.~\eqref{eq:aGamma_defintion}, we obtain the quartic equation 
\begin{equation}
     \left(\frac{a_\Gamma}{a_i}\right)^4 + c\left(\frac{a_\Gamma}{a_i}\right) + d = 0, \\\label{eq:quartic}
\end{equation}
where
\begin{align}
     &d = -\left(\frac{H_i}{\Gamma_Y}\right)^2\left(R_\Gamma + 1\right) ,\\
     &c=d\times
    \begin{cases}
        \left(\frac{\rho_{m,i}}{\rho_{r,i}}\right) & \text{if } a_\Gamma < a_p\\
        \left[\frac{\rho_{m,i}}{\rho_{r,i}} + (1-\epsilon)\left(\frac{a_i}{a_p} \right)  \right] & \text{if } a_\Gamma > a_p\\ 
    \end{cases}.
\end{align}
Equation \eqref{eq:quartic} can be solved analytically. Therefore, under the assumption of $a_i$ being deep in radiation domination such that $H_i$ is completely determined by $\rho_{r,i}$, a value for $a_\Gamma/a_i$ can be determined when equipped with values for $R_\Gamma$, $\Gamma_Y$, $\rho_{m,i}$, and $\rho_{r,i}$. From there, $\rho_{Y,i}$ is determined via Eq.~\eqref{eq:rho_Yi}. Finally, solving Eq.~\eqref{eq:rho_dr} under the assumption of radiation domination such that $H(a) \approx H_i (a/a_i)^{-2}$ leads to $\rho_{\dr,i} = (\Gamma_Y/3H_i) \rho_{Y,i}$. We modify \textsc{CLASS} to derive these initial conditions for $\rho_Y$ and $\rho_{\dr}$ when given values for $R_\Gamma$ and $\Gamma_Y$. Note that Eq.~\eqref{eq:rho_ncdm} is only used when relating $R_\Gamma$ and $\Gamma_Y$ to initial conditions; we do not modify the default calculations for $\rho_{ncdm}$ in \textsc{CLASS}.

\vspace{1cm}
\section{Initial Conditions} \label{sec:appendix_initial_conditions}

\subsection{Initial Conditions for DR Perturbations} \label{sec:appendix_boltzmann_perturbations}
We can derive the attractor initial condition for $\delta^\sync_\dr$, where the superscript (s) denotes synchronous gauge, by analytically solving the Boltzmann equations. The scalar perturbation equations for the $Y$ particle and DR follow from the continuity and Euler equations and, in synchronous gauge, these result in
\begin{align}
    \delta'^\sync_{\dr} &= -\frac43\theta^\sync_{\dr} -\frac43 \left(\frac{h'}{2}\right) + a \Gamma_Y \frac{\rho_Y}{\rho_\dr } \left(\delta^\sync_Y - \delta^\sync_\dr \right) , \label{eq:delta_dr_prime_appendix} \\
    \theta'^\sync_\dr &= \frac{k^2}{4}\delta^\sync_\dr  - a\Gamma_Y \frac{3\rho_Y}{4\rho_\dr} \left( \frac43\theta^\sync_\dr - \theta^\sync_Y  \right). \label{eq:theta_dr_prime_appendix} 
\end{align}
Here, a prime denotes a derivative with respect to conformal time and we have assumed that anisotropic stress is initially negligible. Assuming initial conditions are set during radiation domination, $\delta^\sync_\gamma = -(2/3)C(k\tau)^2$ and $h = C(k\tau)^2$ \cite{ma_cosmological_1995}, where $C$ is an arbitrary constant. As demonstrated in Sec.~\ref{sec:perturbations}, if the evolution of an interacting fluid's energy density is unaffected by its interaction, then $\delta_i/\delta_\gamma$ is the same in both synchronous and conformal Newtonian gauge, where $\delta_i$ is the density perturbation of the interacting fluid. Since $\rho_Y \propto a^{-3}$ initially, we know $\delta^\sync_Y = (3/4)\delta^\sync_\gamma$ and $\theta^\sync_Y = 0$. Additionally, if we take $H(a) = H_i (a/a_i)^{-2}$, then the initial evolution of $\rho_\dr$ can be determined by solving Eq.~\eqref{eq:rho_dr} analytically: $\rho_\dr(a) = (\Gamma_Y \rho_{Y,i}/3H_i) (a/a_i)^{-1}$. It follows that 
\begin{equation}
     \Gamma_Y \frac{\rho_Y}{\rho_\dr} = 3 H .\\ \label{eq:rho_Y/rho_dr_initial}
\end{equation}
For adiabatic initial conditions, the DR perturbations will take the same form as standard radiation in that the first non-vanishing terms in a power series expansion in ($k\tau$) for $\delta_\dr^\sync$ and $\theta_\dr^\sync$ are 
\begin{equation}
\delta^\sync_\dr = D (k\tau)^2, \hspace{1cm}
\theta^\sync_\dr = E k^4 \tau^3,
\end{equation}
where $D$ and $E$ are constants. Recalling that \mbox{$h = C(k\tau)^2$}, \mbox{$\delta^\sync_Y = (3/4)\delta_\gamma^\sync = -(1/2)C(k\tau)^2$}, \mbox{$\theta^\sync_Y=0$}, and that $\rho_Y/\rho_\dr$ is given by Eq.~\eqref{eq:rho_Y/rho_dr_initial}, it follows that Eqs.~\eqref{eq:delta_dr_prime_appendix} and \eqref{eq:theta_dr_prime_appendix} reduce to the algebraic equations of
\begin{align}
    2Dk^2\tau &= -\frac{4}{3}Ek^4\tau^3 - \frac{2}{3}(2Ck^2\tau) -3k^2\tau\left(\frac{C}{2} + D \right),\label{eq:delta_intermediate} \\
    3Ek^4\tau^2 &= \frac{1}{4}Dk^4\tau^2 - 3Ek^4\tau^2.  \label{eq:theta_intermediate}
\end{align}
Solving Eqs.~\eqref{eq:delta_intermediate} and \eqref{eq:theta_intermediate}, we find that
\begin{align}
    D &= 24 E, \\
    E &= -\frac{17C}{8(90 + k^2\tau^2)} \approx -\frac{17C}{720}.
\end{align}
Therefore, $\delta^\sync_\dr = -(17/30)C(k\tau)^2$ and $\delta^\sync_\dr/\delta^\sync_\gamma = (17/20)$. We enforce this initial condition in the perturbation module of CLASS. Note that while this derivation does not formally prove $\delta^\sync_\dr/\delta^\sync_\gamma = (17/20)$ is an attractor solution, the numerical solution shown in Fig.~\ref{fig:perturbations_initial_conditions} confirms that $\delta^\sync_\dr$ quickly converges to $(17/20)\delta^\sync_\gamma$.

\vspace{1cm}
\subsection{Generalized initial conditions for non-interacting species } \label{sec:appendix_k2_noninteracting}

We employ an iterative approach to derive a general expression for adiabatic initial conditions of scalar perturbations in conformal Newtonian gauge. We consider three non-interacting species with energy densities $\rho_1$, $\rho_2$, and $\rho_d$, where $\rho_d$ is assumed to dominate the energy density of the Universe. Since these fluids do not exchange energy, the evolution of their energy densities are determined by their respective equation of state parameters $w_1$, $w_2$, and $w_d$ (see Eqs.~\eqref{eq:case1_rhod}-\eqref{eq:case1_rho2}).

The perturbation equations for these species can be derived by perturbing the covariant form of Eqs.~\eqref{eq:case1_rhod}-\eqref{eq:case1_rho2}. We treat each species as a perfect fluid with an energy momentum tensor of
\begin{equation}
    ^{(i)}T^{\mu\nu} = p_i g^{\mu\nu} + \left( \rho_i + p_i \right) u_{(i)}^\mu u_{(i)}^\nu, \label{eq:energy_momentum_tensor}
\end{equation}
were $i$ denotes each individual fluid, $p_i$ is the pressure of the fluid, and $u_{(i)}^\mu \equiv dx^\mu/d\lambda$ is the fluid's four velocity. Since these fluids are non-interacting 
\begin{equation}
    \nabla_\mu \left( ^{(i)}T^\mu_\nu\right) = 0. \label{eq:covariant_noninteracting}
\end{equation}
The perturbation equations are found by evaluating Eq.~\eqref{eq:covariant_noninteracting} with the perturbed conformal Newtonian metric in Eq.~\eqref{eq:Newt_metric} and introducing perturbations to the energy density of each fluid $\rho_i(t, \vec{x}) = \rho_i^0(t)\left[1 + \delta_i(t, \vec{x}) \right]$, where $\rho_i^0(t)$ is the background energy density of each fluid and $\delta_i(t,\vec{x}) \equiv \delta \rho_i/\rho_i^0$ is the fractional density perturbation of a fluid. We also introduce perturbations to the four-velocity of each fluid: $u^0 = (1 - \Psi)$ and $u^j_{(i)} = (1-\Psi)V^j_{(i)}$, where $V^j_{(i)} \equiv dx^j/dt$ is the peculiar velocity of fluid $i$.

The $\nu = 0$ component of Eq.~\eqref{eq:covariant_noninteracting} reduces to 
\begin{equation}
    \frac{\dd\delta_i}{\dd t} + (1+w_i)\frac{\theta_i}{a} + 3(1+w_i)\frac{\dd\Phi}{\dd t} = 0, \label{eq:non-interacting_ddelta_dt}
\end{equation}
where $\theta_i \equiv a \partial_j V^j_{(i)}$ is the divergence of the fluid's conformal velocity. The divergence of the spatial component ($\nu = j$) of Eq.~\eqref{eq:covariant_noninteracting} results in 
\begin{equation}
    \frac{\dd \theta_i}{ \dd t} + (1-3w_i)H\theta_i + \frac{\nabla^2 \Psi}{a} + \frac{w_i}{1+w_i}\frac{\nabla^2 \delta_i}{a} = 0. \label{eq:non-interacting_dtheta_dt}
\end{equation}
We apply Eqs.~\eqref{eq:non-interacting_ddelta_dt} and \eqref{eq:non-interacting_dtheta_dt} to the three non-interacting fluids we are considering and the resulting suite of equations that we solve is
\begin{widetext}
    \begin{align}
    a^2 E(a)\delta'_1(a) + (1+w_1)\tilde{\theta}_1(a) + 3(1+w_1)a^2E(a) \Phi'(a) &= 0, \label{eq:non-interacting_deltaprime_1} \\
    a^2E(a)\tilde{\theta}'_1(a) + (1-3w_1)aE(a)\tilde{\theta}_1(a) + \tilde{k}^2\Phi(a) - \left(\frac{w_1}{1+w_1}\right)\tilde{k}^2\delta_1(a) &= 0, \label{eq:non-interacting_thetaprime_1} \\
    a^2 E(a)\delta'_2(a) + (1+w_2)\tilde{\theta}_2(a) + 3(1+w_2)a^2E(a) \Phi'(a) &= 0, \label{eq:non-interacting_deltaprime_2} \\
    a^2E(a)\tilde{\theta}'_2(a) + (1-3w_2)aE(a)\tilde{\theta}_2(a) + \tilde{k}^2\Phi(a) - \left(\frac{w_2}{1+w_2}\right)\tilde{k}^2\delta_2(a) &= 0, \label{eq:non-interacting_thetaprime_2} \\
    a^2 E(a)\delta'_d(a) + (1+w_d)\tilde{\theta}_d(a) + 3(1+w_d)a^2E(a) \Phi'(a) &= 0, \label{eq:non-interacting_deltaprime_d} \\
    a^2E(a)\tilde{\theta}'_d(a) + (1-3w_d)aE(a)\tilde{\theta}_d(a) + \tilde{k}^2\Phi(a) - \left(\frac{w_d}{1+w_d}\right)\tilde{k}^2\delta_d(a) &= 0, \label{eq:non-interacting_thetaprime_d} \\
    \tilde{k}^2\Phi(a) + 3a^2 E^2(a) \left[ a\Phi'(a) + \Phi(a) \right] &= \frac{3}{2} a^2 \left[\tilde{\rho}_d(a) \delta_d(a) + \tilde{\rho}_1(a) \delta_1(a) + \tilde{\rho}_2(a) \delta_2(a) \right]. \label{eq:non-interacting-time-time}
    \end{align}
\end{widetext}
Here, $E(a) \equiv H(a)/H_i$, $\tilde{k} \equiv k/H_i$, $\tilde{\theta}\equiv\theta/H_i$, \mbox{$\tilde{\rho} \equiv \rho/\rho_{crit,i}$}, and a prime denotes a derivative with respect to scale factor, $a$. Equation \eqref{eq:non-interacting-time-time} is the perturbed time-time component of the Einstein equation. 

We assume perturbations evolve from an initial scale factor of $a_i=1$ and that the initial time is set sufficiently early such that all modes of interest are super-horizon ($\tilde{k} < 1$). In conformal Newtonian gauge, perturbations are constant outside the horizon at zeroth order in $k/aH$ so we begin by setting $\Phi' = 0$. If $\rho_d$ dominates the energy density of the Universe, then $E(a) \approx a^{-\frac{3}{2}(1+w_d)}$ and $\tilde{\rho}_d(a) \approx a^{-3(1+w_d)}$ because $\rho_{d,i}/\rho_{crit,i}\approx 1$. Under these conditions and dropping terms proportional to $\tilde{k}^2$ and $\Phi'$, Eq.~\eqref{eq:non-interacting-time-time} results in $\delta_d(a_i) = 2\Phi(a_i)$ at zeroth order in $k/aH$. This result can then be used to simplify Eq.~\eqref{eq:non-interacting_thetaprime_d}, which has the solution
\begin{equation}
    \tilde{\theta}_d = - \frac{2a^{\frac12 (1+3w_d)}}{3(1+w_d)}  \tilde{k}^2 \Phi_p,  \label{eq:non-interacting_theta_d_solution}
\end{equation}
where $\Phi_p \equiv \Phi(a_i)$. The superhorizon initial conditions for $\delta_1$ and $\delta_2$ at zeroth order in $k/aH$ can be found by combining Eqs.~\eqref{eq:non-interacting_deltaprime_1}, \eqref{eq:non-interacting_deltaprime_2}, and \eqref{eq:non-interacting_deltaprime_d} while neglecting $\tilde{\theta}\propto\tilde{k}^2$ terms. In doing so, we find that $\delta_1 = 2\Phi_p (1+w_1)/(1+w_d)$ and $\delta_2 = 2\Phi_p (1+w_2)/(1+w_d)$. Combining these results with Eqs.~\eqref{eq:non-interacting_thetaprime_1} and \eqref{eq:non-interacting_thetaprime_2} and solving, it follows that $\tilde{\theta}_d=\tilde{\theta}_1=\tilde{\theta}_2$. This result is a manifestation of the fact that adiabatic initial conditions in conformal Newtonian gauge have the same velocity perturbations between all fluids \cite{ma_cosmological_1995}.

While $\Phi' = 0$ at zeroth order in $k/aH$, $\Phi$ can evolve at higher orders in $k/aH$. To determine this evolution, we begin by simplifying Eq.~\eqref{eq:non-interacting-time-time} to 
\begin{equation}
    \frac13 a^{1-3w_d} \tilde{k}^2 \Phi(a) + a\Phi'(a) + \Phi(a) = \frac12 \delta_d. \label{eq:non-interacting_Phi_simple}
\end{equation}
Differentiating Eq.~\eqref{eq:non-interacting_Phi_simple} with respect to scale factor and combining with Eqs.~\eqref{eq:non-interacting_deltaprime_d} and \eqref{eq:non-interacting_theta_d_solution}, we obtain the differential equation 
\begin{align*}
    & 0 = a\Phi''(a) + \Phi'(a) \left[ \frac13 a^{1+3w_d}\tilde{k}^2 + \frac32 (1+w_d)  + 2 \right] \\
    & \hspace{4.65cm}+ \Phi(a)\left[w_d a^{3w_d}\tilde{k}^2 \right].  \stepcounter{equation}\tag{\theequation}\label{eq:non-interacting_Phi_differential}
\end{align*}
Solving Eq.~\eqref{eq:non-interacting_Phi_differential} with the ansatz of $\Phi(\tau) = A + Bk^2\tau^2$, where
\begin{equation}
    \tau = \left(\frac{2}{1+3w_d} \right)\frac{a^{\frac{1}{2}(1+3w_d)}}{H_i}, \label{eq:non-interacting_tau_transform}
\end{equation}
results in
\begin{align*}
    &\Phi = \Phi_p - \left[\frac{2 w_d  a^{1+3w_d }}{4 - 24(1+w_d) + 27(1+w_d)^2 } \right] \tilde{k}^2 \Phi_p \\
    &\hspace{6.4cm} +\mathcal{O}(\tilde{k}^3).\stepcounter{equation}\tag{\theequation} \label{eq:non-interacting_Phi_solution}
\end{align*}
Combining Eqs.~\eqref{eq:non-interacting_theta_d_solution} and \eqref{eq:non-interacting_Phi_solution} with either Eq.~\eqref{eq:non-interacting_deltaprime_1}, \eqref{eq:non-interacting_deltaprime_2}, or \eqref{eq:non-interacting_deltaprime_d}, it follows that the adiabatic initial condition for each of the three fluids takes the form of
\begin{align*}
    &\delta_j \simeq 2\left(\frac{1+w_j}{1+w_d}\right)\Phi_p \\ 
    &\hspace{0.5cm}+ \frac{2}{3}\left(\frac{1+w_j}{1+w_d}\right)\left[ \frac{7 + 18w_d + 9w_d^2 }{7+30w_d + 27w_d^2}  \right] a^{(1+3w_d)}\tilde{k}^2 \Phi_p. \stepcounter{equation}\tag{\theequation} \label{eq:non-interacting_delta_solution}
\end{align*}
Or, applying the transformation of Eq.~\eqref{eq:non-interacting_tau_transform},
\begin{align*}
    &\delta_j \simeq 2\left(\frac{1+w_j}{1+w_d}\right)\Phi_p \\ 
    &\hspace{0.5cm}+\frac{2}{3}\left(\frac{1+w_j}{1+w_d}\right)\left[ \frac{7 + 39w_d + 63w_d^2 + 27w_d^3 }{28 + 36w_d}   \right] (k\tau)^2 \Phi_p. \stepcounter{equation}\tag{\theequation} \label{eq:non-interacting_delta_solution_tau}
\end{align*}
Note that the term in square brackets is only dependent on $w_d$ and is therefore the same between all individual fluids. Even at next-to-leading order in $k\tau$, the ratio of initial conditions between any two non-interacting fluids is $\delta_i/\delta_j = (1+w_i)/(1+w_j)$.


\vspace{1cm}
\subsection{Generalized initial conditions for interacting case } \label{sec:appendix_k2_interacting}
Here we employ a similar approach to that used in Appendix \ref{sec:appendix_k2_noninteracting} in order to derive the superhorizon adiabatic initial conditions for interacting fluids in conformal Newtonian gauge. Let us consider three fluids with energy densities $\rho_1$, $\rho_2$, and $\rho_d$, where $\rho_d$ dominates the energy density of the Universe and species 1 is a massive particle ($w_1=0$) that decays into species 2 with a decay rate $\Gamma$. Equations \eqref{eq:case2_rhod}-\eqref{eq:case2_rho2} describe the energy density evolution of the three species. Each species is assumed to be a perfect fluid with an energy momentum tensor given by Eq.~\eqref{eq:energy_momentum_tensor}, and so the covariant form of Eqs.~\eqref{eq:case2_rhod}-\eqref{eq:case2_rho2} can be written as 
\begin{equation}
    \nabla_\mu \left( ^{(i)}T^\mu_\nu\right) = Q_\nu^{(i)}, \label{eq:covariant_interacting}
\end{equation}
where $i$ specifies each fluid. It follows from Eqs.~\eqref{eq:case2_rhod}-\eqref{eq:case2_rho2} that 
\begin{align}
    Q^{(d)}_\nu &= 0, \label{eq:Qd}\\
    Q^{(1)}_\nu &=  ^{(1)}T_{\mu\nu}u^\mu_{(1)} \Gamma, \label{eq:Q1}\\
    Q^{(2)}_\nu &= - Q^{(1)}_\nu\label{eq:Q2} .
\end{align}
We derive the perturbation equations by evaluating Eq.~\eqref{eq:covariant_interacting} with the perturbed conformal Newtonian metric of Eq.~\eqref{eq:Newt_metric} and enforcing the same perturbations to the fluid energy density and velocity as in Appendix \ref{sec:appendix_k2_noninteracting}: $\rho_i(t, \vec{x}) = \rho_i^0(t)[ 1 + \delta_i(t, \vec{x}) ]$, $u^0 = (1-\Psi)$, and $u^j_{(i)} = (1-\Psi)V^j_{(i)}$. From these perturbations, we find
\begin{align}
    Q_0^{(1)} &= \Gamma \rho_1^0 \left(1 + \delta_1 + \Psi \right), \\
    Q_j^{(1)} &= -\Gamma \rho_1^0 a^2 \delta_{kj} V^k_1 . 
\end{align}
Here, the quantity $Q_0^{(1)}$ is composed of a zeroth-order piece, $Q_0^{(1),(0)} \equiv \Gamma \rho_1^0$, and a first order component, $Q_0^{(1),(1)} \equiv \Gamma \rho_1^0 (\delta_1 + \Psi)$. It follows that the $\nu=0$ component of Eq.~\eqref{eq:covariant_interacting} lends
\begin{equation}
      \frac{\dd\delta_i}{\dd t} + (1+w_i)\frac{\theta_i}{a} + 3(1+w_i)\frac{\dd\Phi}{\dd t} = \frac{1}{\rho^0_i}\left[ Q_0^{(i),(0)} \delta_i - Q_0^{(i),(1)} \right], \label{eq:interacting_ddelta_dt}
\end{equation}
and the divergence of the $\nu = j$ component of Eq.~\eqref{eq:covariant_interacting} results in 
\begin{align*}
    \frac{\dd \theta_i}{ \dd t} + (1-3w_i)H\theta_i + \frac{\nabla^2 \Psi}{a} + \frac{w_i}{1+w_i}\frac{\nabla^2 \delta_i}{a} =& \\
    \frac{1}{\rho_i^0}\left[ \frac{\partial_j Q^{(i)}_j}{a(1+w_i)}  + Q^{(i),(0)}_0 \theta_i  \right]&. \stepcounter{equation}\tag{\theequation}\label{eq:interacting_dtheta_dt}
\end{align*}
Considering the energy exchange between each species described by Eqs.~\eqref{eq:Qd}-\eqref{eq:Q2}, the suite of equations that derive from Eqs.~\eqref{eq:interacting_ddelta_dt} and \eqref{eq:interacting_dtheta_dt} are
\begin{widetext}
    \begin{align}
    a^2 E(a)\delta'_1(a) + \tilde{\theta}_1(a) + 3a^2E(a) \Phi'(a) &= a\tilde{\Gamma}\Phi(a), \label{eq:interacting_deltaprime_1} \\
    a^2E(a)\tilde{\theta}'_1(a) + aE(a)\tilde{\theta}_1(a) + \tilde{k}^2\Phi(a) &= 0, \label{eq:interacting_thetaprime_1} \\
    a^2 E(a)\delta'_2(a) + (1+w_2)\tilde{\theta}_2(a) + 3(1+w_2)a^2E(a) \Phi'(a) &= a\tilde{\Gamma}\frac{\tilde{\rho}_1(a)}{\tilde{\rho}_2(a)} \left[\delta_1(a) - \delta_2(a) - \Phi(a) \right], \label{eq:interacting_deltaprime_2} \\
    a^2E(a)\tilde{\theta}'_2(a) + (1-3w_2)aE(a)\tilde{\theta}_2(a) + \tilde{k}^2\Phi(a) - \left(\frac{w_2}{1+w_2}\right)\tilde{k}^2\delta_2(a) &= a\tilde{\Gamma}\frac{\tilde{\rho}_1(a)}{\tilde{\rho}_2(a)} \left[ \frac{\tilde{\theta}_1(a)}{1+w_2} - \tilde{\theta}_2(a) \right] , \label{eq:interacting_thetaprime_2} \\
    a^2 E(a)\delta'_d(a) + (1+w_d)\tilde{\theta}_d(a) + 3(1+w_d)a^2E(a) \Phi'(a) &= 0, \label{eq:interacting_deltaprime_d} \\
    a^2E(a)\tilde{\theta}'_d(a) + (1-3w_d)aE(a)\tilde{\theta}_d(a) + \tilde{k}^2\Phi(a) - \left(\frac{w_d}{1+w_d}\right)\tilde{k}^2\delta_d(a) &= 0, \label{eq:interacting_thetaprime_d} \\
    \tilde{k}^2\Phi(a) + 3a^2 E^2(a) \left[ a\Phi'(a) + \Phi(a) \right] &= \frac{3}{2} a^2 \left[\tilde{\rho}_d(a) \delta_d(a) + \tilde{\rho}_1(a) \delta_1(a) + \tilde{\rho}_2(a) \delta_2(a) \right]. \label{eq:interacting-time-time}
    \end{align}
\end{widetext}
Here, $\tilde{\Gamma} \equiv \Gamma/H_i$ and again Eq.~\eqref{eq:interacting-time-time} is the perturbed time-time Einstein equation.

At zeroth order in $k/aH$, $\Phi'=0$ outside the horizon in conformal Newtonian gauge. As in Appendix \ref{sec:appendix_k2_noninteracting}, we assume $\tilde{\rho}_d \gg \tilde{\rho}_1,\tilde{\rho}_2$ such that $E(a) \approx a^{-\frac{3}{2}(1+w_d)}$ and $\tilde{\rho}_d(a) \approx a^{-3(1+w_d)}$. Applying the superhorizon limit ($\tilde{k}^2, \Phi' \rightarrow 0$) to Eq.~\eqref{eq:interacting-time-time} results in $\delta_d(a_i) = 2\Phi(a_i)$ at zeroth order in $k/aH$. This solution can be combined with Eq.~\eqref{eq:interacting_thetaprime_d} which results in $\tilde{\theta}_d$ being equivalent to Eq.~\eqref{eq:non-interacting_theta_d_solution}. The initial condition at zeroth order in $k/aH$ for $\delta_1$ can be derived by combining Eqs.~\eqref{eq:interacting_deltaprime_1} and \eqref{eq:interacting_deltaprime_d} and dropping $\tilde{\theta} \propto \tilde{k}^2$ terms. Additionally, we assume that the initial time is set sufficiently early such that $\Gamma/H_i$ is initially small. This leads to $\delta_1 = 2\Phi_p/(1+w_d)$ at zeroth order in $k/aH$, which mirrors the non-interacting case of Appendix \ref{sec:appendix_k2_noninteracting} since $w_1 = 0$. To solve for the initial condition of $\delta_2$ at zeroth order in $k/aH$, we combine the zeroth-order result for $\delta_1$ with Eq.~\eqref{eq:interacting_deltaprime_2} while neglecting any $\tilde{\theta} \propto \tilde{k}^2$ terms, and enforcing that $\delta'_2 = \Phi' = 0$ on superhorizon scales. Doing so results in $\delta_2 = \Phi_p [(1- w_d)/(1+w_d)]$ at zeroth order in $k/aH$. Inserting these zeroth-order results for $\delta_1$ and $\delta_2$ in Eqs.~\eqref{eq:interacting_thetaprime_1} and \eqref{eq:interacting_thetaprime_2} leads to $\tilde{\theta}_1 = \tilde{\theta}_2 = \tilde{\theta}_d$ all being equivalent to Eq.~\eqref{eq:non-interacting_theta_d_solution}. This confirms that, even in the presence of energy transfer, the initial velocity perturbation for all fluids is universal for adiabatic perturbations in conformal Newtonian gauge. 

Since the evolution of $\Phi(a)$ on superhorizon scales is completely determined by the dominant species of the Universe, and $\rho_d$ is still non-interacting, it follows that the superhorizon evolution of $\Phi(a)$ at second order in $k/aH$ is still given by Eq.~\eqref{eq:non-interacting_Phi_solution}. This would still be true if $\rho_1$ were the dominant energy density instead of $\rho_d$; even though species 1 is interacting, the evolution of $\rho_1$ is unaffected by this interaction and so its influence on $\Phi(a)$ is equivalent to that of a non-interacting dominant fluid. 

Equipped with $\tilde{\theta}(a)$ and $\Phi(a)$ at second order in $k/aH$, we can derive the adiabatic initial conditions for $\delta_1$, $\delta_2$, and $\delta_d$ at order $(k/aH)^2$. Combining Eqs.~\eqref{eq:non-interacting_theta_d_solution} and \eqref{eq:non-interacting_Phi_solution} with either Eq.~\eqref{eq:interacting_deltaprime_1} or Eq.~\eqref{eq:interacting_deltaprime_d}, and applying Eq.~\eqref{eq:non-interacting_tau_transform} to report in terms of $\tau$, the density perturbations for species 1 is
\begin{align*}
    &\delta_1 \simeq 2\left(\frac{1}{1+w_d}\right)\Phi_p \\ 
    &\hspace{0.5cm}+\frac{2}{3}\left(\frac{1}{1+w_d}\right)\left[ \frac{7 + 39w_d + 63w_d^2 + 27w_d^3 }{28 + 36w_d}   \right] (k\tau)^2 \Phi_p,\stepcounter{equation}\tag{\theequation} \label{eq:interacting_delta1_solution}
\end{align*}
and the density perturbation for the dominant species is
\begin{align*}
    &\delta_d \simeq 2\Phi_p +\frac{2}{3}\left[ \frac{7 + 39w_d + 63w_d^2 + 27w_d^3 }{28 + 36w_d}   \right] (k\tau)^2 \Phi_p.\stepcounter{equation}\tag{\theequation} \label{eq:interacting_deltad_solution}
\end{align*}
Both Eq.~\eqref{eq:interacting_delta1_solution} and Eq.~\eqref{eq:interacting_deltad_solution} are equivalent to Eq.~\eqref{eq:non-interacting_delta_solution_tau} in Appendix \ref{sec:appendix_k2_noninteracting} (since $w_1 = 0$). For species 2, inserting Eqs.~\eqref{eq:non-interacting_theta_d_solution} and \eqref{eq:non-interacting_Phi_solution} into Eq.~\eqref{eq:interacting_deltaprime_2} and applying the transformation Eq.~\eqref{eq:non-interacting_tau_transform} results in 
\begin{equation}
    \delta_2 \simeq \left(\frac{1 - w_d}{1+w_d}\right)\Phi_p + \mathcal{W} (k\tau)^2 \Phi_p, \label{eq:interacting_delta2_solution}
\end{equation}
where $\mathcal{W}$ is a constant equal to 
%
\begin{align*}
    %
     \mathcal{W} &= \frac{1+3w_d}{6 (5 + 6w_2 + 9w_d)(7 + 16w_d + 9w_d^2)}\times \\
     &\hspace{1.5cm} \Big[35 + 162w_d + 225w_d^2 + 90w_d^3  \\
     &\hspace{2.5cm} + 2w_2(28 + 102w_d + 99w_d^2 + 27w_d^3 ) \Big] . \stepcounter{equation}\tag{\theequation}\label{eq:bigW}\\
\end{align*}
%

\vspace{1cm}
\section{Post-decay \texorpdfstring{$\Neff$}{Neff}} \label{sec:appendix_Neff}
Decays in the short-lived regime are primarily constrained by their influence on the CMB anisotropies via changes in $\Delta\Neff$ from injected DR. Here, we derive a mapping between the $\Delta\Neff$ resulting from a decay and the decay parameters $R_\Gamma$ and $\Gamma_Y$. 

\begin{figure}[t]
\centering
  \includegraphics[width=\linewidth]{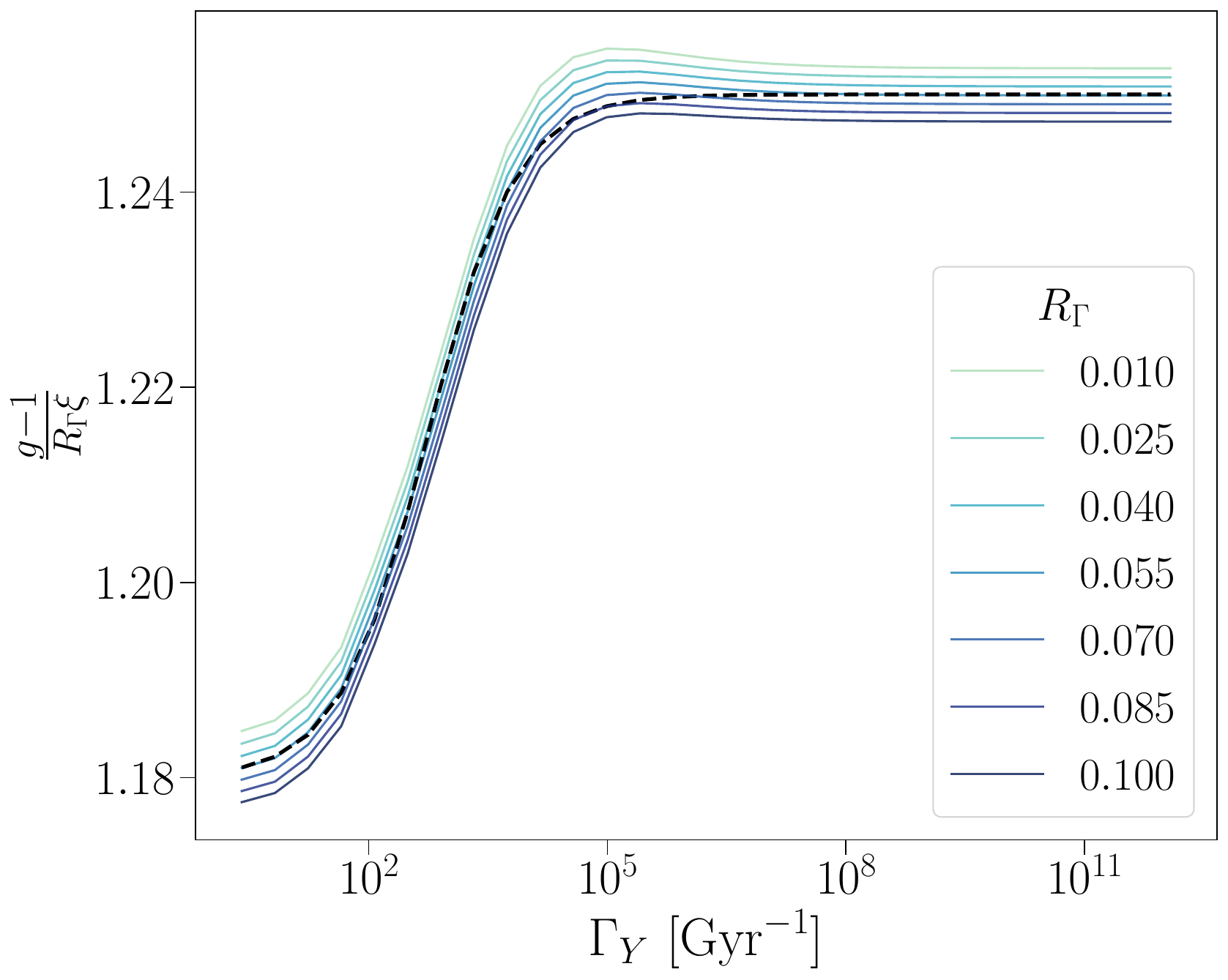}
  \caption{\footnotesize Dependence of Eq.~\eqref{eq:gprime_appendix} as a function of both $R_\Gamma$ and $\Gamma_Y$. Deep in radiation domination (large $\Gamma_Y$), this quantity has no dependence on $\Gamma_Y$ and only a minimal dependence on $R_\Gamma$. The dashed black line shows the numerical fit of Eq.~\eqref{eq:gprime_fit_appendix}.}
  \label{fig:gprime_fit}
\end{figure}

Let us define the ratio of the comoving radiation energy density before and after the decay as
\begin{equation}
    g \equiv \frac{(\rho_{sr,f} + \rho_{\dr,f}) a_f^4}{ \rho_{sr,i} a_i^4},
\end{equation}
where $\rho_{sr}$ is the combined radiation energy density of photons and massless neutrinos, $\rho_\dr$ is the energy density of DR, and the $i$ and $f$ subscripts denote before and after significant production of DR from the decay, respectively. To fit $g$ as a function of $R_\Gamma$ and $\Gamma_Y$, we apply the sudden decay approximation in which all of the $Y$ particle's energy is instantaneously converted to DR at a scale factor of $a_{SD}$. Under this approximation, we have
\begin{equation}
g-1 = \frac{\rho_{sr,f}a_f^4 + \rho_{\dr,f}a_f^4 - \rho_{sr,i}a_i^4 }{\rho_{sr,i}a_i^4 } \approx \frac{\rho_{\dr}(a_{SD})}{\rho_{sr}(a_{SD})}, \\
\end{equation}
where we have assumed that the standard comoving radiation is unchanged (i.e. $\rho_{sr,i} a_i^4 = \rho_{sr,f} a_f^4$ ) and that the decay is instantaneous such that $a_i = a_f = a_{SD}$. Taking $\rho_\dr(a_{SD}) = \rho_Y(a_{SD})$, it follows that 
\begin{equation}
    g-1 \approx \left(\frac{a_{SD}}{a_i} \right)\frac{\rho_{Y,i}}{\rho_{sr,i}} \approx \left(\frac{a_{SD}}{a_i} \right)\frac{\rho_{Y,i}}{\rho_{r,i}} \epsilon^{-1}, \label{eq:g1_sudden_decay}
\end{equation}
where $\epsilon \equiv \rho_{sr,i}/\rho_{r,i}$ and $\rho_{r,i} = \rho_{sr,i} + \rho_{ncdm,i}$. Inserting Eq.~\eqref{eq:rho_Yi} into Eq.~\eqref{eq:g1_sudden_decay} gives
\begin{equation}
\frac{g-1}{R_\Gamma \xi} \approx  \frac{a_{SD}}{a_\Gamma} , \label{eq:gprime_appendix}
\end{equation}
where
\begin{equation}
\xi = \frac{1}{\epsilon} \times
    \begin{cases}
        \left(1 + \frac{a_\Gamma}{a_i}\frac{\rho_{sm,i}}{\rho_{r,i}} \right), & a_\Gamma < a_p\\
        \left[\epsilon + \frac{a_\Gamma}{a_i}\frac{\rho_{sm,i}}{\rho_{r,i}} + (1-\epsilon)\left(\frac{a_i}{a_p}\right)\frac{a_\Gamma}{a_i}\right], & a_\Gamma > a_p\\ 
    \end{cases}.
\end{equation}
Both $a_{SD}$ and $a_\Gamma$ parametrize the scale factor at which $\Gamma_Y \approx H$, so we expect these quantities to be similar in all decay scenarios. The left-hand side of Eq.~\eqref{eq:gprime_appendix} should therefore have a simple functional form that minimally depends on $R_\Gamma$ and $\Gamma_Y$. Figure \ref{fig:gprime_fit} shows $(g-1)/(R_\Gamma \xi)$ as a function of $R_\Gamma$ and $\Gamma_Y$. Here we see a plateau at large $\Gamma_Y$, corresponding to cases in which the $Y$ particle decays deep in radiation domination, and another plateau at smaller $\Gamma_Y$ corresponding to scenarios in which the decay predominately occurs during matter domination. 

The small dependencies that $(g-1)/(R_\Gamma\xi)$ has on $R_\Gamma$ and $\Gamma_Y$ in Fig.~\ref{fig:gprime_fit} are the result of inaccuracies in the sudden decay approximation, which incorrectly assumes the decay to be instantaneous. While the duration of the decay depends on proper time, $t$, the amount that $\rho_Y$ changes while $\rho_\dr$ is being produced is dependent on scale factor. Therefore, the height of the plateaus change depending on if the decay occurs predominately in radiation or matter domination because the mapping between $t$ and $a$ is different for these two regimes. At large $\Gamma_Y$, there is a slight dependence on $R_\Gamma$ because an increase in $R_\Gamma$ leads to a larger contribution of $\rho_Y$ and thus a deviation from radiation domination. Towards smaller $\Gamma_Y$ the dependence on $R_\Gamma$ stems from a deviation from matter domination; an increase in $R_\Gamma$ leads to more DR production and a further deviation from matter domination. 

The dashed black line in Fig.~\ref{fig:gprime_fit} shows a numerical fit described by 
\begin{equation}
    \frac{g-1}{R_\Gamma \xi} = \frac{x + y\left(x^{-1}\Gamma_Y\right)^z}{1 + \left(x^{-1}\Gamma_Y\right)^z}, \label{eq:gprime_fit_appendix}
\end{equation}
for $x = 1.18$, $y = 1.25$, and $z = 0.78$. From this fit, we can determine $g(R_\Gamma, \Gamma_Y)$.
\begin{figure}[t]
\centering
  \includegraphics[width=\linewidth]{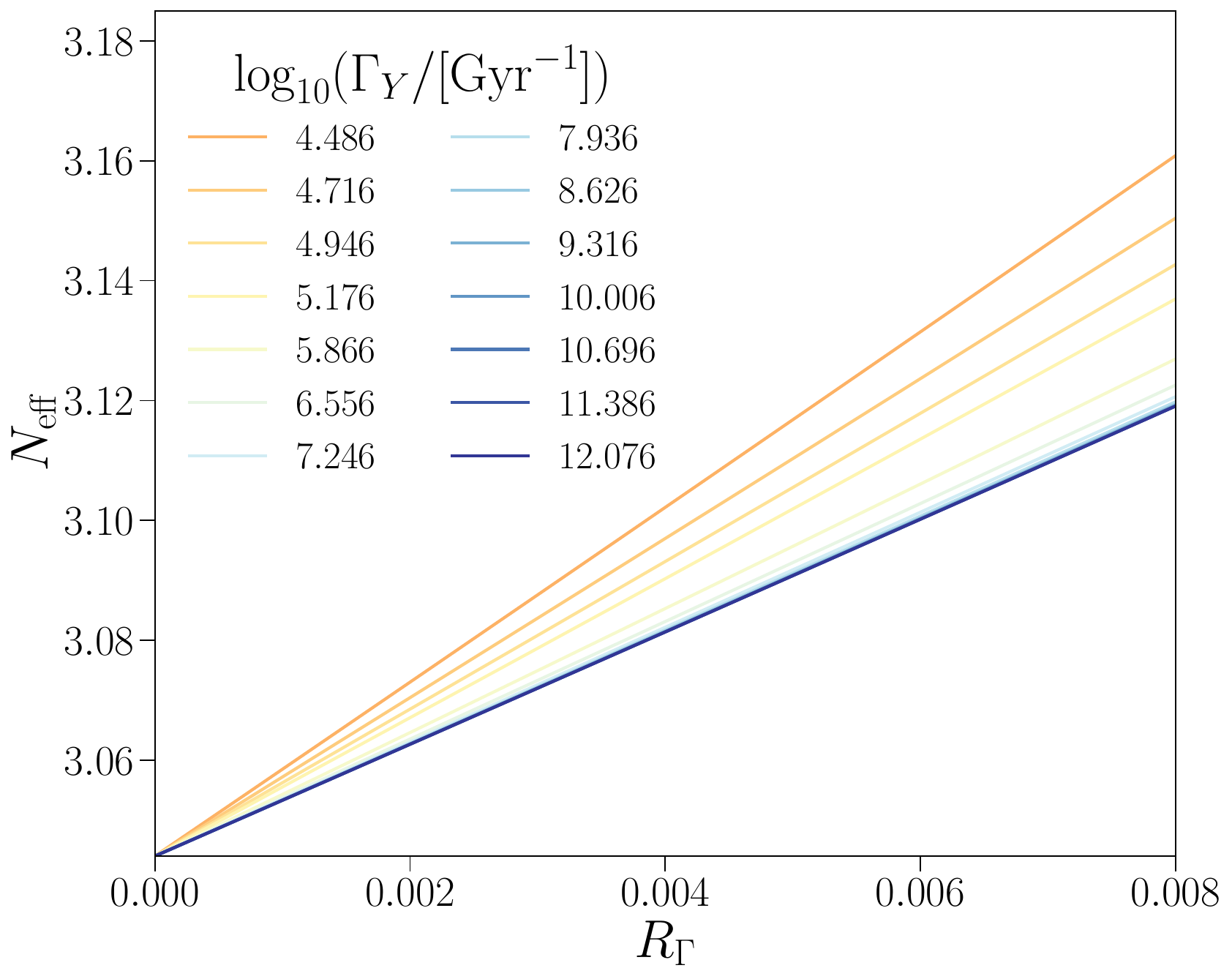}
  \caption{\footnotesize Post-decay $\Neff$ for varying values of $R_\Gamma$ and $\Gamma_Y$. For short-lived cases in which the $Y$ particle decays during radiation domination, $\Delta\Neff$ is simply a function of $R_\Gamma$. Once the energy density of non-relativistic matter begins making a non-negligible contribution to the total energy density, $\Delta\Neff$ also becomes sensitive to $\Gamma_Y$.}
  \label{fig:Neff_appendix}
\end{figure}
If $N_{ur}$ is the effective number of relativistic species before injection, excluding the massive neutrino, and $N'_{ur}$ is the corresponding number after the injection of DR, then it follows that 
\begin{equation}
    g = \frac{(\rho_{sr,f} + \rho_{\dr,f}) a_f^4}{ \rho_{sr,i} a_i^4} = \frac{ \rho_{\gamma,f} a_f^4 \left[ 1 + \frac{7}{8}N'_{ur}\left(\frac{4}{11}\right)^{4/3}  \right]  }{  \rho_{\gamma,i} a_i^4 \left[ 1 + \frac{7}{8}N_{ur}\left(\frac{4}{11}\right)^{4/3}  \right]  }.
\end{equation}
The comoving photon energy density does not change and so this reduces to
\begin{equation}
    N'_{ur} = N_{ur} + \left(g-1\right)\left[\frac{8}{7}\left(\frac{11}{4} \right)^{4/3} + N_{ur}\right]. \label{eq:Nur_prime}
\end{equation}
The contribution that the single massive neutrino species makes to $\Neff$ is $N_{ncdm} = (11/4)^{4/3}(0.71611)^4 = 1.0132$, and we enforce that $\Neff = N_{ur} + N_{ncdm}$ is equal to 3.044 before the injection of any DR. The post-decay $\Neff$ is then determined via Eq.~\eqref{eq:Nur_prime}. Figure \ref{fig:Neff_appendix} shows the post-decay $\Neff$ values for a range of $R_\Gamma$ and $\Gamma_Y$. For a fixed $\Gamma_Y$, a larger $R_\Gamma$ corresponds to larger $\Delta\Neff$, as expected. If a decay scenario occurs deep in radiation domination, then $\Delta\Neff$ is solely a function of $R_\Gamma$. However, as we approach $Y$ particle lifetimes that extend to matter domination, $\Delta\Neff$ picks up an additional dependence on $\Gamma_Y$. This dependence on $\Gamma_Y$ stems from how $R_\Gamma$ is defined. $R_\Gamma$ is a measure of the energy density of the $Y$ particle compared to the total energy density of the Universe at the scale factor of $a_\Gamma$. In other words, $R_\Gamma \sim \rho_Y(a_\Gamma)/\rho_{d}(a_\Gamma) \sim \rho_\dr(a_\Gamma)/\rho_{d}(a_\Gamma)$, where $\rho_{d}$ is the dominant species. Therefore, if $a_\Gamma \ll a_{eq}$, $R_\Gamma \sim \rho_\dr(a_\Gamma)/\rho_{sr}(a_\Gamma)$ and thus $\Delta\Neff$ is solely dependent on $R_\Gamma$. However, if $a_\Gamma > a_{eq}$, then $R_\Gamma \sim \rho_\dr(a_\Gamma)/\rho_{m}(a_\Gamma) \sim [\rho_\dr(a_\Gamma)/\rho_{sr}(a_\Gamma)]\times[\rho_{sr}(a_\Gamma)/\rho_{m}(a_\Gamma)]$ and so $\Delta\Neff \sim R_\Gamma \times [\rho_m(a_\Gamma)/\rho_{sr}(a_\Gamma)] $ for $Y$ particle lifetimes extending into matter domination.

\bibliography{Comprehensive_Constraints_on_Dark_Radiation_Injection_After_BBN}{}

\end{document}